\documentclass[12pt]{report}
\usepackage[francais]{babel}
\usepackage{amsmath}
\usepackage{fancyhdr}
\usepackage{a4wide}
\usepackage{color}
\usepackage[ansinew]{inputenc}
\usepackage{amsfonts}
\usepackage{amssymb}
\usepackage{amstext}
\usepackage{amsthm}
\usepackage{newlfont}
\usepackage{graphicx}
\usepackage{titlesec}
\usepackage[dvipdfm]{hyperref}
\makeatletter
\def\ps@headings{\def\@oddfoot{}%
\def\@oddhead{\makebox[\textwidth][l]{\underline{\hbox to \textwidth{\bf
\firstmark\hfill\thepage}}}}%
\def\@evenfoot{}%
\def\@evenhead{\makebox[\textwidth][r]{\underline{\hbox to \textwidth{\bf
\thepage\hfill\@lhead}}}}%
\def\chaptermark##1{\mark{}\def\@lhead{##1}}}%
\newcommand{\A}{\`{A}\,}

\titleformat{\chapter}[display]
 {\normalfont\Large\filcenter\bf}
 {
   \vspace{1pt}%
   \vspace{1pc}%
   \raggedright\LARGE{\chaptername} \thechapter}
 {1pc}
 {\titlerule
    \vspace{1pc}%
   \Huge}[{\vspace{1pc}\titlerule}]

\begin{document}
\begin{titlepage}
\begin{center}
{ {\bf République Algérienne Démocratique et Populaire}
\vspace{0.1cm}
\\{\bf Ministère de l'Enseignement et de la Recherche Scientifique
\vspace{0.5cm}}
\\{\bf Université A.MIRA de Béja\"{\i}a}\vspace{0.1cm}
\\{\bf Faculté des Sciences et des Sciences de
l'Ingénieur}\vspace{0.1cm}
\\{\bf Département de Physique}\vspace{2cm}
\\
{\Huge{\bf{Mémoire}\vspace{0.5cm}
\begin{center}
     {\large \textbf{Présenté par} Mr. BELABBAS
     Abdelmoumene}\\
     {\large En vue de l'obtention du diplôme de Magistère en physique}
\end{center}
{\large Option: Physique Théorique}\\
\vspace{0.5cm}{\Huge\bf Thème \vspace{0.5cm}
\hrule\vspace{\baselineskip} {\Huge Les Potentiels non
Gravitationnels}\vspace{0.2cm} {\Huge et la Structure de
l'Espace-Temps}\vspace*{0.9cm} }}}}
\hrule\vspace{\baselineskip}\vspace{0.5cm}
\end{center}

\bf {Soutenu publiquement devant le jury suivant:}\vspace{0.1cm}

\begin{center}
\begin{tabular}{lllll}
\hspace{-0.5cm}{\bf Président} & Mr~~BOUDRAHEM.S & Professeur & U.A.M&
Béja\"{\i}a\\
\hspace{-0.5cm}{\bf Rapporteur} & Mr~~BOUDA.A & Professeur & U.A.M &
Béja\"{\i}a\\
\hspace{-0.5cm}{\bf Examinateur} & Mr~~BELKHIR.M.A& Professeur & U.A.M &
Béja\"{\i}a\\
\hspace{-0.5cm}{\bf Invité} & Mr~~MOHAMED~MEZIANE.A & Docteur &
U.A.M &
Béja\"{\i}a\\
\end{tabular}\vspace{0.5cm}
\end{center}
\begin{center}
{\bf Université A.MIRA de Béja\"{\i}a, Décembre 2006}
\end{center}
\end{titlepage}
\pagestyle{empty}

\begin{flushright}
A mes très chers parents et à mon frère $\cdots$ \\
\end{flushright}
\newpage
\noindent\textbf{\Large Résumé}\\\\
\noindent L'objet de ce mémoire consiste en l'analyse d'une proposition récente, avancée par C.C.Barros, selon laquelle les interactions non gravitationnelles peuvent affecter la métrique de l'espace-temps de façon analogue à la gravité. En effet, dans le contexte de la solution de Schwarzschild, l'atome d'hydrogène est décrit de façon tout à fait inédite : au lieu d'adopter la démarche habituelle pour décrire l'électron  sous l'action du potentiel coulombien en utilisant le couplage minimal, l'interaction " proton-électron " est plutôt incorporée dans la métrique. Dans le cadre de cette approche, nous retrouvons dans ce mémoire les équations prévues par la théorie de Dirac à la limite des faibles potentiels. Au terme de notre analyse, nous affirmons que, contrairement à ce qui est avancé par Barros en prétendant apporter une correction insignifiante aux niveaux de l'électron, cette nouvelle approche a le mérite de reproduire exactement le spectre relativiste de la théorie de Dirac. Ces résultats spectaculaires nous interpellent et nous incitent à nous poser des questions sur le rôle du Principe d'Equivalence dans les fondements de la théorie de la Relativité Générale.

\vspace{3cm}
\noindent\textbf{\Large Abstract}\\\\
\noindent The subject of this dissertation consists in analyzing a recent proposition, advanced by C.C.Barros, in which the non gravitational interactions can affect the space-time metric as in gravity. In fact, in the context of the Schwarzschild solution, the hydrogen atom is described in a completely new way : instead of following the usual approach to describe the electron under the Coulomb potential by using the minimal coupling, the "proton-electron" interaction is rather incorporated in the metric. In this context, we reproduce in this dissertation the same equations as predicted in Dirac theory for the weak potential approximation. Contrary to the statement made by Barros, claiming that he brought an insignificant correction to the electron levels, at the end of our analysis, we assert that this new approach has the merit of reproducing the relativistic spectrum as known in the Dirac theory. These spectacular results incite us to wonder about the role of the Principle of Equivalence in the foundations of the general theory of relativity.

\newpage
\pagestyle{empty} {\Huge \textbf{Remerciements}}

\vspace{3cm} Je tiens à exprimer ma gratitude et ma sincère
reconnaissance à Mr Bouda, professeur à l'université A.MIRA, pour
sa confiance et pour l'honneur qu'il m'a fait en acceptant de
diriger ce travail. Mr Bouda a non seulement le mérite d'avoir
assuré une grande partie de notre formation, mais il nous a aussi
fait découvrir un domaine de recherche des plus passionnants: la
mécanique quantique déterministe; c'est pourquoi
je ne le remercierais jamais assez pour tout ce qu'il a fait.\\
\\
\indent Je remercie Mr BOUDRAHEM.S, Professeur à l'université A.MIRA, pour l'honneur
qu'il me fait en acceptant de présider le jury de soutenance. Qu'il veuille agréer
l'expression
de ma haute considération.\\
\\
\indent Que Mr BELKHIR.M.A, Professeur à l'université A.MIRA, puisse trouver ici
l'expression de ma profonde gratitude pour l'honneur qu'il me fait, en acceptant de
juger ce
travail.\\
\\
\indent Je remercie vivement Mr MOHAMED~MEZIANE.A, Chargé de Cours à l'université
A.MIRA, pour ses encouragements, sa disponibilité et pour l'intérêt qu'il a porté à ce
travail. Qu'il puisse agréer l'expression de ma reconnaissance.\\
\\
\indent Un grand merci à tous les enseignants qui ont assuré ma
formation au sein du département de physique ainsi qu'à tous mes condisciples.

\tableofcontents
\newpage
\pagestyle{fancy} \lhead{chapitre\;1}\rhead{Introduction}
\chapter{Introduction}
La  théorie  de  la  relativité générale et la mécanique quantique sont
incontestablement  les  piliers  les  plus  importants de la physique               du
20eme              siècle.

\A  la  fin  du  19eme  siècle  régnait  une telle satisfaction
des avancées  de  la physique, que la plupart des éminents
physiciens pensaient  que  l'édifice  de la physique était presque
achevé, et qu'il  ne restait que quelques petits problèmes, qui
devaient être rapidement
surmontés.

Cependant,  ce  grand  climat  de satisfaction n'a pas empêché des esprits critiques,
tel Einstein et Planck, de déceler deux sérieux problèmes  qui consistent, d'une part,
en l'expérience négative de Michelson  et  Morley, et d'autre part, au profond
désaccord entre l'expérience  et  la  courbe  théorique  de Rayleigh-Jeans dans le
domaine  ultraviolet.  Le mérite de ces grands physiciens consiste d'abord  en  la
constatation  de  la  faillite  de la physique de l'époque  face  à ces deux
problèmes, et ensuite en la proposition de  solutions  inédites qui vont donner par la
suite naissance aux théories  de  la relativité générale et de la mécanique quantique.

La     relativité    a    connu    deux    grandes    évolutions.

La première est la relativité restreinte \cite{relativrestreinte}, qui   est
applicable  à  des  référentiels  d'inertie  animés  de mouvements rectilignes et
uniformes les uns par rapport aux autres ;  c'est  une  théorie qui repose sur deux
postulats \cite{garov}, énoncés       par       A.       Einstein      en      1905 :
\begin{itemize}
    \item L'invariance de la vitesse de la lumière dans tous les référentiels d'inertie.
    \item La covariance des lois de la nature au passage d'un référentiel d'inertie à un autre.
\end{itemize}

La théorie de la gravitation de Newton ne répond pas aux exigences
de  la  relativité restreinte. En effet, la formulation de Newton
repose  sur  une  hypothèse  implicite, selon laquelle l'influence
gravitationnelle   qui   s'exercerait   entre   deux masses est
instantanée  \cite{boratav}.  Ceci  est en contradiction avec l'un
des principes fondamentaux  de  la  relativité restreinte, où on
considère  la  vitesse de la lumière comme la vitesse limite de la
propagation   d'une   information quelconque \cite{garov}.  Par
contre,  la théorie électromagnétique de Maxwell était en parfait
accord  avec  la relativité restreinte. L'analogie frappante entre
la  loi de  Newton et la loi de Coulomb a orienté les physiciens à
tenter   de   rendre conforme  la gravitation  à  la  relativité
restreinte,  à  travers l'ajout d'un nouveau champ qui jouerait un
rôle  analogue  au  champ  magnétique  dans  la théorie de Maxwell
\cite{boratav}. Toutes les tentatives, proposées dans ce sens, ont
échoué. Einstein a compris qu'il ne fallait pas chercher à rendre
conforme  la théorie de la gravitation newtonienne à la relativité
restreinte,  mais qu'il fallait plutôt généraliser la relativité à
tous  les  référentiels,  quels que soient leurs états de
mouvements. Ainsi la relativité restreinte serait un cas
particulier de cette nouvelle   théorie, applicable   dans   le
cas  où  le  champ gravitationnel est     très     faible, voir
inexistant.

La  seconde étape est donc la relativité générale. C'est une œuvre
exclusive d'Einstein, se basant sur un postulat fondamental selon
lequel : “Toutes les lois de la nature prennent la même forme dans
tous les référentiels, quels que soient leurs états de mouvement”.
La relativité  générale  décrit la gravité dans un cadre
géométrique, en reliant le champ gravitationnel à la courbure de
l'espace-temps .  Cette  relation est une conséquence du postulat
d'équivalence énoncé  par  Einstein en interprétant convenablement
l'égalité des masses           gravitationnelle et d'inertie.

Alors  qu'on disposait d'une description théorique cohérente de la
structure  du monde à grande échelle, il fallait en faire de même
pour  explorer  l'infiniment petit.  La période qui s'étend entre
1920 et 1933 marque l'une des plus belles conquêtes de la physique
moderne   :   la   mécanique   quantique   fait   son apparition
\cite{histoire}.  C'est  une  discipline  qui  vise  à explorer la
matière jusqu'à  son  niveau  le  plus  infime.  \A cette échelle,
l'intuition  humaine,  basée sur  une  expérience  journalière  à
l'échelle  macroscopique,  est  désormais inopérante  ;  ce qui a
conduit  nécessairement la mécanique quantique à s'appuyer sur des
principes         inédits,        comme        par        exemple:
\begin{itemize}
    \item       La       quantification       de      l'énergie,
    \item          La          dualité          onde-corpuscule,
    \item     Le     principe    d'incertitude    d'Heisenberg,
    \item     Le     principe     de     correspondance     ...
\end{itemize}

L'élaboration de la mécanique quantique a connu des épisodes très
mouvementés   et  a vu  l'émergence  de  deux  grandes  tendances
d'interprétations.

L'interprétation de l'école de Copenhague. Les partisans de cette
école   prônent une description   probabiliste   et  renoncent
définitivement à une description causale des processus à l'échelle
quantique  \cite{nielsbohr}.  Au  lieu  de  cela, tout  état d'un
corpuscule  ou  d'un  système  doit  pouvoir, à tout instant, être
représenté par  une  fonction  d'onde,  de  carré  sommable, dont
l'évolution  est déterminée par  l'équation  de Schrödinger ; le
carré  du  module de cette fonction d'onde est interprété comme la
probabilité  pour  qu'une  observation  permette  de localiser le
corpuscule  en  question  \cite{broglie}.  On introduit dans cette
théorie  un formalisme où les variables cinématiques et dynamiques
de  la mécanique classique sont remplacées par des symboles soumis
à  une  algèbre  non  commutative \cite{nielsbohr}. Désormais, des
observations faites simultanément, au cours de la même expérience,
ne peuvent  jamais  nous  permettre  d'avoir  sur  les  grandeurs
canoniquement conjuguées, liées à un corpuscule, des connaissances
plus  précises  que ne  le permettent les inégalités d'Heisenberg
\cite{broglie}.  L'origine  de  cette limitation fondamentale des
connaissances  accessibles  par  expérience  n'est  pas due à une
limitation  de  précision des appareils de mesure, mais trouve son
origine profonde  dans  l'existence  même du quantum d'action. Le
processus  de  mesure acquiert  une place très importante dans la
théorie  quantique,  du  fait  de l'impossibilité d'effectuer une
mesure  sans  la  provocation  d'une perturbation incontrôlable du
système considéré \cite{nielsbohr}; une indétermination intervient
au moment de la mesure de sorte que son issue n'est pas certaine.
Désormais,  on  ne peut qu'évaluer  les différentes probabilités
susceptibles    d'être    prises    par la grandeur   mesurée.

Cette   démarche   de   prévision   a   été   pleinement  vérifiée expérimentalement,
par contre les fondements philosophiques d'une telle   version   sont  contestés  par
certains  physiciens,  tel Einstein,  De  Broglie…,  notamment  le  fait  de  définir
l'état quantique   uniquement   sur   les  informations  accessibles  par
l'expérience.  Ceci  a  conduit  à nier l'existence d'une grandeur avant  sa  mesure
: c'est la mesure qui créé la valeur ; ainsi il est  impossible  d'attribuer une
trajectoire à une particule. Elle ne  peut  avoir  que  des  localisations  isolées
sans  positions intermédiaires \cite{broglie}.

Les   partisans   d'une  interprétation  déterministe  tentent  de reformuler  la
version  actuelle de la mécanique quantique qui ne cherche  plus  à  décrire les
faits, mais seulement à les prévoir. Einstein  en  particulier  a affirmé plusieurs
fois que la version actuelle  de  la  théorie,  qui  est  parfaitement exacte dans ses
prévisions,    ne    serait   que   l'aspect   statistique   d'une représentation
plus  profonde  qui  rétablirait l'existence d'une réalité   objective
\cite{broglie}   ;  réalité  qui  doit  être indépendante   de  nos  connaissances.
Les  tentatives  les  plus significatives  visant à la construction d'une mécanique
quantique déterministe                         sont                        :
\begin{enumerate}
    \item La théorie de l'onde pilote \cite{broglie}, dans laquelle L. de Broglie émet
    l'hypothèse  qu'un  corpuscule est astreint à suivre l'une des lignes de courant de
    Madelung, correspondant à la propagation des fonctions d'ondes de
la  mécanique  quantique  selon une représentation hydrodynamique. L'argument  majeur
contre cette théorie est la volonté d'aboutir à une description causale du mouvement
d'un corpuscule en le faisant dépendre  d'une  onde  qui  se  propage dans un espace
abstrait de configuration  et non dans un espace physique ; onde qui subit, de plus,
une  profonde  modification de structure lors d'une mesure.
    \item La théorie de Bohm\cite{unBohm,deuxBohm}, qui reprend la forme de
    la fonction d'onde de la théorie précédente, en la remplaçant dans l'équation
    de Schrödinger, ce qui conduit d'une part à une équation de continuité du fluide
    de probabilité, et d'autre part à une équation dite d'Hamilton-Jacobi quantique
    qui ne diffère de l'équation classique que par un nouveau terme: le potentiel quantique
    qui s'ajoute au potentiel classique et qui expliquerait les effets quantiques. Cette
    théorie présente un problème dans le cas des fonctions d'ondes réelles, pour
    lesquelles on prévoit des vitesses nulles pour les particules.
    \item L'approche de Floyd \cite{floyd,unfloyd,deuxfloyd,mathonefloyd,mathdeuxfloyd},
    développée dans un espace à 1 dimension et dans un régime stationnaire, où il propose
    de maintenir la forme des fonctions d'ondes complexes et d'utiliser une nouvelle forme,
    trigonométrique, pour les fonctions d'ondes réelles. Il retrouve l'équation
    d'Hamilton-Jacobi quantique, qui est une équation différentielle non linéaire d'ordre 3,
    pour laquelle il propose une solution construite par deux solutions réelles et
    indépendantes de l'équation de Schrödinger correspondante. De plus, il propose
    l'utilisation de la procédure classique, basée sur le théorème de Jacobi, pour
    retrouver les trajectoires admissibles (trajectoires de Floyd). Dans le cas des
    états liés, il prévoit l'existence d'états possibles, appelés micro-états, que la
    fonction d'onde ne détecte pas. Il conclut ainsi que la fonction d'onde de Schrödinger
    ne  décrit  pas  de  façon  exhaustive  les systèmes physiques.
    L'approche de Floyd comporte certaines insuffisances, telles l'utilisation de formes
    différentes de fonctions d'ondes selon les cas, l'utilisation du théorème
    de Jacobi, utilisé classiquement pour une équation d'Hamilton-Jacobi d'ordre 1, et
    surtout la dépendance des trajectoires du choix mathématique des solutions de
    l'équation de Schrödinger lors de la résolution de l'équation d'Hamilton-Jacobi
    quantique.
    \item   Le   postulat   d'équivalence  de  Faraggi  et  Matone
    \cite{unfarggi,deuxfarggi,resumefarggi}, qui ressemble étrangement au postulat
    fondamental de relativité générale, stipule qu'il est toujours possible de connecter
    deux états quantiques différents par des transformations de coordonnées.
    Dans le cadre de ce postulat, on retrouve l'équation d'Hamilton-Jacobi quantique,
    mais avec une forme unifiée de la fonction d'onde. On retrouve aussi la quantification
    de l'énergie et on explique l'effet tunnel sans passer par une interprétation
    probabiliste, ce qui rend cette piste l'une des plus prometteuses dans l'effort
    d'une interprétation déterministe de la mécanique quantique. Farragi et Matone
    maintiennent toujours les trajectoires de Floyd, obtenues par une utilisation
    a priori non justifiée du théorème de Jacobi. De nouvelles
propositions d'obtention de
    trajectoires  quantiques  ont  été  avancées à l'université de
    Béjaia \cite{boudaprob,boudalagrang,boudareply,boudarelativ,boudanewton,boudatrajectory}.
\end{enumerate}

La  physique  actuelle  connaît  une course vers l'unification des
quatre interactions  fondamentales. Le problème majeur qui bloque
ce  grand projet est la réconciliation de la relativité générale
qui décrit la gravité en terme géométrique de la structure de
l'espace-temps, et   de  la  mécanique  quantique,  qui  décrit
les interactions: électromagnétique,  forte et faible. Une étape
intermédiaire a été effectuée,  dans  ce sens, par Dirac en
formulant sa théorie basée sur  l'espace-temps plat  de la
relativité restreinte. Avec cette formulation,  le  spin et les
antiparticules apparaissent de façon naturelle  dans  la  théorie.
Par  contre cette théorie n'est pas valable  dans  le  cas  où
règne un champ gravitationnel intense. Plusieurs  auteurs ont
essayé de quantifier la gravité sans succès du  fait  de
l'apparition  de problèmes insurmontables, notamment
l'impossibilité   de   réduire les   infinités   (théories   non
renormalisables).

Les raisons profondes de cet échec trouvent leurs origines dans le fait   que  ces
deux  théories  s'appuient  sur  des  conceptions philosophique  très  différentes  ;
la  première  est une théorie déterministe,   alors   que   la   seconde   est
essentiellement probabiliste,   d'où  l'intérêt  de  l'élaboration  d'une  théorie
déterministe        de        la        mécanique        quantique.

 Le  sujet  de  ce mémoire consiste en l'analyse d'une proposition
 récente, avancée par C.C.Barros \cite{barros,barrosbis,barrosquark,barrosstability}, qui
 stipule que, d'une façon analogue à l'interaction gravitationnelle, les autres
 interactions peuvent aussi se manifester à travers les propriétés
de  l'espace-temps.  L'idée  fondamentale  consiste  à décrire une particule  dans
une  région plongée dans un potentiel (non gravitationnel), à symétrie sphérique, qui
affecte la métrique de l'espace-temps et les effets de  la  métrique  dans  ce
domaine subatomique seront étudiés.

Ce        mémoire       est       organisé       comme       suit:
\begin{itemize}
    \item Le second chapitre est un rappel sur les idées de base de la
théorie         de         la         relativité         générale.
    \item Le troisième chapitre est consacré à la description d'une
particule évoluant dans un espace courbe et soumise à un potentiel
affectant  la métrique  de  Schwarzschild.  Dans un premier temps,
l'analyse  vectorielle en coordonnées curvilignes va nous
permettre d'étendre  le  principe  de  correspondance à un espace
courbe. En second  lieu,  on  détermine  les principales grandeurs
dynamiques relatives  à  un  corpuscule,  dans la métrique de
Schwarzschild, à savoir: l'énergie,  les  impulsions, l'expression
de l'invariant relativiste, et surtout on établit un lien entre le
potentiel (non  gravitationnel)  et la structure de l'espace-temps
à travers l'expression du coefficient de structure $\xi(r)$ . La
combinaison de  tous  ces  résultats va  permettre  de définir les
opérateurs $(E,\overrightarrow{p},\overrightarrow{p}^{2})$,
indispensables à l'élaboration d'équations  d'ondes  quantiques
dans  un  espace courbe.
    \item   Le   quatrième  chapitre  consiste en  la  proposition
    d'équations d'ondes quantiques pour des particules de spin 0 et $1/2$, dans
    un espace courbe répondant à une métrique de Schwarzschild. De
    plus, une méthode de résolution de ces équations, basée sur la
    séparation des variables, au régime stationnaire est développée.
    \item Dans le cinquième chapitre, une application à l'atome d'hydrogène est
    envisagée. Cette application va permettre d'une part d'illustrer la théorie,
    et d'autre  part de vérifier sa  validité, à travers une comparaison de ces
    implications avec les résultats bien connus en mécanique quantique habituelle.
    \item Le dernier chapitre est consacré à la discussion des résultats.
\end{itemize}

\newpage
\pagestyle{fancy} \lhead{chapitre\;2}\rhead{Rappels sur la théorie de relativité
générale}

\chapter{Rappels sur la théorie de relativité générale}
Le présent chapitre est un rappel sur la théorie de la relativité générale. Il ne sera
question, ici, que du contenu physique de la théorie, et aucun développement du cadre
mathématique n'est
envisagé. L'attention sera portée sur les points suivants:\\

\begin{itemize}
    \item La nécessité d'étendre le principe de relativité à tous
    les référentiels, quels que soient leurs états de mouvements.
    \item Les postulats d'équivalence et de relativité générale.
    \item La nécessité d'un cadre non-euclidien de l'espace-temps.
\end{itemize}

\section{Difficultés de définir un référentiel d'inertie}

Le  principe  de relativité restreinte permet d'affirmer que les lois  générales de la
nature ont exactement la même forme, pour tout  référentiel en mouvement relatif
uniforme par rapport à un référentiel d'inertie.

Il est évident que ce principe s'appuie sur une existence
préalable  d'un référentiel d'inertie. La discussion qui suit,
soulevée par Einstein \cite{einsteinenfeld}, permet de montrer que
la définition d'un tel référentiel d'inertie n'est pas évidente.
Pour ce faire, examinons comment est défini ce référentiel en
mécanique classique.

\newtheorem{theoreme}{Définition}[chapter]
 \begin{theoreme}
Un référentiel d'inertie est un système de coordonnées dans lequel un corps qui ne
subit aucune force extérieure, se meut de façon rectiligne et uniforme.
 \end{theoreme}


Que  signifie  l'expression : “ne subit aucune force extérieure”?

Un corps qui n'est pas soumis à une force extérieure ne subit
aucune  accélération (loi fondamentale de dynamique),  c'est à
dire que sa vitesse ne varie pas au cours du  temps. Donc
l'expression entre guillemets signifie tout simplement que  le
corps se meut uniformément dans le système de coordonnées. Il est
clair que la définition précédente comporte une ambiguïté
implicite, due à une équivalence entre la condition et la
conséquence de la proposition.

Comment éliminer toutes les influences extérieures sur notre corps?

Ce cas de figure pourrait s'obtenir, dans la pratique, si nous
pouvons nous éloigner assez de tous les corps matériels pour nous
libérer de toute interaction. Néanmoins, cette procédure ne
permet" que de réduire ces influences extérieures, sans pour
autant les éliminer complètement. Pour une élimination totale des
ces influences, il faut éliminer tous les corps, de sorte qu'il ne
reste dans l'univers qu'un seul corps, formant notre système de
coordonnées. En résolvant le problème d'élimination des forces
extérieures, on crée un autre problème, car parler du mouvement
d'un seul corps dans l'univers est une aberration, vu le caractère
relatif de la notion de mouvement.

Cette petite discussion a révélé l'existence d'un certain nombre de difficultés
d'ordre pratique pour définir un référentiel d'inertie, susceptible de servir comme
cadre pour des lois physiques décrivant la nature.

\section{Les référentiels accélérés par rapport à un référentiel d'inertie}
Attirons  l'attention  sur  un  autre  problème  qui  a interpellé Einstein  et  qui
se  résume  comme  suit.

Pour la description physique  des  événements  de  la  nature, en relativité
restreinte, aucun référentiel se déplaçant uniformément par rapport à un référentiel
inertiel ne se distingue  de l'autre, ils sont tous valables et équivalents       pour
servir de cadre à l'établissement des lois de la physique.

La validité du principe de relativité n'était admise que pour
cette classe restreinte de référentiels. Mais qu'en est-il des
référentiels accélérés  ou dotés de mouvements quelconques   par
rapport   à   un   référentiel  d'inertie  ? Est-il possible de
formuler les lois de la physique de telle manière  qu'elles soient
valables pour tous les systèmes  de coordonnées,   sans aucune
exception quant à leurs mouvements relatifs ?

\section{Référentiels accélérés et mouvement absolu}
Les corps accélérés acquièrent des propriétés mécaniques différentes des corps  non
accélérés, ceci est dû à l'apparition des  forces à caractère inertiel. Illustrons ce
propos par un exemple \cite{coudec}.

Tant  qu'un  passager  est  transporté  dans  un bus effectuant un mouvement
uniforme,  il  ne  remarque  rien. Il peut tout aussi affirmer que le bus est au repos
alors que des immeubles, situés au bord  de  la route, sont en mouvement, comme il
peut affirmer que le bus  est  en mouvement alors que les immeubles sont au repos.
Mais dès  que  le conducteur freine, le passager se trouve projeté vers l'avant, ce
qui le pousse à attribuer au mouvement non uniforme une espèce  de  réalité  absolue.
Il affirmera  qu'il est en mouvement   alors   que   les   immeubles sont  au repos.

On est obligé d'admettre que le cours des événements n'aurait pas été  le même si le
bus avait poursuivi uniformément sa route. Le fait que le  système (bus) perde son
caractère galiléen, suite à une accélération,  n'échappe  pas au passager. C'est
pourquoi, il paraît  impossible  que  les deux systèmes, uniforme et accéléré, soient
équivalents pour décrire les lois de la mécanique.

Est-il possible de construire une physique réellement relativiste,
valable  pour tous les systèmes de coordonnées, une physique où il
n'y  aurait  plus de place  pour le mouvement absolu, mais
seulement pour          le          mouvement relatif ?

La  théorie  de  la relativité générale se propose de formuler les lois  de  la
physique pour n'importe quel système de coordonnées, contrairement  à  la relativité
restreinte qui n'admet pour cadre, afin d'exprimer les lois décrivant la nature, que
les référentiels inertiels.

Les  lois  physiques  valables  pour  n'importe  quel système  de coordonnées  doivent
se  réduire,  dans le cas spécial de système inertiel, aux lois déjà formulées en
relativité restreinte. Ainsi la théorie  de la relativité générale doit, en principe,
englober la théorie       de      la relativité restreinte.
\section{Identité des masses inertielle et gravitationnelle}

\subsection{Masse d'inertie}
C'est une grandeur qui exprime la capacité d'un corps à résister à toute  modification
de  son  état de mouvement, dans lequel il se trouve  ;  elle  nous renseigne sur la
plus ou moins facilité avec laquelle  un  corps  répond  à  l'appel  d'une force
extérieure.

En soumettant, par exemple, deux corps différents, initialement au repos,  à  la même
force extérieure, le corps le plus massif va manifester  une  plus  grande  opposition
à la modification de son état dynamique (repos) et va acquérir une vitesse moins
importante que        le        corps        de        petite        masse.

La  masse  d'inertie intervient dans la $2^{eme}$ lois de Newton :
\begin{equation}\label{loinewton}
    \overrightarrow{F}=m\,\overrightarrow{a}.
\end{equation}
\subsection{Masse                                gravitationnelle}
C'est  une grandeur qui exprime la capacité d'un corps à interagir avec  un  autre
lorsqu'ils  sont  séparés par une distance ; elle
figure    dans    la   loi   de   Newton   de   gravitation   :\\
\begin{equation}\label{gravitnewton}
    \overrightarrow{F}=-G\,\displaystyle\frac{m_{g}\,m_{g}^{'}}{r^{2}}\,\overrightarrow{e_{r}}.
\end{equation}
\subsection{Incompatibilité de la gravitation de Newton et de la relativité restreinte}
Remarquons  qu'une  description  de l'interaction gravitationnelle
avec la loi de Newton présente un sérieux problème vis-à-vis de la
relativité restreinte. En effet, supposons que nos deux masses
soient séparées par une très grande distance, soit des milliers
d'années-lumière, et que sous l'effet d'une quelconque
perturbation extérieure, une des deux masses soit déplacée  ;
alors la  loi de Newton nous dit que la seconde masse  "  ressent
" instantanément la nouvelle force de gravitation,  due à la
nouvelle configuration des deux masses.

La  loi  de  Newton  est basée  sur  une hypothèse, implicitement admise, qui stipule
que la vitesse de propagation de l'interaction gravitationnelle  est infinie, ce qui
est en contradiction avec l'un des piliers incontournables de la théorie de la
relativité restreinte  qui dit qu'une information ne peut voyager plus vite que
la vitesse        de        la        lumière.
\subsection{Champ de gravitation}
Une  autre  façon de décrire l'interaction gravitationnelle est de
regarder  la  masse comme  source  d'un  champ  de  gravitation,
celle-ci  engendre  dans  son  voisinage immédiat un champ qui se
propage    à    une    vitesse   finie   dans   tout l'espace.
N'importe   quelle   masse  qui  se  trouverait  dans  le  domaine
d'influence   de   la   source,   “ressentirait”  une  force
gravitationnelle proportionnelle à sa masse et à la valeur du
champ en                          ce point.

Contrairement  aux  champs électrique et magnétique, les champs de
gravitation   jouissent   d'une   propriété   fondamentale \cite{relativrestreinte}:\\

 \textbf{“Tous les corps se déplaçant uniquement sous
l'influence d'un champ   de  gravitation,  sont  soumis  à  la même accélération,
indépendamment  de  leurs masses et de leurs
états physiques”}\\

Cette  propriété  remarquable  a été établie expérimentalement par Galilée qui, du
haut de la tour de Pise, laissait tomber des corps de  masses  et  de  natures
différentes (suffisamment denses pour éliminer     les     effets     de
résistance de l'air).

\subsection{Identité des masses inertielle et gravitationnelle}
Newton  était déjà conscient de la différence des deux définitions de  la  masse  d'un
corps. Conformément à la loi de mouvement d'un corps  en  chute  libre  dans  le champ
de gravitation terrestre :
\begin{equation}
    (\textrm {force})=(\textrm{masse
    inerte})\times(\textrm{accélération}),
\end{equation}
et  d'autre  part, la loi de gravitation permet d'exprimer la même force   qui
s'exerce   sur   le   corps   en   chute   libre   :
\begin{equation}
    (\textrm {force})=(\textrm {masse pesante})\times(\textrm {intensité du champ de
    gravitation}),
\end{equation}
ce qui permet d'écrire, après égalisation :
\begin{equation}
    (\textrm {accélération})=\Big(\displaystyle\frac{\textrm {masse pesante}}{\textrm {masse inerte}}\Big)\times(\textrm {intensité du champ de
    gravitation}).
\end{equation}

Mais comme l'accélération des corps soumis au champ de gravitation est  la  même,
conformément à la propriété énoncée en (2.4.4), il faut admettre que le rapport de la
masse pesante à la  masse inerte est une constante égale pour tous les corps. En
choisissant convenablement les unités, il est possible de rendre le rapport égal à
l'unité.

Déjà  en  Mécanique classique, on a exprimé le principe d'identité des  masses  inerte
et  pesante,  sans pour autant lui donner une interprétation  profonde, par  contre,
en théorie de la relativité générale,   il   forme   la   base   de   toute
l'argumentation.

\section{Principe d'équivalence}
Ce fut Einstein qui, dans son effort de généralisation du principe de  relativité  à
tous  les  systèmes de coordonnées, s'est rendu compte  de l'importance de l'égalité
des masses inerte et pesante. En effet, l'interprétation d'une telle égalité est
l'argument crucial qui va lui permettre de construire sa théorie de la relativité
générale.

Profondément influencé par la pensée de Mach, qui prône l'adoption d'un   point  de
vue  positiviste  pour  étudier  les  phénomènes physiques (ne considérer une grandeur
physique dans sa théorie que s'il  est  possible  de  définir  une  expérience
permettant de la mesurer),  et  particulièrement  par  son  étude  critique  de  la
mécanique     classique     où     il     affirme     que \cite{boratav}:\\

\textbf{“La  masse inertielle d'un objet est due à l'interaction
gravitationnelle  de cet objet avec les autres masses de l'univers”}\\

Einstein  reconnaît  le  fait  que  la  même qualité d'un corps se manifeste  suivant
les circonstances, comme “inertie” ou comme “poids”.

\A travers   une   analyse   très  poussée,  par  l'intermédiaire
d'expériences  de pensée  mémorables, il a su montrer qu'il était
possible  d'édifier  une  théorie où le principe de relativité est
valable  pour  tous  les  référentiels  en  mouvement relatif. Une
telle théorie a été réalisée en reliant la gravitation à la
géométrie de l'espace-temps.

\subsection{Expérience de pensée (connexion Relativité - Gravitation) }
Supposons  une  boîte qui se trouve dans une région si éloignée de
toute   masse importante  qu'on  pourrait  négliger  toutes  les
influences extérieures. \A l'intérieur de cette boite se trouve un
homme muni d'appareils de mesures. Un système de référence affecté
à   une   telle   boite   est   un   référentiel d'inertie
\cite{relativrestreinte}.

En  soumettant  la  boite  à une force constante, elle sera animée
d'un  mouvement uniformément accéléré (par  rapport  à un autre
référentiel d'inertie). Pour l'observateur  interne,
l'accélération  de la boite lui est transmise   par
l'intermédiaire   du   plancher sous  forme  de contre-pression
qu'il  pourrait absorber en se mettant debout sur ses jambes,
exactement comme le ferait un homme à la surface de la terre,   où
il   est  soumis  à  un  champ  de  gravitation. Si à présent
l'homme lâche sans vitesse initiale deux corps qu'il tenait à la
main, par exemple  :  une  montre  et  une plume, dans ce cas
l'accélération provoquée par la force constante n'est plus
communiquée aux objets par  l'intermédiaire  de la main, ainsi, le
plancher de la boîte se rapprocherait  des  deux  corps  avec  le
même mouvement relatif accéléré. Comme  les  deux  objets
toucheraient le plancher au même instant (lâchés  au  même
instant), l'observateur interne serait convaincu que
l'accélération  d'un  corps  quelconque vers le plancher est
toujours  la  même, indépendamment  de sa substance. Il serait en
droit  de  supposer  qu'il  se trouve dans un champ de gravitation
constant.

\subsection{Enoncé du Postulat d'équivalence}
\A  la  lumière  de  ce  raisonnement  Einstein  énonça le
principe
d'équivalence:\\

\textbf{“Un  référentiel  uniformément  accéléré  est localement équivalent  à  un
référentiel  inertiel  plongé  dans un champ de
gravitation”}\\

\begin{figure}[\here]
   \centering
  \includegraphics[width=10cm]{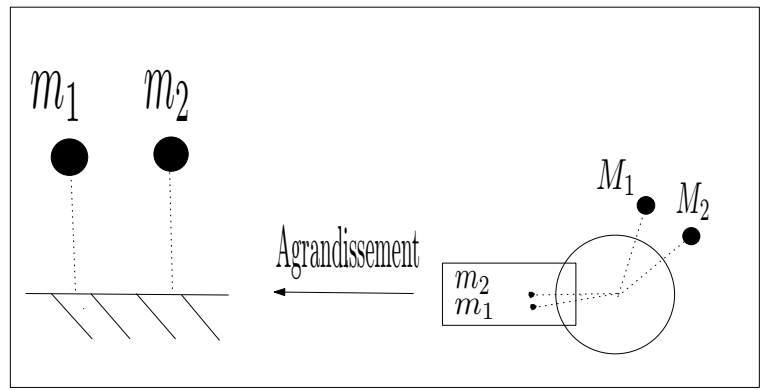}\\
  \caption{Représentation schématique de la validité locale du principe
   d'équivalence}\label{fig1}
\end{figure}

Soulignons que cette équivalence est locale, dans la mesure où les trajectoires  de
deux corps convergeant vers la source du champ de gravitation  peuvent  être
considérées comme parallèles dans une petite région de l'espace, alors que deux objets
dans un référentiel uniformément accéléré auraient des trajectoires parallèles  ( voir
\ref{fig1}).

Il  est  à  noter  qu'une  telle  équivalence  entre référentiels accélérés   et
référentiels  inertiels  soumis  à un  champ  de gravitation  repose  sur  la
propriété fondamentale du champ de gravitation  à  communiquer  la même accélération à
tous les corps, ce  qui revient à dire qu'elle repose sur le principe d'égalité des
masses inerte                       et pesante. Grâce   au principe d'égalité  des
masses gravitationnelle  et inertielle, il est possible à un observateur accéléré
d'expliquer les  effets inertiels par l'hypothèse  d'un  champ de gravitation, et  il
lui est possible de supposer que son corps de référence est “immobile”.

Cette propriété de communiquer la même accélération à tous les corps, quelles que
soient leurs substances, est une caractéristique exclusive du champ de gravitation.
Aucun autre champ connu ne pourra être introduit pour expliquer les effets inertiels.

\section{Postulat de la relativité générale}
Soit un observateur se trouvant dans un ascenseur et imaginons que le câble soutenant
celui-ci soit rompu alors, en vertu de la propriété fondamentale relative au champ de
gravitation, l'ascenseur et l'observateur seront soumis à la même accélération, de
telle sorte qu'ils ne subissent aucun mouvement relatif \cite{einsteinenfeld}.

Si l'observateur lâche des corps de sa main, il remarque que ceux-ci ne subissent
aucun mouvement relativement à lui, ils flottent devant avec lui. Il serait en droit
de conclure que le système de coordonnées affecté à son ascenseur est un référentiel
d'inertie.

Un observateur extérieur, se trouvant à la surface de la terre, remarque que le
mouvement de l'ascenseur et de tous les corps s'y trouvant à l'intérieur est en
parfait accord avec la loi de gravitation de Newton. Pour lui le mouvement est
uniformément accéléré.

De cet exemple il est clair qu'une description cohérente des phénomènes physiques dans
deux systèmes de coordonnées dotés d'un mouvement relatif accéléré est possible. Ceci
constitue un argument en faveur du principe de la relativité générale qui
stipule que \cite{relativrestreinte}:\\

\textbf{"Tous les corps de référence, quels que soient leurs états de mouvement, sont
équivalents pour la description des lois de la
nature"}\\

Une description cohérente dans les différents systèmes de coordonnées n'est rendue
possible qu'en tenant compte de la gravitation. Ainsi, pour l'observateur extérieur,
il y a mouvement de l'ascenseur dans un champ de gravitation, alors que pour
l'observateur interne il y a repos et champ de gravitation inexistant.

Pour Einstein, le champ de gravitation constitue une sorte de " pont " permettant le
passage d'un système de coordonnée à un autre \cite{einsteinenfeld}.

Abordons à présent le problème de la connexion entre la théorie de la relativité
générale et la géométrie.

\section{Nécessité d'un cadre non-euclidien de l'espace-temps}
\subsection{Référentiels inertiel et non inertiel}
\begin{figure}[\here]
   \centering
  \includegraphics[width=5cm]{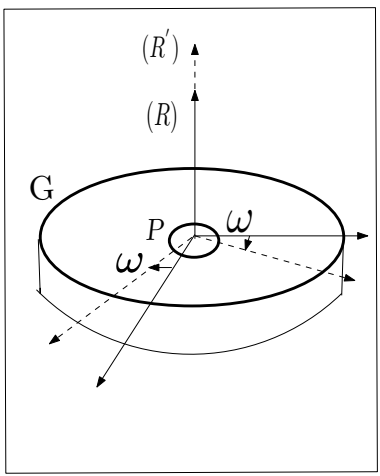}\\
  \caption{Représentation de deux cercles concentriques sur un disque tournant}\label{fig3}
\end{figure}

Imaginons  un grand disque sur lequel sont tracés deux cercles (voir \ref{fig3}), l'un
petit ($P$)  et  l'autre  très grand ($G$), de telle sorte que leurs centres
coïncident sur l'axe  de symétrie du disque \cite{einsteinenfeld}.

Dans un premier temps, supposons que le disque est au repos (il ne subit pas de
rotation). Dans ce cas, le référentiel ($R$) attaché à l'axe de symétrie est galiléen.
Soit un observateur se référant à ce référentiel d'inertie. Il mesure les
circonférences et les rayons de deux cercles identiques aux précédents, en utilisant
une petite règle. Il s'aperçoit que le rapport des circonférences aux diamètres
respectifs   des deux cercles  est :
\begin{equation}
    \displaystyle\frac{C_{G}}{2R_{G}}=\displaystyle\frac{C_{P}}{2R_{P}}=\pi.
\end{equation}

Un  tel résultat est une preuve que le cadre géométrique auquel se réfère
l'observateur utilisant un référentiel inertiel, est la géométrie
euclidienne.

Dans un deuxième temps, supposons que le disque tourne autour de son axe de symétrie
avec une vitesse constante $\omega$. Un référentiel ($R^{'}$) attaché à un observateur
se trouvant sur un tel disque n'est pas un référentiel d'inertie, puisque le mouvement
de rotation donne naissance à des effets inertiels, sous forme de forces centrifuges.
Quel est le cadre géométrique servant de support pour décrire les phénomènes
physiques, dans un référentiel non inetiel?

Pour répondre à cette question, imaginons que l'observateur se
référant à ($R^{'}$) mesure, à son tour, les circonférences et les
rayons des deux cercles respectifs ($P$) et ($G$). Pour ce faire,
il va utiliser la même règle que celle employée par l'observateur
attaché à ($R$). Il commence par la mesure du rayon et de la
circonférence du petit cercle. Les vitesses mises en jeu à
l'intérieur de ($P$) sont très négligeables devant la vitesse de
la lumière, de sorte que l'observateur pourra se contenter
d'utiliser ses connaissances de mécanique classique pour effectuer
les mesures. Il trouve que le rayon et la circonférence du petit
cercle sont identiques à ceux trouvés dans ($R$). D'autre part,
comme la vitesse à laquelle est soumis un point sur le disque
croîs à mesure que l'on s'éloigne du centre, l'observateur se
trouvant sur le grand cercle ($G$) sera soumis, ainsi, à une
vitesse si importante qu'il ne pourra plus la négliger devant la
vitesse de la lumière. Il sera obligé d'utiliser ses connaissances
de relativité restreinte pour faire les mesures. Il  remarque que
le grand rayon est aussi identique à celui trouvé par
l'observateur extérieur, car   il effectue une mesure
perpendiculaire  au mouvement  (pas de contraction de longueur).
Par  contre,  il est obligé  de faire une mesure tangentielle au
mouvement  pour mesurer la grande circonférence, ce qui fait qu'il
trouverait  une plus  grande valeur comparativement  à celle de
l'observateur extérieur   (du fait que la   règle   posée
tangentiellement au cercle  se trouve dans  la direction  du
mouvement,  ce  qui lui fait subir une contraction). Dans ce cas :
\begin{eqnarray}
    \displaystyle\frac{C_{G}^{'}}{2R_{G}^{'}}>\displaystyle\frac{C_{P}^{'}}{2R_{P}^{'}}\simeq\pi.
\end{eqnarray}

Ce  résultat  frappant  est  une  preuve  que le cadre géométrique euclidien,  valable
pour  les  référentiels inertiels, n'est plus valable  pour  servir  de  cadre à des
référentiels non inertiels.

\subsection{Sur une sphère}
\begin{figure}[\here]
   \centering
  \includegraphics[width=9cm]{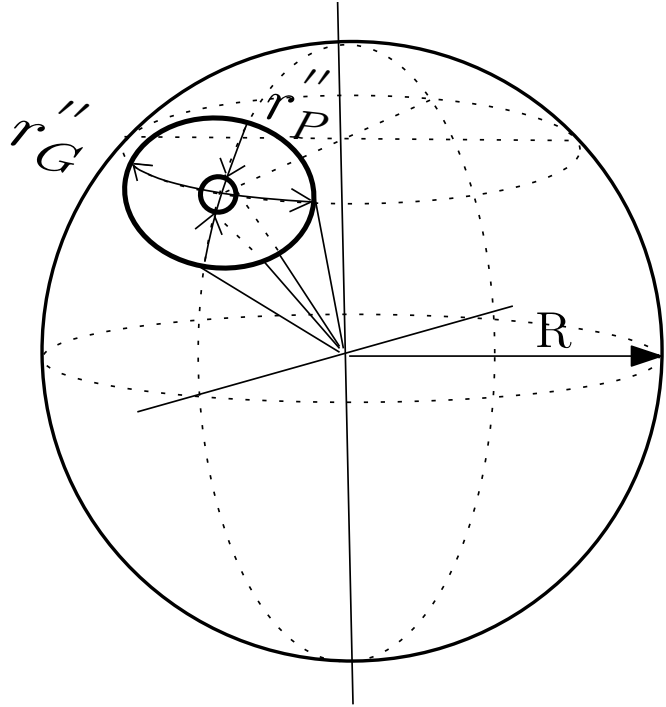}\\
  \caption{Représentation de deux cercles concentriques sur une
  sphère}\label{fig4}
\end{figure}

Il est curieux que le résultat précédent puisse s'obtenir aussi en dessinant  les
deux  cercles  concentriques  sur la surface d'une sphère      à      grand      rayon
\cite{einsteinenfeld}. Remarquons qu'il est possible de dessiner les cercles en
utilisant une  ficelle  au bout  de laquelle  est attaché  un crayon. La longueur   de
la ficelle   détermine le   rayon   du  cercle correspondant (voir \ref{fig4}).

Du  moment  que  les  rayons  des  deux cercles sont fixés par les
longueurs  des ficelles qui ont servi pour le dessin, ils auront
ainsi  les  mêmes  valeurs  que ceux  précédemment  établis  dans
l'exemple                       du disque. Vu la  petitesse  du
rayon  du  petit cercle, il est possible de considérer avec une
bonne approximation qu'il est dessiné dans un plan  à  2
dimensions (du fait que la courbure de la sphère tend à s'annuler
localement), dans  ce  cas  de  figure, la géométrie euclidienne
constitue  une  très  bonne  approximation  du  cadre géométrique
(analogie   avec   le  référentiel  inertiel).

Par contre, l'importance de la courbure de la sphère ne peut plus être négligée, quand
on se place sur le grand cercle, de sorte que sa circonférence est différente de celle
d'un cercle de même rayon et dessiné dans un plan,  ainsi le rapport des
circonférences aux rayons des deux cercles (le petit et le grand)  n'est  pas le même.
Le résultat dépendrait d'une part du rayon    de    la sphère et du rayon   du cercle
\cite{relativrestreinte} :
\begin{eqnarray}
\displaystyle\frac{C_{G}^{''}}{2r_{G}^{''}}=\pi\,\displaystyle\frac{\sin\Big(\displaystyle\frac{r_{G}^{''}}{R}\Big)}{\Big(\displaystyle\frac{r_{G}^{''}}{R}\Big)},\\
\displaystyle\frac{C_{P}^{''}}{2r_{P}^{''}}=\pi\,\displaystyle\frac{\sin\Big(\displaystyle\frac{r_{P}^{''}}{R}\Big)}{\Big(\displaystyle\frac{r_{P}^{''}}{R}\Big)},
\end{eqnarray}
Mais pour le petit cercle, on peut écrire: $$\lim_{\displaystyle r_{P}^{''}\rightarrow
0}\displaystyle\frac{\sin\Big(\displaystyle\frac{r_{P}^{''}}{R}\Big)}{\Big(\displaystyle\frac{r_{P}^{''}}{R}\Big)}=1.$$
Ce qui conduit à:
\begin{equation}
    \displaystyle\frac{C_{G}^{''}}{2r_{G}^{''}}\neq
    \displaystyle\frac{C_{P}^{''}}{2r_{P}^{''}}\simeq\pi.
\end{equation}

Ce  résultat  prouve  que le cadre géométrique de la surface d'une sphère  est  un
cadre  non euclidien: c'est un espace courbe à 2 dimensions.
\section{Conclusion}
Des deux exemples précédents, il est possible de dégager les
remarques suivantes :\\
\begin{itemize}
    \item Le cadre géométrique de la relativité générale est
non-eucildien. \item Le cadre géométrique de la relativité restreinte (pour les
référentiels inertiels) est un cadre euclidien. \item L'observateur se trouvant sur le
petit cercle n'est pas soumis à d'importantes forces centrifuges; il pourrait
expliquer ces effets inertiels, en vertu du principe d'équivalence, par la présence
d'un champ gravitationnel de faible intensité. Or la géométrie euclidienne est
suffisante pour servir de cadre à la description des phénomènes rapportés à un
référentiel inertiel, ce qui permet d'affirmer que:\\\\
\textbf{La description des phénomènes faisant intervenir des champs de gravitation
faibles se fait par le biais d'une géométrie euclidienne.}\\ \item Par contre, décrire
les phénomènes faisant intervenir des champs de gravitation intenses se fait par le
recours à la théorie de relativité générale qui se décrit dans un cadre géométrique
non-euclidien; l'espace ne peut plus être considéré comme plat, il est courbé suite à
une déformation causée par le champ de
gravitation intense:\\\\
\textbf{La présence de la matière (source du champ de gravitation) fait courber
l'espace-temps environnant et cette courbure est d'autant plus importante que la
densité de matière l'est aussi.}
 \end{itemize}

\newpage
\pagestyle{fancy} \lhead{chapitre\;3}\rhead{Métrique et Dynamique de Schwarzschild}

\chapter{ Métrique et Dynamique de Schwarzschild}
La  suite  du  mémoire est consacrée à l'analyse d'une proposition récente,  avancée
par C.C.Barros \cite{barros,barrosbis,barrosquark,barrosstability}, qui stipule que
d'une façon analogue à l'interaction gravitationnelle, les autres interactions peuvent
aussi se manifester   à travers  la structure  de l'espace-temps. L'idée fondamentale
consiste à décrire une particule évoluant dans une  région,  en présence d'un
potentiel non-gravitationnel qui affecterait  la  métrique  de l'espace-temps.  Les
effets  de la métrique    dans    le domaine subatomique   seront   étudiés.

Comme   première  étape,  un  système  à  symétrie  sphérique  est considéré,  mais
l'auteur affirme que l'idée de base pourrait être généralisée   à  des  systèmes
soumis à des potentiels arbitraires. En adoptant une métrique similaire à celle de
Schwarzschild, on détermine les grandeurs dynamiques relatives à une particule
évoluant dans un espace courbe, à savoir: les impulsions, l'énergie, l'expression de
l'invariant relativiste. On exprime, aussi, comment la métrique est affectée par le
potentiel non-gravitationnel. Enfin, l'utilisation de l'analyse vectorielle en
coordonnées curvilignes va permettre d'étendre le principe de correspondance à
l'espace courbe en question. Ceci permettra de déterminer les opérateurs
$(E,\overrightarrow{p},\overrightarrow{p}^{2})$, indispensables à l'écriture
d'équations d'ondes quantiques dans un espace courbe.

\section{Position du problème}
Avant d'aller plus loin, il est utile de faire quelques remarques qui vont permettre
de situer et de comprendre la démarche de Barros.

Premièrement, dans la théorie de la relativité générale, le champ
gravitationnel est décrit en terme de structure géométrique de
l'espace-temps. Cette description découle de l'application du
postulat d'équivalence. L'argument crucial qui justifie l'adoption
de ce postulat est l'égalité des masses inertielle et
gravitationnelle; égalité qui se traduit par le fait que tous les
corps se déplaçant uniquement sous l'influence d'un champ de
gravitation, sont soumis à la même accélération, indépendamment de
leurs masses, de leurs substances et de leurs états physiques.
Cette propriété fondamentale caractérise exclusivement le champ
gravitationnel, c'est à dire que des corps de masses différentes,
se mouvant sous l'action d'un champ non gravitationnel ne seront
pas soumis à la même accélération. C'est pour cette raison qu'en
théorie de la relativité générale, on ne décrit pas les
interactions: électromagnétique, forte et faible en terme de
structure d'espace-temps. Il n'y a donc, a priori, pas d'arguments
physiques qui justifient la proposition de Barros.

D'autre part, l'utilisation d'une métrique de Schwarzschild n'est pas également
justifiée. En effet, une telle solution est obtenue grâce à une utilisation des
équations de champs d'Einstein, déterminées exclusivement pour la description du champ
de gravitation. Ainsi, le recours à cette métrique n'est sûrement pas justifié, en
théorie de la relativité générale, pour servir à la description d'une particule
évoluant sous l'action d'un champ non-gravitationnel.

Barros ignore complètement ces considérations, il fait abstraction de tout le
cheminement qui a poussé Einstein à décrire le champ de gravitation en terme de
structure de l'espace-temps. Son point de
vue est plutôt le suivant:\\

\textbf{Comme point de départ, il adopte une métrique similaire à celle de
Schwarzschild, de telle sorte à la faire dépendre du potentiel non-gravitationnel.
Alors quelles sont toutes les conséquences, qu'il peut tirer, pour la description
d'une particule soumise à un champ non-gravitationnel, en utilisant le formalisme de
la relativité générale?}

\section{La métrique de Schwarzschild}
Soit, donc, une particule évoluant sous l'action d'un champ
non-gravitationnel, caractérisé par un potentiel central $V(r)$.
La source du champ possède une distribution sphérique, décrite par
un tenseur énergie-impulsion $T_{\mu\nu}\neq0$, dans une certaine
région limitée de l'espace. \A l'extérieur de la source, où
$T_{\mu\nu}=0$, l'auteur adopte une métrique similaire à celle de
Schwarzschild \cite{weinbergs,landauchamps}:
\begin{equation}\label{metrique}
    ds^{2}=c^{2}\xi(r)dt^{2}-r^{2}(d\theta^{2}+\sin^{2}\theta
    d\phi^{2})-[\xi(r)]^{-1}dr^{2},
\end{equation}
où $\xi(r)$ est une fonction de l'unique paramètre $r$, et
indépendante du temps. D'une manière analogue à la théorie de la
relativité générale, cette fonction est déterminée par le
potentiel $V(r)$, c'est-à-dire que la métrique est affectée par le
champ non-gravitationnel à travers cette fonction.

Par définition, le carré de l'intervalle, séparant deux points infiniment voisins, est
relié à la métrique à travers la forme quadratique fondamentale:
\begin{equation}\label{quadratiquefond}
    ds^{2}=\displaystyle\sum_{\mu=0}^{3}
    \displaystyle\sum_{\nu=0}^{3}g_{\mu\nu}\,dx^{\mu}dx^{\nu},
\end{equation}

En définissant le quadrivecteur position comme:
\begin{equation}\label{positionquadriv}
    x^{\mu}(ct,r,\theta,\phi)=(ct,\overrightarrow{x}),
\end{equation}
il est clair que d'après (\ref{metrique}) et (\ref{quadratiquefond}), le tenseur
métrique $(g_{\mu\nu})$ est diagonal,
\begin{equation}\label{tenseurmetrique}
    (g_{\mu\nu})=\left(%
\begin{array}{cccc}
  \xi & 0 & 0 & 0 \\
  0 & -\xi^{-1} & 0 & 0 \\
  0 & 0 & -r^{2} & 0 \\
  0 & 0 & 0 & -r^{2}\sin^{2}\theta \\
\end{array}%
\right).
\end{equation}
Or
\begin{equation}\label{produitdesmetriques}
    \displaystyle\sum_{\mu=0}^{3}g_{\mu\nu}\,g^{\nu\alpha}=\delta_{\mu}^{\alpha},
\end{equation}
ce qui implique que:
\begin{equation}\label{tenseurmetriquecontravar}
    (g^{\mu\nu})=\left(%
\begin{array}{cccc}
  \xi^{-1} & 0 & 0 & 0 \\
  0 & -\xi & 0 & 0 \\
  0 & 0 & -r^{-2} & 0 \\
  0 & 0 & 0 & -r^{-2}\sin^{-2}\theta \\
\end{array}%
\right).
\end{equation}
Les éléments de cette matrice peuvent se mettre sous forme:
\begin{equation}\label{lienmetriqueavecplatte}
     g^{\mu\nu}=h_{\mu}^{-2}\,\eta^{\mu\nu},
\end{equation}
où
\begin{equation}\label{metrique platte}
    (\eta_{\mu\nu})=\left(%
\begin{array}{cccc}
  1 & 0 & 0 & 0 \\
  0 & -1 & 0 & 0 \\
  0 & 0 & -1 & 0 \\
  0 & 0 & 0 & -1 \\
\end{array}%
\right),
\end{equation}
et
\begin{equation}\label{mateginv}
    \eta_{\mu\nu}=\eta^{\mu\nu}.
\end{equation}
En effet, à partir des équations (\ref{tenseurmetriquecontravar}),
(\ref{lienmetriqueavecplatte}), (\ref{metrique platte}) et
(\ref{mateginv}), on peut vérifier que les coefficients $h_{\mu}$
s'écrivent:
\begin{eqnarray}\label{coefhmunu}
    &&h_{0}=\sqrt{\xi(r)},\\
    &&h_{1}=1/\sqrt{\xi(r)},\\
    &&h_{2}=r,\\
    &&h_{3}=r\sin\theta.
\end{eqnarray}

Ces coefficient vont jouer un rôle très important dans la généralisation des
expressions d'opérateurs vectoriels, initialement établis dans un espace plat. Par
exemple: gradient et rotationnel d'un vecteur, laplacien d'un scalaire,... Dans la
section (\ref{a9}), les expressions de tels opérateurs seront établies.

\subsection{Principe de correspondance en coordonnées curvilignes
de Schwarzschild} Le     principe     de     correspondance    est un    principe
empirique   permettant   de   retrouver  les expressions  des opérateurs   de   la
mécanique   quantique,  à partir  des expressions             classiques
correspondantes.

Dans un espace-temps plat, le principe de correspondance relatif à l'impulsion et à
l'énergie s'écrit comme suit \cite{broglie}, \cite{landauquantique}:
\begin{equation}\label{correspondanceimpulsionclassique}
\left\{%
\begin{array}{ll}
    p_{j}=-i\hbar\,\displaystyle\frac{\partial}{\partial x^{j}},\\
    j=1,2,3\\
\end{array}%
\right.
\end{equation}

\begin{equation}\label{correspondancenrjclassique}
    E=+i\hbar\,\frac{\partial}{\partial t}.
\end{equation}

La forme de ces expressions trouve son origine dans les lois de conservation de
l'impulsion et de l'énergie, liées à l'homogénéité de l'espace et du temps, tout en
ayant recours au passage à la limite classique:
\begin{enumerate}
    \item  Opérateur énergie:\\ Le postulat fondamental de la mécanique quantique
    stipule que l'état d'un système physique est complètement déterminé par une fonction
    d'onde $\psi$. La connaissance d'une telle fonction à un instant donné
    $t_{0}$,
    doit non seulement nous renseigner sur l'état actuel du système, mais doit en plus
    nous permettre de connaître l'état du système pour n'importe quel instant
    ultérieur $t>t_{0}$, ce qui
    se traduit mathématiquement:
    \begin{equation}
     \displaystyle\frac{\partial \psi}{\partial t}=F(\psi).
    \end{equation}
    En vertu du principe de superposition, indispensable pour
    expliquer les phénomènes ondulatoires (coexistence
    ondes-corpuscules), cette dépendance doit être linéaire \cite{landauquantique}:
    \begin{equation}
     i\;\displaystyle\frac{\partial \psi}{\partial t}=A\;\psi,
    \end{equation}
    où $A$ est un opérateur linéaire et le coefficient $i$ est
    introduit par commodité, pour faciliter la démonstration d'hermicité de
    $A$. Pour déterminer à quelle grandeur classique correspond
    l'opérateur $A$, il faut effectuer le passage à la limite
    classique, où on fait correspondre au mouvement rectiligne et
    uniforme d'une particule d'énergie $E$ et de quantité de
    mouvement $m\overrightarrow{v}$, la propagation d'une onde
    monochromatique, ayant la fréquence $(E/h)$ et la
    longueur d'onde $(h/mv)$ \cite{broglie}. Chaque fois que
    l'optique géométrique sera valable pour la propagation de
    l'onde $\psi$, nous pouvons poser \cite{landauquantique}:
    \begin{equation}\label{ondelimite}
       \psi=a\;e^{\left({\frac{i\,S}{\hbar}}\right)},
     \end{equation}
     et les trajectoires prévues par la dynamique classique du
     point matériel ne seront autre chose que les rayons de l'onde
     $\psi$.\\
     Dans ce cas: $$ i\;\displaystyle\frac{\partial \psi}{\partial t}=
     \left(\displaystyle\frac{-1}{\hbar}\displaystyle\frac{\partial S}{\partial t}\right)\psi,$$
     ce qui montre que l'opérateur linéaire $A$ se réduit, à la
     limite classique, à une simple multiplication. Or en
     mécanique classique, $-(\partial S/\partial
     t)$, n'est autre que la fonction d'Hamilton du système, par conséquent, $(\hbar
     A)$ est l'opérateur correspondant en mécanique quantique, c'est
     l'hamiltonien $H$ qui vérifie l'équation d'onde:
     \begin{equation}
          i\hbar\;\displaystyle\frac{\partial \psi}{\partial
          t}=H\;\psi.
     \end{equation}
     Le passage de l'équation aux valeurs propres:
     \begin{equation}
                 H\;\psi=E\;\psi,
     \end{equation}
     à l'équation d'onde précédente se fait en adoptant le
     principe de correspondance (\ref{correspondancenrjclassique}).
    \item Opérateur impulsion:\\ Considérons un système de
    particules ($j:1\rightarrow N$) non soumis à un champ
    extérieur \cite{landauquantique}. L'homogénéité de l'espace se
    traduit par une invariance des propriétés dynamiques du système, en particulier
    l'hamiltonien $H$, sous n'importe quelle translation arbitraire. Dans une
    translation infiniment petite et arbitraire
    $\overrightarrow{\delta r}$, les rayons vecteurs
    $\overrightarrow{r_{j}}$ de toutes les particules prennent le
    même accroissement $\overrightarrow{r_{j}}\rightarrow \overrightarrow{r_{j}}+
    \overrightarrow{\delta r}$.
    Sous cette transformation, la fonction d'onde du système
    s'écrit:
     \begin{eqnarray}
      \psi(\overrightarrow{r_{1}}+\overrightarrow{\delta r},
      \overrightarrow{r_{2}}+\overrightarrow{\delta r},...)&\simeq&
      \psi(\overrightarrow{r_{1}},\overrightarrow{r_{2}},...)+\overrightarrow{\delta
      r}.\sum_{j=1}^{N}\overrightarrow{\nabla_{j}}\;\psi,\nonumber\\
      &=&\left(1+\overrightarrow{\delta
      r}.\sum_{j=1}^{N}\overrightarrow{\nabla_{j}}\right)\;\psi(\overrightarrow{r_{1}},\overrightarrow{r_{2}},...),\\
      &=&T\;\psi(\overrightarrow{r_{1}},\overrightarrow{r_{2}},...),
     \end{eqnarray}
     où $$\overrightarrow{\nabla_{j}}=\overrightarrow{e_{x}}\displaystyle\frac{\partial}{\partial x^{j}}+
     \overrightarrow{e_{y}}\displaystyle\frac{\partial}{\partial y^{j}}+
     \overrightarrow{e_{z}}\displaystyle\frac{\partial}{\partial
     z^{j}}$$ et $T$ l'opérateur de translation infinitésimale. L'invariance de l'équation
     d'onde sous la translation $T$ conduit à la relation de commutation: $$[\;H,T\;]=0.$$
     Or $[1,H]=0,$ ainsi la constante de mouvement est:
     $$[\;H,\displaystyle\sum_{j=1}^{N}\overrightarrow{\nabla_{j}}\;]=
     [\;H,\overrightarrow{\nabla}\;]=0.$$
     En mécanique classique, la grandeur qui se conserve quand le
     système est invariant sous une translation arbitraire est la
     quantité de mouvement $\overrightarrow{P}$, ceci suggère que
     $\overrightarrow{P}\sim \overrightarrow{\nabla} $.
     Pour déterminer le coefficient de proportionnalité, appliquons
     l'opérateur $\overrightarrow{P}$ sur (\ref{ondelimite}):
    \begin{equation}
         \overrightarrow{P}\;\psi=cte\times\displaystyle\frac{i}{\hbar}
         \left(\overrightarrow{\nabla}S\right)\;\psi.
     \end{equation}
     Or $(\overrightarrow{\nabla}S)$ représente l'impulsion de la
     particule, en mécanique classique. Ce qui implique que
     $(cte\times i/\hbar)=1$, ou encore $cte=-i\hbar$.
     Finalement: $$\overrightarrow{P}=-i\hbar\;\overrightarrow{\nabla},$$
     ce qui confirme (\ref{correspondanceimpulsionclassique}).
\end{enumerate}

Dans un espace-temps courbe répondant à la métrique de Schwarzschild,   les
expressions du principe  de  correspondance précédentes  doivent  être  modifiées  du
fait que l'expression du gradient  et  de  la  dérivée par rapport au temps sont
modifiées.

\subsubsection{a. L'opérateur impulsion dans un espace courbe}
En s'inspirant de (\ref{correspondanceimpulsionclassique}), il est possible d'étendre
le principe de correspondance relatif à l'impulsion, en remplaçant le gradient de
l'espace plat par celui de l'espace courbe. Le principe s'écrit:
\begin{equation}\label{correspondance}
    \overrightarrow{p}=-i\hbar\overrightarrow{\nabla},
\end{equation}
tel que (\ref{gradientscalaire}):
\begin{equation}\label{gradient}
    \nabla_{i}=h_{i}^{-1}\frac{\partial}{\partial x^{i}}.
\end{equation}
En appliquant (\ref{correspondance}) et (\ref{gradient}), on peut écrire l'opérateur
impulsion en métrique de Schwarzschild:
\begin{equation}\label{impulsioncourbe}
    \overrightarrow{p}=-i\hbar\Bigg[\overrightarrow{e_{1}}h_{1}^{-1}\frac{\partial}{\partial x^{1}}+\overrightarrow{e_{2}}h_{2}^{-1}\frac{\partial}{\partial x^{2}}+\overrightarrow{e_{3}}h_{3}^{-1}\frac{\partial}{\partial x^{3}}\Bigg]
\end{equation}
et en utilisant les expressions des coefficients $h_{\mu}$, établies pour la métrique
de Schwarzschild, finalement:
\begin{equation}\label{impulscorres}
    \overrightarrow{p}=-i\hbar\Bigg[\overrightarrow{e_{r}}\,\sqrt{\xi(r)}\,\frac{\partial}{\partial r}+
    \overrightarrow{e_{\theta}}\frac{1}{r}\frac{\partial}{\partial \theta}+
    \overrightarrow{e_{\phi}}\frac{1}{r\sin\theta}\frac{\partial}{\partial
    \phi}\Bigg].
\end{equation}

Remarquons que dans une région où $V(r)=0$
($\xi(r)=1$),c'est-à-dire dans une région assimilée à un
espace-temps plat, on retombe sur l'expression bien connue en
coordonnées sphériques:
\begin{equation}\label{impulsionspherique}
 \overrightarrow{p}=-i\hbar\Bigg[\overrightarrow{e_{r}}\frac{\partial}{\partial r}+
 \overrightarrow{e_{\theta}}\frac{1}{r}\frac{\partial}{\partial \theta}+\overrightarrow{e_{\phi}}\frac{1}{r\sin\theta}\frac{\partial}{\partial \phi}\Bigg]
\end{equation}

\subsubsection{b. L'opérateur énergie dans un espace courbe}
Il ne reste à présent qu'à retrouver l'expression de l'opérateur de l'énergie dans un
espace courbe, doté d'une métrique de Schwarzschild. Nous pouvons nous inspirer de
(\ref{correspondancenrjclassique}), tout en prenant soin de modifier la dérivée
partielle temporelle qui doit s'écrire en tenant compte de $h_{0}$:
\begin{equation}\label{correspondancenrj}
    E=+i\hbar\,\nabla_{0},
\end{equation}
tel que (\ref{gradientscalaire}):
\begin{equation}\label{deriveetemporellecourbe}
    \nabla_{0}=h_{0}^{-1}c\frac{\partial}{\partial x^{0}}=h_{0}^{-1}\frac{\partial}{\partial t},
\end{equation}
tel que: $x^{0}=ct$ (voir (\ref{positionquadriv})). Finalement, après remplacements,
l'opérateur énergie recherché s'écrit:
\begin{equation}\label{correspondancenrjcourbe}
 E=+\displaystyle\frac{i\hbar}{\sqrt{\xi(r)}}\,\frac{\partial}{\partial
 t}.
\end{equation}

Remarquons aussi que pour une région où $V(r)=0$ ($\xi(r)=1$), on retombe aussi sur
l'expression de l'opérateur d'énergie dans un espace-temps plat:
\begin{equation}
    E=+i\hbar\,\frac{\partial}{\partial t}
\end{equation}

\subsection{Les opérateurs laplacien et
carré du module de l'impulsion en métrique de Schwarzschild}

\subsubsection{a. Laplacien en métrique de Schwarzschild}
On utilise (\ref{laplaciencourbe}), tout en veillant à utiliser les coefficients
$h_{i},(i=1,2,3)$, relatifs à la métrique de Schwarzschild. Ce qui donne:
\begin{equation}
\overrightarrow{\nabla}^{2}=\Bigg(\displaystyle\frac{r^{2}\sin\theta}{\sqrt{\xi(r)}}\Bigg)^{-1}\Bigg[\frac{\partial}{\partial
r}\Bigg(r^{2}\sin\theta\sqrt{\xi(r)}\;\frac{\partial}{\partial
r}\Bigg)+\frac{\partial}{\partial
\theta}\Bigg(\displaystyle\frac{\sin\theta}{\sqrt{\xi(r)}}\;\frac{\partial}{\partial
\theta}\Bigg) +\frac{\partial}{\partial
\phi}\Bigg(\displaystyle\frac{1}{\sqrt{\xi(r)}\;\sin\theta}\,\frac{\partial}{\partial
\phi}\Bigg)\Bigg]\nonumber
\end{equation}
\begin{equation}
\overrightarrow{\nabla}^{2}=\Bigg(\displaystyle\frac{r^{2}\sin\theta}{\sqrt{\xi(r)}}\Bigg)^{-1}\Bigg[\sin\theta\,\frac{\partial}{\partial
r}\Bigg(r^{2}\sqrt{\xi(r)}\frac{\partial}{\partial
r}\Bigg)+\displaystyle\frac{1}{\sqrt{\xi(r)}}\,\frac{\partial}{\partial
\theta}\Bigg(\sin\theta\,\frac{\partial}{\partial
\theta}\Bigg)+\displaystyle\frac{1}{\sqrt{\xi(r)}\sin\theta}\,\frac{\partial^{2}}{\partial
\phi^{2}}\Bigg],\nonumber
\end{equation}
Finalement:
\begin{equation}\label{laplacienexplicite}
\overrightarrow{\nabla}^{2}=\displaystyle\frac{\sqrt{\xi(r)}}{r^{2}}\;\frac{\partial}{\partial
r}\Bigg(r^{2}\sqrt{\xi(r)}\,\frac{\partial}{\partial
r}\Bigg)+\displaystyle\frac{1}{r^{2}\sin\theta}\frac{\partial}{\partial
\theta}\Bigg(\sin\theta\,\frac{\partial}{\partial \theta}\Bigg)
+\displaystyle\frac{1}{r^{2}\sin^{2}\theta}\frac{\partial^{2}}{\partial \phi^{2}},
\end{equation}
ou encore, après développement et action sur une fonction $f(r,\theta,\phi)$:
\begin{eqnarray}\label{laplamomentcin}
 &&\overrightarrow{\nabla}^{2}=\Bigg[\frac{2\xi}{r}\,\frac{\partial}{\partial r}+\frac{1}{2}\frac{\partial \xi}{\partial r}\frac{\partial}{\partial r}+\xi\,\frac{\partial^{2}}{\partial r^{2}}\Bigg]
 +\frac{1}{r^{2}}\underbrace{\Bigg[\frac{1}{\tan\theta}\,\frac{\partial}{\partial \theta}+\frac{\partial^{2}}{\partial \theta^{2}}+\frac{1}{\sin^{2}\theta}\frac{\partial^{2}}{\partial
 \phi^{2}}\Bigg]}_{\displaystyle\sim\overrightarrow{L}^{2}},
\end{eqnarray}
tel que $\overrightarrow{L}^{2}$ est le moment cinétique en coordonnées sphériques.

On remarque que seule la partie radiale du laplacien de l'espace courbe est changée
par rapport au laplacien de l'espace plat, exprimé en coordonnées sphériques, et que
les parties angulaires sont identiques. Ceci va permettre d'exploiter la conservation
du moment cinétique et de la parité dans la proposition de la forme des solutions.

\subsubsection{b. Carré du module de l'impulsion}
\A présent, il est possible de retrouver l'expression de
l'opérateur carré de l'impulsion. En effet, (\ref{correspondance})
implique que:
\begin{equation}\label{carreimpulsion}
    |\overrightarrow{P}^{2}|=-\hbar^{2}\,\overrightarrow{\nabla}^{2}.
\end{equation}
En y remplaçant l'expression (\ref{laplacienexplicite}):
\begin{equation}\label{carreimpulsionexplicit}
 |\overrightarrow{P}^{2}|=-\hbar^{2}\,\Bigg[\displaystyle\frac{\sqrt{\xi(r)}}{r^{2}}\;\frac{\partial}{\partial r}\Bigg(r^{2}\sqrt{\xi(r)}\,\frac{\partial}{\partial r}\Bigg)+\displaystyle\frac{1}{r^{2}\sin\theta}\frac{\partial}{\partial \theta}\Bigg(\sin\theta\,\frac{\partial}{\partial \theta}\Bigg)
+\displaystyle\frac{1}{r^{2}\sin^{2}\theta}\frac{\partial^{2}}{\partial
\phi^{2}}\Bigg],
\end{equation}
ou encore:
\begin{equation}\label{impulcinet}
 |\overrightarrow{P}^{2}|=-\hbar^{2}\,\Bigg[\displaystyle\frac{\sqrt{\xi(r)}}{r^{2}}\;\frac{\partial}{\partial r}\Bigg(r^{2}\sqrt{\xi(r)}\,\frac{\partial}{\partial
 r}\Bigg)\Bigg]-\displaystyle\frac{\hbar^{2}\overrightarrow{L}^{2}}{r^{2}}.
\end{equation}

\subsection{Relations de commutations}\label{bbk}
D'après (\ref{impulscorres}), les opérateurs d'impulsion sont:
\begin{eqnarray}
 &&\overline{P_{r}}=-i\hbar\,\sqrt{\xi(r)}\,\frac{\partial}{\partial r}\\
 &&\overline{P_{\theta}}=-i\hbar\,\frac{1}{r}\,\frac{\partial}{\partial \theta}\\
 &&\overline{P_{\phi}}=-i\hbar\,\frac{1}{r\sin\theta}\,\frac{\partial}{\partial \phi}
\end{eqnarray}
Les différents commutateurs sont donnés, après calculs:
\begin{eqnarray}\label{commutateurs}
    &&[\overline{P_{r}},r]=-i\hbar\,\sqrt{\xi(r)},\\
    &&[\overline{P_{\theta}},\theta]=-i\hbar\,\frac{1}{r},\\
    &&[\overline{P_{\phi}},\phi]=-i\hbar\,\frac{1}{r\sin\theta},\\
    &&[\overline{P_{r}},\theta]=[\overline{P_{r}},\phi]=0,\\
    &&[\overline{P_{\theta}},r]=[\overline{P_{\theta}},\phi]=0,\\
    &&[\overline{P_{\phi}},\theta]=[\overline{P_{\phi}},r]=0,\\
    &&[\overline{P_{r}},\overline{P_{\theta}}]=\frac{\hbar^{2}}{r^{2}}\,\frac{\partial^{2}}{\partial \theta^{2}},\\
    &&[\overline{P_{r}},\overline{P_{\phi}}]=\displaystyle\frac{\hbar^{2}\,\sqrt{\xi(r)}}{r^{2}\sin\theta}\,\frac{\partial}{\partial \phi},\\
    &&[\overline{P_{\theta}},\overline{P_{\phi}}]=\displaystyle\frac{\hbar^{2}\,\cot\theta}{r^{2}\sin{\theta}}\,\frac{\partial}{\partial
    \phi}.
\end{eqnarray}

Il est clair que, d'une part, ces opérateurs ne vérifient pas les relations de
commutations bien connues:
\begin{eqnarray}
    &&[P_{i},q_{j}]=i\hbar\,\delta_{ij},\\
    &&[P_{i},P_{j}]=0,
\end{eqnarray}
et d'autre part, les opérateurs d'impulsion ne sont pas hermitiques. En effet, par
définition, un opérateur $A$ est hermitien si et seulement si pour toutes fonctions
complexes $f$ et $g$ définies, uniformes et continues dans un domaine $D$, nous avons
\cite{broglie}:
\begin{equation}\label{conditionhermicite}
    \int_{D}f^{*}\,A(g)\,d\tau=\int_{D}g\,A^{*}(f^{*})\,d\tau.
\end{equation}

Prenons comme exemple l'opérateur $\overline{P_{r}}$, et montrons qu'il n'est pas
hermitien:
\begin{eqnarray}
\int_{D}f^{*}(r)\,\overline{P_{r}}\;[g(r)]\,dr&=&-i\hbar\,\int_{D}f^{*}(r)\;
\sqrt{\xi(r)}\;\frac{\partial g}{\partial r}\;dr\nonumber\\
          &=&-i\hbar\,\int_{D}dr\Bigg[\overbrace{\displaystyle\frac{\partial}{\partial r}
          \left(f^{*}(r)\,\sqrt{\xi(r)}\,g(r)\right)}^{=0}
          -\displaystyle\frac{\partial \sqrt{\xi(r)}}{\partial r}\,f^{*}(r)\,g(r)
          -\sqrt{\xi}\,g(r)\,\frac{\partial f^{*}(r)}{\partial r}\Bigg]\nonumber\\
          &=&i\hbar\,\int_{D}dr\,\displaystyle\frac{\xi^{'}(r)}{2\sqrt{\xi(r)}}\,
          f^{*}(r)\,g(r)+i\hbar\,\int_{D}dr\,g(r)\,\sqrt{\xi(r)}\,
          \frac{\partial f^{*}}{\partial r}\nonumber\\
          &=&i\hbar\,\int_{D}dr\,\displaystyle\frac{\xi^{'}(r)}{2\sqrt{\xi(r)}}\,
          f^{*}(r)\,g(r)+\int_{D}dr\,g(r)\,\overline{P_{r}}\;^{*}(f^{*})\nonumber\\
          &\neq&\int_{D}dr\,g(r)\,\overline{P_{r}}\;^{*}[f^{*}(r)].\nonumber
\end{eqnarray}
Ce qui prouve, compte tenu de (\ref{conditionhermicite}), que $\overline{P_{r}}$ n'est
pas un opérateur hermitien.


\section{Dynamique de Schwarzschild}
Afin d'obtenir les équations d'onde, deux expressions sont nécessaires: l'énergie et $\xi(r)$.\\
Pour se familiariser avec les nouvelles notations, il est utile de faire un petit
aperçu des plus importants résultats de la dynamique dans la métrique de
Schwarzschild, et de voir comment les quantités qui nous intéressent peuvent être
exprimées.

\subsection{Temps propre et quadrivecteur position}
Le temps propre, associé à une particule, est relié au carré d'intervalle, séparant
deux événements infiniment proches, par la formule:
\begin{equation}
     ds^{2}=c^{2}d\tau_{0}^{2},
\end{equation}
ce qui permet d'écrire, en utilisant (\ref{metrique}):
\begin{eqnarray}\label{temps propre}
    d\tau_{0}&=c^{-1}&\sqrt{ds^{2}}\nonumber\\
             &=&dt\displaystyle\sqrt{\xi(r)-r^{2}\bigg(\frac{d\theta}{cdt}\bigg)^{2}-r^{2}\sin^{2}\theta\bigg(\frac{d\phi}{cdt}\bigg)^{2}-[\xi(r)]^{-1}\bigg(\frac{dr}{cdt}\bigg)^{2}}\nonumber\\\nonumber \\
             &=&dt\displaystyle\sqrt{\xi(r)-[\xi(r)]^{-1}\bigg(\frac{dr}{cdt}\bigg)^{2}-r^{2}\bigg[\bigg(\frac{d\theta}{cdt}\bigg)^{2}+\bigg(\frac{d\phi}{cdt}\bigg)^{2}\sin^{2}\theta}\bigg].
\end{eqnarray}

Rappelons que le quadrivecteur position est défini, dans cette
métrique, comme étant:
\begin{equation}
    x^{\mu}(ct,r,\theta,\phi)=(ct,\overrightarrow{x}),
\end{equation}
donc le temps propre s'écrit:
\begin{equation}\label{tempsproprecompact}
   d\tau_{0}=\displaystyle\frac{dt}{\gamma_{s}},
\end{equation}
tel que:
\begin{equation}\label{gammas}
   \gamma_{s}=\Bigg[\xi(r)-[\xi(r)]^{-1}\displaystyle\frac{\beta_{r}^{2}}{c^{2}}-r^{2}\displaystyle\frac{\beta_{t}^{2}}{c^{2}}\Bigg]^{-1/2},
\end{equation}
et où
\begin{eqnarray}\label{beta}
    \overrightarrow{\beta}&=&\frac{d\overrightarrow{x}}{dt},\\
    \beta_{r}&=&\frac{dr}{dt},\\
    \beta_{t}&=&\displaystyle\sqrt{\bigg(\frac{d\theta}{dt}\bigg)^{2}
    +\bigg(\frac{d\phi}{dt}\bigg)^{2}\sin^{2}\theta},
\end{eqnarray}
sont respectivement: la vitesse, la vitesse radiale et la vitesse transversale de la
particule.
\subsection{Action et Lagrangien}
L'action d'un système à une particule est par définition \cite{landauchamps}:
\begin{eqnarray}\label{action}
   S\equiv\int L\,dt
    =-m_{0}c\int ds
    =-(m_{0}c^{2})\int \frac{dt}{\gamma_{s}},
\end{eqnarray}
ce qui permet de déduire l'expression du lagrangien:
\begin{equation}\label{lagrangien}
    L=\frac{-(m_{0}c^{2})}{\gamma_{s}}=-(m_{0}c^{2})\displaystyle\sqrt{\xi(r)-[\xi(r)]^{-1}\bigg(\frac{dr}{cdt}\bigg)^{2}
    -r^{2}\bigg[\bigg(\frac{d\theta}{cdt}\bigg)^{2}+\bigg(\frac{d\phi}{cdt}\bigg)^{2}\sin^{2}\theta\bigg]}.
\end{equation}

\subsection{Quadrivecteur vitesse}
Le quadrivecteur vitesse est par définition:
\begin{equation}\label{defvitesequadriv}
   \beta^{\mu}\equiv\frac{dx^{\mu}}{d\tau_{0}}=\bigg(\frac{cdt}{d\tau_{0}},\frac{dr}{d\tau_{0}},\frac{d\theta}{d\tau_{0}},\frac{d\phi}{d\tau_{0}}\bigg).
\end{equation}
Compte tenu de (\ref{tempsproprecompact}):
\begin{eqnarray}\label{quadrivectvitesse}
    \beta^{\mu}=\gamma_{s}\bigg(c,\frac{dr}{dt},\frac{d\theta}{dt},\frac{d\phi}{dt}\bigg).
\end{eqnarray}

\subsection{Quadrivecteurs énergie-impulsion}
\subsubsection{a. Quadrivecteur énergie-impulsion contravariant}
La multiplication par la masse au repos de (\ref{quadrivectvitesse}) permet de
retrouver l'expression du quadrivecteur énergie-impulsion contravariant:
\begin{equation}\label{quadnrjimpulsioncontr}
    P^{\mu}\equiv
    m_{0}\beta^{\mu}=m_{0}\gamma_{s}\bigg(c,\frac{dr}{dt},\frac{d\theta}{dt},\frac{d\phi}{dt}\bigg).
\end{equation}

\subsubsection{b. Quadrivecteur énergie-impulsion covariant}
Le passage d'un quadrivecteur contravariant au quadrivecteur covariant se fait à
l'aide du tenseur métrique:
\begin{equation}\label{quadnrjimpulsioncov}
    P_{\mu}=\displaystyle\sum_{\nu=0}^{3}g_{\mu\nu}P^{\nu}.
\end{equation}
En remplaçant (\ref{quadnrjimpulsioncontr}) et (\ref{tenseurmetrique}) dans
(\ref{quadnrjimpulsioncov}):
\begin{equation}\label{pmucov}
    P_{\mu}=(P_{0},P_{1},P_{2},P_{3})=m_{0}\gamma_{s}\bigg(c\xi(r),-[\xi(r)]^{-1}\frac{dr}{dt},-r^{2}\frac{d\theta}{dt},-r^{2}\sin^{2}\theta\frac{d\phi}{dt}\bigg).
\end{equation}
Par ailleurs \cite{landauchamps}:
\begin{eqnarray}\label{composantesimpulsioncov}
    P_{r}&\equiv&\displaystyle\frac{\partial L}{\partial\bigg(\displaystyle \frac{dr}{dt}\bigg)}=m_{0}\gamma_{s}[\xi(r)]^{-1}\frac{dr}{dt}=-P_{1}\\\nonumber\\
    P_{\theta}&\equiv&\displaystyle\frac{\partial L}{\partial\bigg( \displaystyle\frac{d\theta}{dt}\bigg)}=m_{0}\gamma_{s}r^{2}\frac{d\theta}{dt}=-P_{2}\\\nonumber\\
    P_{\phi}&\equiv&\displaystyle\frac{\partial L}{\partial\bigg(\displaystyle \frac{d\phi}{dt}\bigg)}=m_{0}\gamma_{s}r^{2}\sin^{2}\theta\frac{d\phi}{dt}=-P_{3}
\end{eqnarray}

\subsection{Energie}
Par définition l'énergie est \cite{landauchamps}:
\begin{equation}\label{defenergie}
    E\equiv
    P_{r}\dot{r}+P_{\theta}\dot{\theta}+p_{\phi}\dot{\phi}-L.
\end{equation}
En remplaçant les expressions précédentes de ($P_{r}$,$P_{\theta}$,$P_{\phi}$) et en
utilisant (\ref{lagrangien}), l'énergie s'écrit:
\begin{eqnarray}
E&=&m_{0}\gamma_{s}\xi^{-1}(r)\bigg(\frac{dr}{dt}\bigg)^{2}+m_{0}\gamma_{s}r^{2}\bigg(\frac{d\theta}{dt}\bigg)^{2}+m_{0}\gamma_{s}r^{2}\sin^{2}\theta\bigg(\frac{d\phi}{dt}\bigg)^{2}+\frac{m_{0}c^{2}}{\gamma_{s}}\nonumber\\
 &=&m_{0}c^{2}\gamma_{s}\Bigg[\xi^{-1}(r)\bigg(\frac{dr}{cdt}\bigg)^{2}+r^{2}\bigg(\frac{d\theta}{cdt}\bigg)^{2}+r^{2}\sin^{2}\theta\bigg(\frac{d\phi}{cdt}\bigg)^{2}+\gamma_{s}^{-2}\Bigg]\nonumber\\
 &=&m_{0}c^{2}\gamma_{s}\Bigg[\xi^{-1}(r)\bigg(\frac{dr}{cdt}\bigg)^{2}+r^{2}\bigg(\frac{d\theta}{cdt}\bigg)^{2}+r^{2}\sin^{2}\theta\bigg(\frac{d\phi}{cdt}\bigg)^{2}+\Bigg(\xi(r)
-[\xi(r)]^{-1}\bigg(\frac{dr}{cdt}\bigg)^{2}\nonumber\\
&&\hspace{2cm}-r^{2}\bigg(\frac{d\theta}{cdt}\bigg)^{2}-r^{2}\sin^{2}\theta\bigg(\frac{d\phi}{cdt}\bigg)^{2}\Bigg)\Bigg].\nonumber
\end{eqnarray}
Finalement:
\begin{eqnarray}\label{energie}
    E&=&(m_{0}c^{2})\gamma_{s}\,\xi(r)=cP_{0},\nonumber\\
    E&=&\displaystyle\frac{(m_{0}c^{2})\xi(r)}{\displaystyle\sqrt{\xi(r)-[\xi(r)]^{-1}\bigg(\frac{dr}{cdt}\bigg)^{2}-r^{2}\bigg(\frac{d\theta}{cdt}\bigg)^{2}-r^{2}\sin^{2}\theta
    \bigg(\frac{d\phi}{cdt}\bigg)^{2}}}.
\end{eqnarray}
Pour $V(r)=0\;(\xi(r)=1)$, on retrouve l'expression de l'énergie:
$$E=\displaystyle\frac{m_{0}c^{2}}{\displaystyle\sqrt{1-
\bigg(\frac{dr}{cdt}\bigg)^{2}-r^{2}\bigg(\frac{d\theta}{cdt}\bigg)^{2}-
r^{2}\sin^{2}\theta\bigg(\frac{d\phi}{cdt}\bigg)^{2}}}=\displaystyle\frac{m_{0}c^{2}}
{\sqrt{1-\frac{\overrightarrow{\beta}^{2}}{c^{2}}}}$$
\subsection{Invariant relativiste}
Il est possible de former une grandeur relativiste invariante à partir du produit
scalaire de deux quadrivecteurs:
\begin{equation}
   \displaystyle\sum_{\mu=0}^{3}P_{\mu}P^{\mu}=
   (m_{0}\gamma_{s})^{2}\bigg(c\xi(r),-[\xi(r)]^{-1}\frac{dr}{d\tau},-r^{2}\frac{d\theta}
   {d\tau},-r^{2}\sin^{2}\theta\frac{d\phi}{d\tau}\bigg)\,
    .\bigg(c,\frac{dr}{d\tau},\frac{d\theta}{d\tau},\frac{d\phi}{d\tau}\bigg),\nonumber\\
\end{equation}
\begin{equation}
    \displaystyle\sum_{\mu=0}^{3}P_{\mu}P^{\mu}=(m_{0}\gamma_{s})^{2}
    \Bigg\{c^{2}\xi(r)-[\xi(r)]^{-1}\bigg(\frac{dr}{dt}\bigg)^{2}-r^{2}
    \bigg[\bigg(\frac{d\theta}{dt}\bigg)^{2}+
    \bigg(\frac{d\phi}{dt}\bigg)^{2}\sin^{2}\theta\bigg]\Bigg\},\nonumber\\
\end{equation}
Finalement,
\begin{equation}\label{egali}
     \displaystyle\sum_{\mu=0}^{3}P_{\mu}P^{\mu}=m_{0}^{2}c^{2}.
\end{equation}
Par ailleurs,
\begin{eqnarray}
    \displaystyle\sum_{\mu=0}^{3}P_{\mu}P^{\mu}&=&\displaystyle\sum_{\mu=0}^{3}
    P_{\mu}\Bigg(\displaystyle\sum_{\nu=0}^{3}g^{\mu\nu}P_{\nu}\Bigg)
    =\displaystyle\sum_{\mu=0}^{3}\displaystyle\sum_{\nu=0}^{3}
    g^{\mu\nu}P_{\mu}P_{\nu}\nonumber\\
  &=&g^{00}P_{0}P_{0}+g^{11}P_{1}P_{1}+g^{22}P_{2}P_{2}+g^{33}P_{3}P_{3},\nonumber
\end{eqnarray}
en tenant compte de (\ref{tenseurmetriquecontravar}) et (\ref{energie}), il vient:
\begin{equation}\label{egal}
\displaystyle\sum_{\mu=0}^{3}P_{\mu}P^{\mu}=
\displaystyle\frac{\Big(\displaystyle\frac{E^{2}}{c^{2}}\Big)}
{\xi(r)}-\xi(r)P_{r}^{2}-\displaystyle\frac{P_{\theta}^{2}}{r^{2}}-
\displaystyle\frac{P_{\phi}^{2}}{r^{2}\sin^{2}\theta}.
\end{equation}
Finalement, compte tenu de (\ref{egali}) et (\ref{egal}), nous avons:
\begin{equation}
    \displaystyle\frac{E^{2}}{c^{2}\xi(r)}-\xi(r)P_{r}^{2}-\displaystyle\frac{P_{\theta}^{2}}{r^{2}}-\displaystyle\frac{P_{\phi}^{2}}{r^{2}\sin^{2}\theta}=m_{0}^{2}c^{2},\nonumber
\end{equation}
ou encore:
\begin{equation}\label{invariantrelativiste}
    \displaystyle\frac{E^{2}}{\xi(r)}=\Bigg[\xi(r)P_{r}^{2}+\displaystyle\frac{P_{\theta}^{2}}{r^{2}}+\displaystyle\frac{P_{\phi}^{2}}{r^{2}\sin^{2}\theta}\Bigg]c^{2}+m_{0}^{2}c^{4}.
\end{equation}
Il est aussi utile d'écrire l'invariant relativiste sous forme:
\begin{equation}\label{invariantracine}
    \displaystyle\frac{E}{\sqrt{\xi(r)}}=\displaystyle\sqrt{\Bigg[\xi(r)P_{r}^{2}+\displaystyle\frac{P_{\theta}^{2}}{r^{2}}+\displaystyle\frac{P_{\phi}^{2}}{r^{2}\sin^{2}\theta}\Bigg]c^{2}+m_{0}^{2}}c^{4}.
\end{equation}

\subsection{Détermination de $\xi(r)$}
Le référentiel du centre de masse est un référentiel où:
$\overrightarrow{P}=\overrightarrow{0}$ ($P_{r}=P_{\theta}=P_{\phi}=0$). Dans un tel
référentiel, l'équation (\ref{invariantracine}) devient:
\begin{equation}\label{cous}
    E(\overrightarrow{P}=\overrightarrow{0})=(m_{0}c^{2})\sqrt{\xi(r)}=E_{0}\sqrt{\xi(r)}.
\end{equation}

D'autre part l'énergie totale, dans le référentiel de centre de masse, s'écrit comme
la somme de l'énergie au repos et de l'énergie potentielle, étant donné que l'énergie
cinétique est nulle:
\begin{equation}\label{couscous}
  E(\overrightarrow{P}=\overrightarrow{0})=E_{0}+V(r).
\end{equation}
La combinaison de (\ref{cous}) et (\ref{couscous}) permet d'écrire:
\begin{equation}\label{racinxi}
 \sqrt{\xi(r)}=\displaystyle\frac{E_{0}+V(r)}{E_{0}}=1+\frac{V(r)}{m_{0}c^{2}},
\end{equation}
ou encore:
\begin{equation}\label{xi}
    \xi(r)=\Bigg(1+\frac{V(r)}{m_{0}c^{2}}\Bigg)^{2}=1+\frac{2V(r)}{m_{0}c^{2}}+\frac{V^{2}(r)}{m_{0}^{2}c^{4}}.
\end{equation}
Pour les faibles potentiels, caractérisés par
$\displaystyle\frac{V(r)}{m_{0}c^{2}}\ll1$, l'expression (\ref{xi}) se réduit à:
\begin{equation}
 \xi(r)\simeq1+\frac{2V(r)}{m_{0}c^{2}}.
\end{equation}

Avec (\ref{xi}), on retrouve l'expression usuelle en relativité générale, pour
l'interaction gravitationnelle (voir \cite{weinbergs} pour comparaison):
\begin{equation}\label{xigravit}
    \xi_{G}=\Bigg(1-\displaystyle\frac{GM}{rc^{2}}\Bigg)^{2}\sim1-\displaystyle\frac{2GM}{rc^{2}}.
\end{equation}

\section{Conclusion}
Dans ce chapitre, le principe de correspondance a été étendu dans
le cas d'un espace courbe, doté d'une métrique de Schwarzschild,
ce qui a permis d'exprimer les opérateurs
$(E,\overrightarrow{p},\overrightarrow{p}^{2})$. De plus, les
principales grandeurs dynamiques, relatives à une particule
évoluant dans un tel espace, sous l'action d'un potentiel
non-gravitationnel $V(r)$, ont été déterminées, à savoir:
l'énergie, les impulsions, l'invariant relativiste. On a aussi
déterminé l'expression de la fonction $\xi(r)$, qui établit le
lien entre la métrique et l'interaction non-gravitationnelle en
question.

La prochaine étape sera consacrée à l'écriture des équations d'ondes quantiques, pour
des particules de spin $0$ et $1/2$, évoluant dans un space courbe, doté d'une
métrique de Schwarzschild.

\newpage
\pagestyle{fancy} \lhead{chapitre\;4}\rhead{Equations d'ondes quantiques dans un
espace courbe}
\chapter{Equations d'ondes quantiques dans un espace courbe}\label{chap4}
Des équations d'ondes quantiques ont été établies, dans le cadre
de l'espace plat de Minkowski. Celles-ci sont l'équation de
Klein-Gordon pour des particules libres de spin $0$, et l'équation
de Dirac pour des particules de spin $1/2$. Ces équations d'ondes
n'expriment, en fait, que des propriétés liées aux exigences
générales de symétrie spatio-temporelle (lois de conservation).
Elles ne permettent pas de rendre compte de l'interaction de
plusieurs particules, où les phénomènes de la disparition et de la
création des particules ont lieu. Pour tenir compte de la création
et disparition des particules, il faut recourir à la seconde
quantification.

En mécanique quantique non relativiste, les relations
d'incertitude d'Heisenberg imposent des restrictions fondamentales
à l'existence simultanée des grandeurs, canoniquement conjuguées,
relatives à un corpuscule. Une circonstance primordiale, dans le
cadre de la théorie quantique non relativiste, permet d'introduire
la notion de la fonction d'onde. C'est le fait de pouvoir mesurer
séparément chacune des variables dynamiques, liées à un
corpuscule, avec une précision arbitrairement grande, et ceci dans
un laps de temps arbitrairement petit (mesure arbitrairement
précise et rapide). En théorie quantique relativiste, l'existence
d'une vitesse limite (la vitesse de la lumière $c$) impose de
nouvelles restrictions fondamentales aux possibilités de mesure de
diverses grandeurs dynamiques \cite{landauquantrelativ}, de telle
sorte que les notions de localisation et d'impulsion d'une
particule ne gardent rigoureusement leurs sens que dans le cas des
particules libres. Ainsi, dans le cadre de la théorie des champs
quantiques on renonce à toute description temporelle du phénomène
d'interaction des particules. Seules seront observables les
caractéristiques des particules libres. On considère que toute
interaction entre particules peut se réduire à une collision de
particules libres, qui après un temps suffisamment long
($t\rightarrow+\infty$) et loin de la région d'interaction,
peuvent être considérées, à nouveau, comme des particules libres.
Les états initiaux et finaux, libres, sont déterminés par les
équations d'ondes quantique citées ci-dessus.

Dans ce chapitre, on envisage d'écrire des équations d'ondes
quantiques, dans le cas d'une seule particule de spin $0$ ou de
spin $1/2$, soumise à un potentiel non-gravitationnel $V(r)$ et
évoluant dans un espace courbe. Attirons l'attention sur le fait
que dans tout ce qui suit, l'interaction gravitationnelle est
négligée, vu qu'on s'intéresse à des systèmes mettant en jeu de
très petites masses. Ainsi, la courbure de l'espace-temps, dont il
est question ici, est causée par des interactions non
gravitationnelles. Pour garantir que le nombre de particules ne
varie pas, on suppose que la particule est incapable
d'interactions, avec le champ, mettant en jeu une importante
énergie, capable d'atteindre un seuil d'énergie de création ou de
disparition d'une quelconque autre particule
\cite{landauquantrelativ}.

Au lieu d'introduire l'interaction, à laquelle est soumise la particule, sous forme
d'un champ extérieur, par des considérations de symétrie \footnote{pour déduire
l'équation d'onde d'un électron soumis à un champ électromagnétique
$A_{\mu}=(\phi,\overrightarrow{A})$, l'invariance de jauge permet d'effectuer une
transformation, sur l'équation de Dirac libre, dite couplage minimal:
$P_{\mu}\rightarrow P_{\mu}-qA_{\mu}$}(couplage minimal) \cite{landauquantrelativ},
l'interaction sera contenue dans la structure de l'espace-temps \cite{barros}.
Celui-ci est affecté par le potentiel non gravitationnel à travers la fonction
$\xi(r)$ \footnote{voir (\ref{metrique}) et (\ref{xi})}.

\section{Equation d'onde pour une particule
de spin $0$}
\subsection{Etablissement de l'équation}
On veut établir l'équation d'onde quantique pour une particule
sans spin. Celle-ci évolue sous l'influence d'un potentiel non
gravitationnel, affectant la métrique. La procédure à suivre est
similaire à celle utilisée par Klein et Gordon, dans le cadre de
l'espace plat de Minkowski. Cette procédure est basée sur
l'utilisation de l'expression de l'invariant relativiste
(\ref{invariantrelativiste}), combinée aux principes de
correspondance
(\ref{correspondancenrjcourbe}),(\ref{carreimpulsionexplicit}).

Conformément à (\ref{correspondancenrjcourbe}), on peut écrire:
\begin{eqnarray}\label{s}
    E^{2}&=&\left(\displaystyle\frac{i\hbar}{\sqrt{\xi(r)}}\right)^{2}
    \frac{\partial^{\,2}}{\partial t^{\,2}},\nonumber\\
    E^{2}&=&\displaystyle\frac{-\hbar^{2}}{\xi(r)}
    \;\frac{\partial^{\,2}}{\partial t^{\,2}},\nonumber\\
    \displaystyle\frac{E^{2}}{\xi(r)}&=&\displaystyle\frac{-\hbar^{2}}{\xi^{2}(r)}
    \;\frac{\partial^{\,2}}{\partial
    t^{\,2}}.
\end{eqnarray}
Exprimons ensuite la même grandeur précédente, en utilisant
(\ref{invariantrelativiste}),
\begin{equation}
    \displaystyle\frac{E^{2}}{\xi(r)}=\left[\xi(r)P_{r}^{2}+\displaystyle\frac{P_{\theta}^{2}}{r^{2}}
    +\displaystyle\frac{P_{\phi}^{2}}{r^{2}\sin^{2}\theta}\right]c^{2}+m_{0}^{2}\,c^{4}.
\end{equation}
En définissant les composantes ordinaires de l'impulsion, conformément à
(\ref{coordonneesordinaires}):
\begin{eqnarray}\label{lespbar}
    \overline{P_{r}}&=&\sqrt{\xi(r)}\,P_{r},\\
    \overline{P_{\theta}}&=&\frac{P_{\theta}}{r},\\
    \overline{P_{\phi}}&=&\frac{P_{\phi}}{r\sin\theta},
\end{eqnarray}
alors, avec cette nouvelle notation, l'expression de l'invariant
relativiste s'écrit,
\begin{eqnarray}\label{nassim}
\displaystyle\frac{E^{2}}{\xi(r)}&=&\left[\overline{P_{r}}^{2}+\overline{P_{\theta}}^{2}
+\overline{P_{\phi}}^{2}\right]c^{2}+m_{0}^{2}\,c^{4},\nonumber\\
\displaystyle\frac{E^{2}}{\xi(r)}&=&\overrightarrow{P}^{2}c^{2}+m_{0}^{2}\,c^{4}.
\end{eqnarray}
En utilisant (\ref{carreimpulsionexplicit}), l'expression
précédente se met sous forme,
\begin{eqnarray}
 &&\displaystyle\frac{E^{2}}{\xi(r)}=-\hbar^{2}c^{2}\,
 \Bigg[\displaystyle\frac{\sqrt{\xi(r)}}{r^{2}}\;\frac{\partial}{\partial r}
 \left(r^{2}\sqrt{\xi(r)}\,\frac{\partial}{\partial r}\right)\nonumber\\
&&\hspace{4cm}+\displaystyle\frac{1}{r^{2}\sin\theta}\frac{\partial}{\partial
\theta}\left(\sin\theta\,\frac{\partial}{\partial \theta}\right)
+\displaystyle\frac{1}{r^{2}\sin^{2}\theta}\frac{\partial^{2}}{\partial
\phi^{2}}\Bigg]+m_{0}^{2}\,c^{4}.\label{ss}
\end{eqnarray}
Finalement, la combinaison de (\ref{s}) et (\ref{ss}) permet
d'écrire l'équation d'onde pour une particule de spin $0$,
\begin{eqnarray}
&&\Bigg\{\displaystyle\frac{-\hbar^{2}}{\xi^{2}(r)}\,\frac{\partial^{2}}{\partial
t^{2}}\Bigg\}\,\Psi(\overrightarrow{r},t)=\Bigg\{-\hbar^{2}c^{2}
\bigg[\displaystyle\frac{\sqrt{\xi(r)}}{r^{2}}\;\frac{\partial}{\partial
r}\Big(r^{2}\sqrt{\xi(r)}\,\frac{\partial}{\partial r}\Big)\nonumber\\
&&\hspace{5cm}+\displaystyle\frac{1}{r^{2}\sin\theta}\frac{\partial}{\partial \theta}
\Big(\sin\theta\,\frac{\partial}{\partial \theta}\Big)
+\displaystyle\frac{1}{r^{2}\sin^{2}\theta}\frac{\partial^{2}}{\partial
\phi^{2}}\bigg]+m_{0}^{2}\,c^{4}\Bigg\}\,\Psi(\overrightarrow{r},t).\nonumber\\
\end{eqnarray}
En tenant compte de (\ref{impulcinet}), l'équation précédente peut
également se mettre sous la forme:
\begin{eqnarray}\label{sss}
    &&\Bigg\{\displaystyle\frac{-\hbar^{2}}{\xi^{2}(r)}\,\frac{\partial^{2}}
    {\partial t^{2}}\Bigg\}\,\Psi(\overrightarrow{r},t)=
    \Bigg\{-\hbar^{2}c^{2}\,
    \Bigg[\displaystyle\frac{\sqrt{\xi(r)}}{r^{2}}\;\frac{\partial}{\partial r}\Big(r^{2}
    \sqrt{\xi(r)}\,\frac{\partial}{\partial r}\Big)\Bigg]\nonumber\\
    &&\hspace{8cm}-\,\hbar^{2}c^{2}\,\displaystyle\frac{\overrightarrow{L}^{2}}{r^{2}}
    +m_{0}^{2}\,c^{4}
    \;\Bigg\}\,\Psi(\overrightarrow{r},t).
\end{eqnarray}
C'est l'équation d'onde quantique d'une particule, sans spin,
évoluant sous l'influence d'un potentiel non gravitationnel
$V(r)$.

\subsection{Résolution de l'équation}
Pour proposer une solution de l'équation (\ref{sss}), on exploite le fait que l'hamiltonien de notre
système physique commute avec l'opérateur carré du moment cinétique orbital. En effet,
en utilisant, d'une part, l'expression (\ref{ss}) et, d'autre part, le fait que
$[r,\overrightarrow{L}^{2}]=0$, alors
\begin{eqnarray}
    \Bigg[\displaystyle\frac{H^{2}}{\xi(r)},\overrightarrow{L}^{2}\Bigg]&=&
    \Bigg[\;\bigg\{-\hbar^{2}c^{2}\,\bigg(\displaystyle
    \frac{\sqrt{\xi(r)}}{r^{2}}\;\frac{\partial}{\partial r}\Big(r^{2}\sqrt{\xi(r)}\,
    \frac{\partial}{\partial r}\Big)\bigg)
    -\displaystyle\frac{\hbar^{2}c^{2}\overrightarrow{L}^{2}}{r^{2}}
    +m_{0}^{2}c^{4}\;\bigg\},\overrightarrow{L}^{2}\Bigg]\nonumber\\
    &=&-\hbar^{2}c^{2}\Bigg[\displaystyle\frac{\sqrt{\xi(r)}}{r^{2}}\;
    \frac{\partial}{\partial r}\Big(r^{2}\sqrt{\xi(r)}\,\frac{\partial}{\partial
    r}\Big),\overrightarrow{L}^{2}\Bigg]=0,\nonumber
\end{eqnarray}
ce qui implique, finalement, que $[H,\overrightarrow{L}^{2}]=0$.
Ceci nous amène à déduire que les deux opérateurs $H$ et
$\overrightarrow{L}^{2}$ possèdent une base propre commune. Ainsi,
on recherche les fonctions propres de l'hamiltonien, parmi les
fonctions propres de l'opérateur $\overrightarrow{L}^{2}$, qui
vérifie l'équation aux valeurs propres suivante
$$\overrightarrow{L}^{2}Y_{\ell}^{m}(\theta,\phi)=\ell(\ell+1)\hbar^{2}Y_{\ell}^{m}(\theta,\phi).$$
Donc, la partie angulaire de la fonction d'onde vérifiant
l'équation d'onde (\ref{sss}) sera totalement déterminée. Dans le
cas stationnaire, on choisit une fonction d'onde à variables
séparées, sous forme
\begin{equation}\label{sepdevar}
    \Psi(\overrightarrow{r},t)=U(r)\,Y_{\ell}^{m}(\theta,\phi)\,\displaystyle
    e^{-\frac{iEt}{\hbar}},
\end{equation}
où $U(r)$ est la partie radiale inconnue de la fonction d'onde. En remplaçant
(\ref{sepdevar}) dans (\ref{sss}), on détermine l'équation différentielle que
satisfait cette fonction radiale:
\begin{eqnarray}
&&\displaystyle\frac{E^{2}}{\xi^{2}(r)}\;\psi(\overrightarrow{r},t)=
-\,\hbar^{2}c^{2}\;Y_{\ell}^{m}(\theta,\phi)\;e^{-\frac{iEt}{\hbar}}
\;\displaystyle\frac{\sqrt{\xi(r)}}{r^{2}}\;\frac{d}{dr}\left(r^{2}\sqrt{\xi(r)}
\;\frac{d}{dr}\right)\;U(r)\nonumber\\
     &&\hspace{6cm}+ c^{2}\,\frac{U(r)}{r^{2}}\;e^{-\frac{iEt}{\hbar}}
     \;\overrightarrow{L}^{2}\;Y_{\ell}^{m}(\theta,\phi)
     +m_{0}^{2}\,c^{4}\;\psi(\overrightarrow{r},t),\nonumber
\end{eqnarray}
\begin{eqnarray}
    &&\displaystyle\frac{E^{2}}{\xi^{2}(r)}\;\psi(\overrightarrow{r},t)=
    -\hbar^{2}c^{2}\;Y_{\ell}^{m}(\theta,\phi)\;e^{-\frac{iEt}{\hbar}}
    \;\displaystyle\frac{\sqrt{\xi(r)}}{r^{2}}\;\frac{d}{dr}\left(r^{2}\sqrt{\xi(r)}
    \;\frac{d}{dr}\right)\;U(r)\nonumber\\
     &&\hspace{6cm}+ c^{2}\,\frac{U(r)}{r^{2}}\;e^{-\frac{iEt}{\hbar}}
      \;\ell(\ell+1)\hbar^{2}\;Y_{\ell}^{m}(\theta,\phi)
      +m_{0}^{2}\,c^{4}\;\psi(\overrightarrow{r},t).\nonumber
\end{eqnarray}
Une division des deux membres de l'équation par
$U(r)\;Y_{\ell}^{m}(\theta,\phi)\;\displaystyle
    e^{-\frac{iEt}{\hbar}}$, donne
\begin{eqnarray}
\displaystyle\frac{E^{2}}{\xi^{2}(r)}\;U(r)=
-\hbar^{2}c^{2}\;\displaystyle\frac{\sqrt{\xi(r)}}{r^{2}}
\;\frac{d}{dr}\left(r^{2}\sqrt{\xi(r)}\;\frac{d}{dr}\right)\;U(r)
      + \hbar^{2}c^{2}\;\displaystyle\frac{\ell(\ell+1)}{r^{2}}\;U(r)
      +m_{0}^{2}\,c^{4}\;U(r),\nonumber
\end{eqnarray}
\begin{equation}
    \displaystyle\frac{\sqrt{\xi(r)}}{r^{2}}\;\frac{d}{dr}\left(r^{2}\sqrt{\xi(r)}
    \,\frac{d}{dr}\right)\;U(r)
    +\left[\displaystyle\frac{E^{2}}{\hbar^{2}c^{2}\;\xi^{2}(r)}-
    \displaystyle\frac{m_{0}^{2}\,c^{2}}{\hbar^{2}}
    -\displaystyle\frac{\ell(\ell+1)}{r^{2}}\right]\;U(r)=0.
\end{equation}
Finalement, l'équation différentielle que satisfait la partie radiale de la fonction
d'onde, s'écrit \cite{barros}
\begin{equation}\label{kgordon}
    \xi(r)\,\displaystyle\frac{d^{2}U}{dr^{2}}
    +\left[\frac{2\,\xi(r)}{r}+\frac{1}{2}\,\frac{d\xi}{dr}\right]\frac{dU}{dr}
    +\left[\displaystyle\frac{E^{2}}{\hbar^{2}c^{2}\;\xi^{2}(r)}
    -\displaystyle\frac{m_{0}^{2}c^{2}}{\hbar^{2}}
    -\displaystyle\frac{\ell(\ell+1)}{r^{2}}\right]U(r)=0.
\end{equation}
Remarquons que pour chaque système physique, correspond une
fonction $\xi(r)$ qui dépend de l'énergie potentielle $V(r)$.
Après le remplacement  de la fonction $\xi(r)$, relative au
système en question, on résous l'équation différentielle.

\section{Equation d'onde pour une
particule de spin $1/2$}

\subsection{Etablissement de l'équation}

La prochaine étape consiste à obtenir l'analogue de l'équation de
Dirac pour une particule de spin $1/2$. Dirac l'a fait dans le cas
de l'espace-temps plat de Minkowski, alors que l'équation qu'on
veut obtenir est une équation d'onde dans un espace-temps courbe,
doté d'une métrique similaire à celle de Schwarzschild
(\ref{metrique}). La procédure adoptée est de prendre l'équation
de Dirac et de remplacer la dérivée temporelle et les dérivées
spatiales, respectivement par (\ref{deriveetemporellecourbe}) et
(\ref{gradient}). Ce sont des dérivées qui tiennent compte de la
courbure de l'espace-temps.

\subsubsection{a. Hamiltonien de Dirac}

Pour construire une équation d'onde quantique décrivant une particule de spin $1/2$,
Dirac a été guidé, dans le choix de son hamiltonien, par deux considérations
physiques. D'une part, il a voulu avoir une équation linéaire en $\partial/\partial
t$. Ainsi, la connaissance de l'état initial de la particule va permettre de
déterminer tous les états ultérieurs. D'autre part, pour garantir à son équation
d'être invariante par transformation de Lorentz (Rotation+boost), la relativité impose
à $\partial/\partial t$ d'appartenir à un quadrivecteur
$\partial_{\mu}(\overrightarrow{\nabla},\frac{1}{c}\frac{\partial}{\partial t})$, ce
qui implique que l'hamiltonien est linéaire en $\overrightarrow{\nabla}$ ( ou en
$\overrightarrow{p}$). Il propose un hamiltonien sous forme
\begin{eqnarray}\label{hamiltonien}
    H&=&\displaystyle\sum_{i=1}^{3}\alpha_{i}.\underbrace{p_{i}c}_{\textrm{énergie}}
    +\beta\,\underbrace{m_{0}c^{2}}_{\textrm{énergie}},\nonumber\\
     &=&\overrightarrow{\alpha}.\overrightarrow{p}c+\beta\,m_{0}c^{2},
\end{eqnarray}
où les 4 grandeurs $\beta$ et $\alpha_{i}$ sont des grandeurs sans
dimensions.

Dirac a ensuite imposé à cet hamiltonien de vérifier les deux conditions suivantes:
\begin{enumerate}
\item En élevant au carré l'hamiltonien, il faut retrouver l'expression de l'invariant
relativiste,
\begin{equation}
    H^{2}=\overrightarrow{p}^{2}c^{2}+m_{0}^{2}c^{4}.
\end{equation} \item Hermicité de l'hamiltonien,
\begin{equation}
    H=H^{+}.
\end{equation}
\end{enumerate}
Ces deux exigences conduisent à montrer que les quatre grandeurs,
sans dimension, sont des matrices obéissant à une algèbre de
Clifford, définie par les relations:
\begin{eqnarray}
    \{\alpha_{i},\alpha_{j}\}&=&0=\{\alpha_{i},\beta\}, \hspace{1cm}\textrm{si}\; i\neq
    j,\label{algebrecliff1}
    \\
    \alpha_{i}^{2}&=&\beta^{2}=1,\label{algebrecliff2}
    \\
    \alpha_{i}^{+}&=&\alpha_{i}\;,\;\beta^{+}=\beta.\label{algebrecliff3}
\end{eqnarray}

L'expression de l'invariant relativiste,
\begin{equation}\label{zouk}
    E^{2}=c^{2}\left[p_{1}^{\,2}+p_{2}^{2}+p_{3}^{2}\right]+m_{0}^{2}\,c^{4}.
\end{equation}
a poussé Dirac à proposer un hamiltonien sous la forme
\begin{equation} \label{trufe}
    H=c\left(p_{1}\;\alpha_{1}
    +p_{2}\;\alpha_{2}+p_{3}\;\alpha_{3}\right)+\beta\; m_{0}c^{2}.
\end{equation}
L'hamiltonien précédent peut s'écrire comme suit:
\begin{eqnarray}\label{terf}
   H=\overrightarrow{\alpha}.\overrightarrow{p}\,c+\beta\;m_{0}c^{2},
\end{eqnarray}
tel que
    $\overrightarrow{p}=p_{1}\;\overrightarrow{e_{1}}
    +p_{2}\;\overrightarrow{e_{2}}
    +p_{3}\;\overrightarrow{e_{3}}.\nonumber$
On impose à l'hamiltonien, élevé au carré, de vérifier
(\ref{zouk}). En effet:
\begin{eqnarray}
 H^{2}&=&
 \Bigg[c\bigg(p_{1}\;\alpha_{1}
 +p_{2}\;\alpha_{2}+p_{3}\;\alpha_{3}\bigg)
 +\beta\;m_{0}c^{2}\Bigg]^{2}
 \nonumber\\
&=&\bigg(\displaystyle\sum_{i=1}^{3}p_{i}\;\alpha_{i}\,c +\beta\;
m_{0}c^{2}\bigg)\bigg(\displaystyle\sum_{j=1}^{3}p_{j}\,\alpha_{j}\,c
+\beta\;m_{0}c^{2}\bigg)
\nonumber\\
&=&\bigg(p_{1}^{2}\;\alpha_{1}^{2} +p_{2}^{2}\;\alpha_{2}^{2}
+p_{3}^{2}\;\alpha_{3}^{2}\bigg)c^{2} +m_{0}^{2}c^{4}\;\beta^{2}
\nonumber\\
&&\hspace{1cm}+
\bigg[c^{2}\;p_{1}\;p_{2}\;\{\alpha_{1},\alpha_{2}\}
+c^{2}\;p_{1}\;p_{3}\;\{\alpha_{1},\alpha_{3}\}
+c^{2}\;p_{2}\;p_{3}\;\{\alpha_{2},\alpha_{3}\}
\nonumber\\
&&\hspace{4cm}+m_{0}^{2}\,c^{3}\bigg(p_{1}\{\alpha_{1},\beta\}
+p_{2}\{\alpha_{2},\beta\}
+p_{3}\{\alpha_{3},\beta\}\bigg)\bigg],\hspace{1.5cm}\label{zest}
\end{eqnarray}
où $\{A,B\}=AB+BA$ est l'anticommutateur des deux grandeurs $A$ et $B$. En admettant
les conditions suivantes:
\begin{eqnarray}
    &&\alpha_{1}^{2}=\alpha_{2}^{2}=\alpha_{3}^{2}=\beta^{2}=1,\\
    &&\{\alpha_{1},\alpha_{2}\}=\{\alpha_{1},\alpha_{3}\}=\{\alpha_{2},\alpha_{3}\}=
    \{\beta,\alpha_{1}\}=\{\beta,\alpha_{2}\}=\{\beta,\alpha_{3}\}=0,
\end{eqnarray}
alors, l'expression (\ref{zest}) se réduit à la relation
\begin{equation}
    H^{2}=\left(p_{1}^{2}
    +p_{2}^{2}
    +p_{3}^{2}\right)c^{2}+m_{0}^{2}c^{4}.\nonumber
\end{equation}
Cette relation est en parfait accord avec (\ref{zouk}). De plus,
en imposant à l'hamiltonien d'être hermitien
\begin{equation}
    H=H^{+},
\end{equation}
on montre que les 4 matrices $\beta,\alpha_{i}$ sont aussi
hermitiennes
\begin{eqnarray}
\Bigg\{
    \begin{array}{l}
   \alpha_{i}=\alpha^{+}_{i}\\
   i=1,2,3
    \end{array}
,\hspace{1cm}\beta=\beta^{+}.
\end{eqnarray}
On conclut, finalement, que ces 4 matrices figurant dans
l'hamiltonien de Dirac (\ref{hamiltonien}) vérifient une algèbre
de Clifford (\ref{algebrecliff1}),(\ref{algebrecliff2}) et
(\ref{algebrecliff3}).

\subsubsection{b. Détermination de l'ordre des matrices $\beta$ et $\alpha_{i}$}
Avant d'écrire l'équation d'onde de la particule de spin $1/2$,
évoluant sous l'action de l'interaction non-gravitationnelle, il
est utile de déterminer l'ordre des matrices figurant dans
l'hamiltonien (\ref{trufe}). On suivra la procédure décrite dans
\cite{bjorkendrel} et \cite{fradkin}. La détermination de l'ordre
des matrices $\beta$ et $\alpha_{i}$ va permettre de déduire le
nombre de composantes du spineur décrivant l'état d'une telle
particule, dans le cas relativiste. Pour ce faire:
\begin{enumerate}
    \item On détermine les valeurs propres des matrices
$\beta\,,\alpha_{i}:i=1,2,3$.\\
L'équation aux valeurs propres, relative à $\beta$ (respectivement
des $\alpha_{i}$) s'écrit sous forme
\begin{equation}
    \beta\,\overrightarrow{X}=\lambda\,\overrightarrow{X}.\nonumber
\end{equation}
Une deuxième application de $\beta$ (respectivement des
$\alpha_{i}$) donne, en tenant compte de (\ref{algebrecliff2}):
\begin{eqnarray}
    \beta^{2}\,\overrightarrow{X}=\lambda\,\beta\,\overrightarrow{X}\hspace{1cm}
    &\Rightarrow&\hspace{1cm}1.\overrightarrow{X}=\lambda^{2}\,\overrightarrow{X}\nonumber\\
   \lambda^{2}=1\hspace{1cm}&\Rightarrow&\hspace{1cm}\lambda=\pm1.\nonumber
\end{eqnarray}
Ainsi, les valeurs propres des matrices $\beta$ et $\alpha_{i}$
sont $+1$ ou $-1$.
    \item On montre ensuite que les traces
    $Tr(\beta)=Tr(\alpha_{i})=0$.
     Pour ce faire, on va utiliser, d'une part l'anticommutation des matrices en question,
    et d'autre part, les propriétés, bien connues
    \begin{eqnarray}
       Tr(A\,B)=Tr(B\,A),\nonumber\\
       Tr(\lambda\,A)=\lambda\,Tr( A).
    \end{eqnarray}
    En effet,
    \begin{eqnarray}
      Tr(\alpha_{i})&=&Tr(1.\alpha_{i})
       =Tr(\beta^{2}\,\alpha_{i})
       =Tr\big[\beta(\beta\,\alpha_{i})\big]
       =Tr\big[\beta(-\alpha_{i}\,\beta)\big]\nonumber\\
       &=&-Tr\big[\beta(\alpha_{i}\,\beta)\big]
       =-Tr\big[(\alpha_{i}\,\beta)\beta\big]
       =-Tr\big[\alpha_{i}\,\beta^{2}\big]\nonumber\\
       &=&-Tr(\alpha_{i})\nonumber\\
       &\Rightarrow&\hspace{1cm}Tr(\alpha_{i})=0.
    \end{eqnarray}
    Une démonstration similaire peut se faire pour montrer que $Tr(\beta)=0$.
    \item On va exploiter, d'une part, le fait que les matrices hermitiennes $M$
    sont diagonalisables,
    c'est-à-dire il existe une matrice inversible $S$, telle que
    \begin{eqnarray}
     S\,M\,S^{-1}=M_{D}
=\left(
\begin{array}{ccccc}
\lambda_{1}& ... & 0 & ... & 0\\
\vdots & \ddots & \vdots & \ddots & \vdots\\
0 & ... & \lambda_{i} & ... & 0\\
\vdots & \ddots & \vdots & \ddots & \vdots\\
0 & ... & 0 & ... & \lambda_{n}\\
\end{array}
\right),
    \end{eqnarray}
où les $\lambda_{i}$ sont les valeurs propres de $M$. D'autre part, l'égalité des
traces des deux matrices $M$ et $M_{D}$ est également exploitée. En effet,
$$Tr(M)=Tr\big[S^{-1}(M_{D}S)\big]=Tr\big[(M_{D}S)S^{-1}\big]=Tr(M_{D}).$$
Comme les matrices $\beta$ et $\alpha_{i}$ sont hermitiennes,
conformément à (\ref{algebrecliff3}), alors, il est possible
d'utiliser les propriétés précédentes, qui s'écrivent dans le cas
de $\beta$ et $\alpha_{i}$ comme suit
\begin{eqnarray}
    Tr(\beta)=Tr(\alpha_{i})=0\hspace{1cm}&\Rightarrow&\hspace{1cm}
    Tr(\beta_{D})=Tr\big[(\alpha_{i})_{D}\big]=0\nonumber\\
    &\Rightarrow&\hspace{1cm}\displaystyle\sum_{i=1}^{n}\lambda_{i}=0\nonumber\\
    &\Rightarrow&\hspace{1cm}\underbrace{(1+1-1+..-1+1)}_{\textrm{n termes}}=0.\nonumber
\end{eqnarray}
Pour avoir une somme nulle, il faut que les $+1$ et les $-1$ se
compensent complètement, condition qui n'a lieu que si la
dimension des matrices $\beta_{D},(\alpha_{i})_{D}$ , ou encore
($\beta$ et $\alpha_{i}$),
est paire, i.e. $n=2p$.\\
\\
\textbf{Pour $n=2$}, une base des matrices complexes
$M_{2\times2}$ est l'ensemble des matrices de Pauli, en plus de
l'identité $\{\sigma_{1},\sigma_{2},\sigma_{3},1\}$. Dans ce cas,
il n'y a pas de solution, car confondre les $\alpha_{i}$ avec les
$\sigma_{i}$, conduit à prendre nécessairement $\beta=1$. Or
$\beta$
possède une trace différente de 1 ($Tr(1)=2$), ce qui est absurde.\\
\\
\textbf{Pour $n=4$}, il existe des solutions. Celles-ci s'écrivent
en représentation standard sous la forme
\begin{eqnarray}
    \overrightarrow{\alpha}=\Bigg(
\begin{array}{cc}
0 & \overrightarrow{\sigma}\\
\overrightarrow{\sigma} & 0\\
\end{array}
\Bigg) \hspace{1cm},\hspace{1cm} \beta=\Bigg(
\begin{array}{cc}
\mathbf{1} & 0 \\
0 & -\mathbf{1}\\
\end{array}
\Bigg),
\end{eqnarray}
où $\mathbf{1}$ est la matrice identité $(2\times2)$ et
$\overrightarrow{\sigma}=\overrightarrow{e_{1}}\;\sigma_{1}+
\overrightarrow{e_{2}}\;\sigma_{2}+\overrightarrow{e_{3}}\;\sigma_{3}$. Les 3 matrices
de Pauli sont définies par
\begin{equation}\label{matpauli}
    \sigma_{1}=\Bigg(
\begin{array}{cc}
0 & 1 \\
1 & 0 \\
\end{array}
\Bigg) \hspace{0.5cm},\hspace{0.5cm}\sigma_{2}=\Bigg(
\begin{array}{cc}
0 & -i \\
i & 0 \\
\end{array}
\Bigg) \hspace{0.5cm},\hspace{0.5cm}\sigma_{3}=\Bigg(
\begin{array}{cc}
1 & 0 \\
0 & -1 \\
\end{array}
\Bigg).
\end{equation}
Donc finalement, on peut conclure que les matrices $\beta$ et
$\alpha_{i}$, figurant dans l'hamiltonien (\ref{trufe}), sont
d'ordre $4\times4$. Ainsi, la fonction d'onde, décrivant l'état
d'une particule de spin $1/2$, est un spineur à 4 composantes. Ce
spineur permet de décrire, aussi bien, la particule que
l'antiparticule de spin $1/2$. En représentation standard, il est
d'usage d'utiliser la notation condensée suivante
\begin{equation}\label{spineurstandard}
    \psi=\left(%
\begin{array}{c}
  \varphi \\
  \chi \\
\end{array}%
\right),
\end{equation}
où $\varphi$ et $\chi$ sont deux spineurs à deux composantes, décrivant respectivement
la particule et l'antiparticule.
\end{enumerate}

\subsubsection{c. Equation d'onde}
Après avoir déterminé l'hamiltonien de Dirac, répondant à toutes
les exigences physiques, et possédant un principe de
correspondance adapté à la métrique de Schwarzschild, il est
possible, à présent, de formuler une équation d'onde quantique
pour une particule de spin $1/2$, soumise à un potentiel non
gravitationnel qui affecterait la métrique. La procédure consiste
de reprendre l'équation de Dirac, initialement établie dans un
espace-temps plat de Minkowski, et de remplacer, d'une part la
dérivée temporelle par (\ref{deriveetemporellecourbe}), et d'autre
part le gradient "ordinaire" par (\ref{gradient}). Les dérivées
qu'on a remplacé tiennent compte de la courbure de l'espace-temps,
provoquée par le potentiel non-gravitationnel.

Dans un espace-temps de Minkowski, l'équation de Dirac s'écrit
sous la forme:
\begin{eqnarray}
    \left(\overrightarrow{\alpha}.\overrightarrow{p}c+\beta\,m_{0}c^{2}\right)
    \;\Psi(\overrightarrow{r},t)&=&i\hbar\;\frac{\partial \Psi}{\partial t},\nonumber
 \end{eqnarray}
En tenant compte de (\ref{correspondanceimpulsionclassique}),
l'équation de Dirac s'écrit finalement:
 \begin{eqnarray}
    \left(-i\hbar c\,\overrightarrow{\alpha}.\overrightarrow{\nabla}+\beta\,m_{0}c^{2}\right)
    \;\Psi(\overrightarrow{r},t)&=&i\hbar\;\frac{\partial \psi}{\partial t}.\nonumber
\end{eqnarray}

Par contre dans une métrique de Schwarzschild, l'équation d'onde
quantique pour une particule de spin $1/2$, évoluant sous l'action
d'un potentiel non-gravitationnel affectant la métrique, se déduit
de l'équation de Dirac précédente comme suit:
\begin{eqnarray}
     \left(\overrightarrow{\alpha}.\overrightarrow{P}c+\beta\,m_{0}c^{2}\right)
    \;\Psi(\overrightarrow{r},t)&=&i\hbar\,\nabla_{0}\;\Psi(\overrightarrow{r},t),\nonumber
\end{eqnarray}
où $\overrightarrow{P}$ et $\nabla_{0}$ sont donnés respectivement
par (\ref{impulscorres}) et (\ref{deriveetemporellecourbe}). En
effectuant les remplacements nécessaires, on peut écrire
finalement:

\begin{eqnarray}\label{eqdiracourbe}
&&\Bigg\{\displaystyle\frac{i\hbar}{\sqrt{\xi(r)}}\frac{\partial}{\partial
t}\Bigg\}\;\Psi(\overrightarrow{r},t)=\nonumber
\\
&&\hspace{1cm}\Bigg\{-i\hbar
c\overrightarrow{\alpha}.\bigg[\overrightarrow{e_{r}}\sqrt{\xi(r)}\frac{\partial}{\partial
r}+\overrightarrow{e_{\theta}}\frac{1}{r}\frac{\partial}{\partial
\theta}+\overrightarrow{e_{\phi}}\frac{1}{r\sin\theta}\frac{\partial}{\partial
\phi}\bigg]+\beta
m_{0}c^{2}\Bigg\}\;\Psi(\overrightarrow{r},t).\hspace{2cm}
\end{eqnarray}

\subsection{Résolution de l'équation}
\subsubsection{a. Equation stationnaire}
On s'intéresse au cas stationnaire, pour lequel la fonction d'onde
est de la forme
\begin{equation}\label{foncsepdirac}
    \Psi(\overrightarrow{r},t)=\psi(r,\theta,\phi)\,\displaystyle
    e^{-\frac{iEt}{\hbar}}.
\end{equation}
Pour déterminer l'équation d'onde stationnaire, remplaçons
(\ref{foncsepdirac}) dans (\ref{eqdiracourbe}). En effet,
\begin{equation}
    \Bigg\{\displaystyle\frac{i\hbar}{\sqrt{\xi(r)}}\frac{\partial}{\partial t}\Bigg\}\;
    \psi(r,\theta,\phi)\,\displaystyle
    e^{-\frac{iEt}{\hbar}}=
    \bigg(\overrightarrow{\alpha}.\overrightarrow{P}c+
    \beta m_{0}c^{2}\bigg)\;\Psi(\overrightarrow{r},t).\nonumber
\end{equation}
Appliquons la dérivée temporelle et divisons les deux membres par $\displaystyle
    e^{-\frac{iEt}{\hbar}}$
\begin{equation}
     \displaystyle\frac{E}{\sqrt{\xi(r)}}\;\psi(r,\theta,\phi)=
    \bigg(\overrightarrow{\alpha}.
    \overrightarrow{P}c+\beta m_{0}c^{2}\bigg)\psi(r,\theta,\phi),\nonumber
\end{equation}
ce qui donne, après simplification
\begin{equation}
\Bigg\{\left(\overrightarrow{\alpha}.\overrightarrow{P}c\right)
+\beta m_{0}c^{2}-\displaystyle\frac{E}{\sqrt{\xi(r)}}\Bigg\}\;\psi(r,\theta,\phi)=0.\\
\end{equation}
Finalement, en tenant compte de (\ref{impulsioncourbe}),
l'équation d'onde stationnaire d'une particule de spin $1/2$,
évoluant dans un espace courbe s'écrit
\begin{eqnarray}\label{eqdiracstatio}
\Bigg\{-i\hbar c\bigg[\alpha_{r}\,\sqrt{\xi(r)}\frac{\partial}{\partial
r}+\alpha_{\theta}\,\frac{1}{r}\frac{\partial}{\partial
\theta}+\alpha_{\phi}\,\frac{1}{r\sin\theta}\frac{\partial}{\partial \phi}\bigg]
+\beta m_{0}c^{2}-\displaystyle\frac{E}{\sqrt{\xi(r)}}\Bigg\}\;\psi(r,\theta,\phi)=0.\nonumber\\
\end{eqnarray}

En représentation matricielle standard, l'équation
(\ref{eqdiracstatio}) se met sous la forme
\begin{eqnarray}
    &&\Bigg\{
\Bigg(
\begin{array}{cc}
0 & \sigma_{1}\\
\sigma_{1} & 0\\
\end{array}
\Bigg)\sqrt{\xi(r)}P_{r}\,c+ \Bigg(
\begin{array}{cc}
0 & \sigma_{2}\\
\sigma_{2} & 0\\
\end{array}
\Bigg)\displaystyle\frac{P_{\theta}\,c}{r} +\Bigg(
\begin{array}{cc}
0 & \sigma_{3}\\
\sigma_{3} & 0\\
\end{array}
\Bigg)\displaystyle\frac{P_{\phi}\,c}{r\sin\theta}\nonumber
\\
\nonumber
\\
&&\hspace{4.4cm}+\;\,\Bigg(
\begin{array}{cc}
1 & 0\\
0 & -1\\
\end{array}
\Bigg)m_{0}c^{2}-\Bigg(
\begin{array}{cc}
1 & 0\\
0 & 1\\
\end{array}
\Bigg)\displaystyle\frac{E}{\sqrt{\xi(r)}}\Bigg\}\Bigg(
\begin{array}{c}
\varphi\\
\chi\\
\end{array}
\Bigg)=\Bigg(
\begin{array}{c}
\varphi\\
\chi\\
\end{array}
\Bigg),\nonumber\\
\end{eqnarray}

ou encore,
\begin{eqnarray}
\left(
\begin{array}{cc}
m_{0}c^{2}-\displaystyle\frac{E}{\sqrt{\xi(r)}} & \sigma_{1}.\sqrt{\xi(r)}cP_{r}+\sigma_{2}.\displaystyle\frac{cP_{\theta}}{r}+\sigma_{3}.\displaystyle\frac{cP_{\phi}}{r\sin\theta}\\
\\
\sigma_{1}.\sqrt{\xi(r)}cP_{r}+\sigma_{2}.\displaystyle\frac{cP_{\theta}}{r}+\sigma_{3}.\displaystyle\frac{cP_{\phi}}{r\sin\theta} & -m_{0}c^{2}-\displaystyle\frac{E}{\sqrt{\xi(r)}}\\
\end{array}
\right)\left(
\begin{array}{c}
\varphi\\
\\
\\
\\
\chi\\
\end{array}
\right)=\left(
\begin{array}{c}
0\\
\\
\\
\\
0\\
\end{array}
\right).\nonumber
\\\nonumber
\\
\end{eqnarray}
Finalement, en représentation standard, l'équation d'onde
stationnaire pour une particule de spin $1/2$, évoluant sous
l'action d'un potentiel non gravitationnel, s'écrit
\begin{eqnarray}
\left(
\begin{array}{cc}
m_{0}c^{2}-\displaystyle\frac{E}{\sqrt{\xi(r)}} & \overrightarrow{\sigma}.\overrightarrow{P}\,c\\
\overrightarrow{\sigma}.\overrightarrow{P}\,c & -m_{0}c^{2}-\displaystyle\frac{E}{\sqrt{\xi(r)}}\\
\end{array}
\right)\left(
\begin{array}{c}
\varphi\\
\\
\\
\chi\\
\end{array}
\right)=\left(
\begin{array}{c}
0\\
\\
\\
0\\
\end{array}
\right).
\end{eqnarray}
C'est l'ensemble de deux équations reliant les deux spineurs
$\varphi$ et $\chi$:
\begin{equation}\label{k}
\bigg(\displaystyle\frac{E}{\sqrt{\xi(r)}}-m_{0}c^{2}\bigg)\varphi=c\,
\overrightarrow{\sigma}.\overrightarrow{P}\,\chi,
\end{equation}
\begin{equation}\label{kk}
\bigg(\displaystyle\frac{E}{\sqrt{\xi(r)}}+m_{0}c^{2}\bigg)\chi=c\,
\overrightarrow{\sigma}.\overrightarrow{P}\,\varphi.
\end{equation}

Il est possible de montrer que (\ref{k}) et (\ref{kk}) sont équivalentes. En effet,
d'après (\ref{k})
$$\varphi=\displaystyle\frac{c\,\overrightarrow{\sigma}.\overrightarrow{P}}
{\bigg(\displaystyle\frac{E}{\sqrt{\xi(r)}}-m_{0}c^{2}\bigg)}\,\chi,$$ en remplaçant
l'expression de $\varphi$ dans (\ref{kk}), alors
\begin{eqnarray}
   \bigg(\displaystyle\frac{E}{\sqrt{\xi(r)}}-m_{0}c^{2}\bigg)
   \bigg(\displaystyle\frac{E}{\sqrt{\xi(r)}}+m_{0}c^{2}\bigg)\chi&=&c^{2}\,
(\overrightarrow{\sigma}.\overrightarrow{P})^{2}\,\chi\nonumber\\
    \bigg(\displaystyle\frac{E^{2}}{\xi(r)}-m_{0}^{2}c^{4}\bigg)&=&c^{2}\,
    \big(\overrightarrow{P}^{2}+i\,\overrightarrow{\sigma}.
    (\underbrace{\overrightarrow{P}\times\overrightarrow{P}}_{=0})\big)\nonumber\\
     \displaystyle\frac{E^{2}}{\xi(r)}&=&c^{2}\,\overrightarrow{P}^{2}+m_{0}^{2}c^{4}.
\end{eqnarray}
Cette relation n'est autre que l'expression de l'invariant relativiste (\ref{nassim}). Il est
clair que chaque composante du spineur $\psi$, décrivant la particule de spin $1/2$ dans le
cas relativiste, vérifie l'équation de la particule de spin 0.

\subsubsection{b. Proposition d'une solution}
Dans un mouvement à champ central sont conservés le moment orbital et la parité (par
rapport au centre du champ, pris comme origine des coordonnées). Ceci nous incite à
chercher la fonction d'onde des états stationnaires, dans la représentation standard,
sous forme\footnote{Voir l'annexe (Chapitre B) pour plus de précisions sur les
arguments physiques permettant de justifier cette forme de solution}
\cite{landauquantrelativ}
\begin{equation}\label{spineurlandau}
    \psi=\Bigg(
\begin{array}{c}
\varphi\\
\chi\\
\end{array}
\Bigg)=\Bigg(
\begin{array}{c}
f(r)\,\Omega_{j\,\ell\,m}\\
(-1)^{\frac{1+\ell-\ell\,^{'}}{2}}g(r)\,\Omega_{j\,\ell\,^{'}\,m}\\
\end{array}
\Bigg),
\end{equation}
où les moments cinétiques orbitaux sont reliés au moment cinétique total $j$, par la
condition:
\begin{eqnarray}
    \Bigg\{
\begin{array}{c}
\ell=j\pm1/2,\\
\ell\,^{'}=2j-\ell.\\
\end{array}
\end{eqnarray}
Les $\Omega_{j\,\ell\,m}$ et $\Omega_{j\,\ell\,^{'}\,m}$ sont des spineurs sphériques,
définis par une certaine combinaison linéaire d'harmoniques sphériques et des vecteurs
propres des matrices de Pauli (voir (\ref{zeta})). En fait, un spineur sphérique est
la fonction d'onde décrivant une particule libre de spin $1/2$ et ayant un moment cinétique
orbital $\ell$. Dans le cas qui nous concerne, ces deux spineurs sont reliés par
l'expression suivante \cite{landauquantrelativ}
\begin{equation}\label{yahya}
    \Omega_{j\,\ell\,^{'}\,m}=i^{\ell-\ell\,^{'}}\bigg(\overrightarrow{\sigma}.
    \frac{\overrightarrow{r}}{r}\bigg)\Omega_{j\,\ell\,m},
\end{equation}

Il faut déterminer les fonctions $f(r)$ et $g(r)$, représentant la
partie radiale du spineur, décrivant une particule de spin $1/2$.
Pour ce faire, remplaçons (\ref{spineurlandau}) dans (\ref{k}). En
effet,
\begin{eqnarray}\label{hrissa}
\bigg(\displaystyle\frac{E}{\sqrt{\xi(r)}}-m_{0}c^{2}\bigg)f(r)\,\Omega_{j\,\ell\,m}&=&c\,(\overrightarrow{\sigma}.\overrightarrow{P})(-1)^{\frac{1+\ell-\ell\,^{'}}{2}}g(r)\Bigg[i^{\ell-\ell\,^{'}}\bigg(\overrightarrow{\sigma}.\frac{\overrightarrow{r}}{r}\bigg)\Omega_{j\,\ell\,m}\Bigg]\nonumber\\
&=&c\,(-i)(\overrightarrow{\sigma}.\overrightarrow{P})(\overrightarrow{\sigma}.\overrightarrow{r})\,\frac{g(r)}{r}\,\Omega_{j\,\ell\,m}.
\end{eqnarray}
Pour déterminer l'expression du second membre de l'équation
précédente, il faut calculer le double produit scalaire
$(\overrightarrow{\sigma}.\overrightarrow{P})(\overrightarrow{\sigma}.\overrightarrow{r})$.
Pour ce faire, rappelons que les matrices de Pauli vérifient les
relations d'anticommutation et de commutation suivantes
\cite{landauquantrelativ}:
\begin{eqnarray}
    \sigma_{i}\, \sigma_{j}+\sigma_{j}\, \sigma_{i}&=&2\delta_{ij}\\
    \sigma_{i}\, \sigma_{j}-\sigma_{j}\, \sigma_{i}&=&2i\displaystyle\sum_{k=1}^{3}\,\epsilon^{i\,j\,k}\,\sigma_{k}
\end{eqnarray}
La somme membre à membre, des deux équations précédentes, nous permet d'avoir
l'expression du produit
de deux matrices de Pauli,
\begin{equation}
    \sigma_{i}\, \sigma_{j}=\delta_{ij}+i\displaystyle\sum_{k=1}^{3}\,\epsilon^{i\,j\,k}\,\sigma_{k}.
\end{equation}
\A présent, il est possible de calculer
$(\overrightarrow{\sigma}.\overrightarrow{P})
(\overrightarrow{\sigma}.\overrightarrow{r})$. En effet,
\begin{eqnarray}
 (\overrightarrow{\sigma}.\overrightarrow{P})(\overrightarrow{\sigma}.\overrightarrow{r})&=&
 \displaystyle\sum_{i=1}^{3}\displaystyle\sum_{j=1}^{3}(\sigma_{i}.P_{i})
 (\sigma_{j}.r_{j})\nonumber\\
 &=&\displaystyle\sum_{i=1}^{3}\displaystyle\sum_{j=1}^{3}P_{i}\,r_{j}
 (\sigma_{i}\,\sigma_{j})\nonumber\\
 &=&\displaystyle\sum_{i=1}^{3}\displaystyle\sum_{j=1}^{3}P_{i}\,r_{j}
 \left(\delta_{ij}+i\displaystyle\sum_{k=1}^{3}\epsilon^{i\,j\,k}\,\sigma_{k}\right)\nonumber\\
 &=&\displaystyle\sum_{i=1}^{3}P_{i}\,r_{i}+i\displaystyle\sum_{i=1}^{3}\displaystyle\sum_{j=1}^{3}\displaystyle\sum_{k=1}^{3}
 \sigma_{k}\,\epsilon^{i\,j\,k}\,P_{i}\,r_{j}\nonumber\\
 &=&\displaystyle\sum_{i=1}^{3}P_{i}\,r_{i}+i\displaystyle\sum_{k=1}^{3}\sigma_{k}(\overrightarrow{P}\times \overrightarrow{r})_{k}\nonumber\\
 &=&\overrightarrow{P}.\overrightarrow{r}+i\,\overrightarrow{\sigma}.(\overrightarrow{P}\times
 \overrightarrow{r}).
\end{eqnarray}
Donc l'équation (\ref{hrissa}) peut s'écrire sous forme:
\begin{eqnarray}\label{king}
 \left(\displaystyle\frac{E}{\sqrt{\xi(r)}}-m_{0}\,c^{2}\right)\;f(r)\;\Omega_{j\,\ell\,m}&=&
 c\,(-i)\left[\overrightarrow{P}.\overrightarrow{r}
 +i\,\overrightarrow{\sigma}.(\underbrace{\overrightarrow{P}\times
 \overrightarrow{r}}_{-\overrightarrow{L}})\right]\;\frac{g(r)}{r}\;\Omega_{j\,\ell\,m}\nonumber\\
&=&c\,(-i)\left[\overrightarrow{P}.\overrightarrow{r}
-i\,\overrightarrow{\sigma}.\overrightarrow{L}\right]\;\frac{g(r)}{r}\;\Omega_{j\,\ell\,m},
\end{eqnarray}
où $\overrightarrow{L}$ représente le moment cinétique orbital. Il
reste à déterminer, d'une part, l'action de
$(\overrightarrow{P}.\overrightarrow{r})$ sur la fonction
$g(r)/r$, et d'autre part, l'action de
$(i\,\overrightarrow{\sigma}.\overrightarrow{L})$ sur
$\Omega_{j\,\ell\,m}$.\\
Commençons par la détermination de l'action de
$(\overrightarrow{P}.\overrightarrow{r})$ sur $g(r)/r$. Pour ce
faire, attirons l'attention sur le fait que $\overrightarrow{r}$
et $\overrightarrow{P}$ ne commutent pas. Pour pouvoir faire agir
l'opérateur $\overrightarrow{P}$ sur la fonction $g(r)/r$, il faut
utiliser les relations de commutations. Celles-ci sont clairement
établies en coordonnées cartésiennes, où on peut écrire
\begin{eqnarray}
    \overrightarrow{P}.\overrightarrow{r}&=&P_{x}.x+P_{y}.y+P_{z}.z\nonumber\\
     &=&x.P_{x}-\big[x,P_{x}\big]+y.P_{y}-\big[y,P_{y}\big]+z.P_{z}-\big[z,P_{z}\big]\nonumber\\
     &=&x.P_{x}+y.P_{y}+z.P_{z}-3i\hbar\nonumber\\
     &=&\overrightarrow{r}.\overrightarrow{P}-3i\hbar
\end{eqnarray}
De plus, conformément à (\ref{produitscalaire}), le même
 produit scalaire, s'écrit
\begin{equation}
    \overrightarrow{r}.\overrightarrow{P}=\displaystyle\sum_{i=1}^{3}
    \overline{r_{i}}.\overline{P_{i}},\nonumber
\end{equation}
donc l'action de $\overrightarrow{r}.\overrightarrow{P}$ sur la
fonction $g(r)/r$ s'exprime sous forme,
\begin{eqnarray}
    \left(\overrightarrow{r}.\overrightarrow{P}\right)\,\left(\frac{g(r)}{r}\right)&=&
    \left(\overline{r_{1}}\,\overline{P_{1}}+\overline{r_{2}}\,\overline{P_{2}}
    +\overline{r_{3}}\,\overline{P_{3}}\right)\,\left(\frac{g(r)}{r}\right)\nonumber
    \\
      &=&\left(\overline{r_{1}}\,\overline{P_{1}}\right)\,\left(\frac{g(r)}{r}\right),
\end{eqnarray}
car les opérateurs $\overline{P_{2}}$ et $\overline{P_{3}}$ sont définis,
respectivement, par des dérivées relatives à $\theta$ et $\phi$, de sorte que
\begin{eqnarray}
    \overline{P_{2}}\left(\frac{g(r)}{r}\right)\sim
    \frac{\partial}{\partial \theta}\left(\frac{g(r)}{r}\right)=0,\nonumber
    \\
    \overline{P_{3}}\left(\frac{g(r)}{r}\right)\sim
    \frac{\partial}{\partial \phi}\left(\frac{g(r)}{r}\right)=0.\nonumber
\end{eqnarray}
Ainsi,
\begin{eqnarray}
    \left(\overrightarrow{r}.\overrightarrow{P}\right)\,\left(\frac{g(r)}{r}\right)&=&
    \displaystyle\frac{r}{\sqrt{\xi(r)}}\;\left[-i\hbar\;\sqrt{\xi(r)}\;\frac{\partial}
    {\partial r}\left(\frac{g(r)}{r}\right)\right]\nonumber\\
       &=&-i\hbar\;\left(\frac{dg}{dr}-\frac{g(r)}{r}\right).
\end{eqnarray}
Finalement, l'action de $(\overrightarrow{P}.\overrightarrow{r})$
sur la fonction $g(r)/r$ est complètement déterminée. Elle s'écrit,
  \begin{eqnarray}\label{zztop1}
    \left(\overrightarrow{P}.\overrightarrow{r}\right)\,\left(\frac{g(r)}{r}\right)&=&
    \left(\overrightarrow{r}.\overrightarrow{P}-3i\hbar\right)\;\left(\frac{g(r)}{r}\right)
    \nonumber\\
    &=&-i\hbar\left(\frac{dg}{dr}-\frac{g(r)}{r}\right)-3i\hbar\;\frac{g(r)}{r}.
\end{eqnarray}
Déterminons, ensuite, l'action de
$(\,\overrightarrow{\sigma}.\overrightarrow{L})$ sur
$\Omega_{j\,\ell\,m}$. Pour cela, d'après (\ref{best}), il est
possible d'exprimer les matrices de Pauli en fonction du spin, ce
qui se traduit par
\begin{eqnarray}\label{lss}
    \left(\overrightarrow{\sigma}.\overrightarrow{L}\right)\;\Omega_{j\,\ell\,m}&=&\hbar^{-1}
    \left(2\overrightarrow{S}.\overrightarrow{L}\right)\;\Omega_{j\,\ell\,m}.
\end{eqnarray}
Le moment cinétique total est la somme du moment cinétique orbital
et spin, en effet
\begin{eqnarray}\label{ls}
    \overrightarrow{J}=\overrightarrow{L}+\overrightarrow{S}\hspace{1cm}\Rightarrow
    \hspace{1cm} \overrightarrow{J}^{2}=\overrightarrow{L}^{2}+\overrightarrow{S}^{2}+2\overrightarrow{L}.\overrightarrow{S}
\end{eqnarray}
D'après (\ref{ls}), l'expression (\ref{lss}) se met sous la forme:
\begin{eqnarray}
    \left(\overrightarrow{\sigma}.\overrightarrow{L}\right)\;\Omega_{j\,\ell\,m}&=&
    \hbar^{-1}\left[\overrightarrow{J}^{2}-\overrightarrow{L}^{2}
    -\overrightarrow{S}^{2}\right]\;\Omega_{j\,\ell\,m}\nonumber\\
  &=&\hbar^{-1}\hbar^{2}\;\left[j(j+1)-l(l+1)-\frac{1}{2}\left(\frac{1}{2}+1\right)\right]\;\Omega_{j\,\ell\,m}\nonumber\\
  &=&\hbar\;\left[j(j+1)-l(l+1)-\frac{3}{4}\right]\;\Omega_{j\,\ell\,m}\nonumber\\
  &=&\Bigg\{
\begin{array}{c}
\hbar\,(j-1/2)\;\Omega_{j\,\ell\,m}\hspace{2cm}\textrm {pour}:\,\ell=j-1/2,\\
\hbar\,(-j-3/2)\;\Omega_{j\,\ell\,m}\hspace{1.7cm}\textrm {pour}:\,\ell=j+1/2.\\
\end{array}
\end{eqnarray}
Pour unifier l'écriture pour $\ell=j\pm 1/2$, il est commode
d'introduire la notation \cite{landauquantrelativ}
\begin{eqnarray}\label{xcondense}
    x=\Bigg\{
\begin{array}{l}
-\left(j+1/2\right)=-(\ell+1)\hspace{2cm}\textrm {pour}:\,j=\ell+1/2,\\
+\left(j+1/2\right)=\ell\hspace{3.4cm}\textrm {pour}:\,j=\ell-1/2.\\
\end{array}
\end{eqnarray}
Avec cette nouvelle notation, on peut écrire de façon condensée
\begin{equation}\label{zztop2}
    \left(\overrightarrow{\sigma}.\overrightarrow{L}\right)\;\Omega_{j\,\ell\,m}=-(1+x)
    \hbar\;\Omega_{j\,\ell\,m}.
\end{equation}
Finalement, en tenant compte des deux étapes intermédiaires
précédentes, c'est-à-dire d'après (\ref{zztop1}) et
(\ref{zztop2}), l'équation (\ref{king}) s'écrit:
\begin{eqnarray}
&&\left(\displaystyle\frac{E}{\sqrt{\xi(r)}}-m_{0}\,c^{2}\right)\;f(r)\;\Omega_{j\,\ell\,m}=
(-i)c\,\Bigg[-i\hbar\left(\frac{dg}{dr}-\frac{g(r)}{r}\right)-3i\hbar\;\frac{g(r)}{r}\nonumber
\\
&&\hspace{11cm}+\,i\hbar\;(1+x)\frac{g(r)}{r}\Bigg]\;\Omega_{j\,\ell\,m}\nonumber
\end{eqnarray}
\begin{eqnarray}
\left(\displaystyle\frac{E}{\sqrt{\xi(r)}}-m_{0}\,c^{2}\right)\;f(r)\;\Omega_{j\,\ell\,m}&=&
-\hbar\,c\left[\frac{dg}{dr}+\displaystyle\frac{(1-x)}{r}\;g\right]\;\Omega_{j\,\ell\,m}
\nonumber\\
\left(\displaystyle\frac{E}{\sqrt{\xi(r)}}-m_{0}\,c^{2}\right)\;f(r)&=&
-\hbar\,c\left[\frac{dg}{dr}+\displaystyle\frac{(1-x)}{r}\,g\right],\nonumber
\end{eqnarray}
on aboutit à l'équation différentielle suivante
\begin{equation}\label{z}
    \frac{dg}{dr}+\displaystyle\frac{(1-x)}{r}\;g=-\displaystyle\frac{1}{\hbar\,c}
    \left(\displaystyle\frac{E}{\sqrt{\xi(r)}}-m_{0}\,c^{2}\right)\;f(r)
\end{equation}

Au lieu de refaire le même calcul pour retrouver la deuxième équation différentielle
(remplacer (\ref{spineurlandau}) dans (\ref{kk}) et refaire des calculs similaires),
on peut déduire la deuxième équation à partir de (\ref{z}). En effet, en comparant
(\ref{k}) et (\ref{kk}), on s'aperçoit qu'il suffit d'effectuer les transformations
suivantes \cite{landauquantrelativ}:
\begin{eqnarray}
    f\rightarrow g,\; g\rightarrow -f,\; m_{0}\rightarrow -m_{0},\; \ell\rightarrow \ell\,^{'},\; x\rightarrow
    -x.
\end{eqnarray}
ainsi, la seconde équation différentielle s'exprime comme suit
\begin{equation}\label{zamalek}
    \frac{df}{dr}+\displaystyle\frac{(1+x)}{r}\;f=\displaystyle\frac{1}{\hbar\,c}
    \left(\displaystyle\frac{E}{\sqrt{\xi(r)}}+m_{0}\,c^{2}\right)\;g(r)
\end{equation}
Finalement, les deux fonctions radiales vérifient le système d'équations
différentielles couplées suivant \cite{barros}:
\begin{eqnarray}\label{sysdiff}
\left\{%
\begin{array}{ll}
    \displaystyle\frac{df}{dr}+\displaystyle\frac{(1+x)}{r}\;f=\displaystyle\frac{1}{\hbar\,c}
    \left(\displaystyle\frac{E}{\sqrt{\xi(r)}}+m_{0}\,c^{2}\right)\;g(r)
    \\
    \\
    \displaystyle\frac{dg}{dr}+\displaystyle\frac{(1-x)}{r}\;g=-\displaystyle\frac{1}{\hbar\,c}
    \left(\displaystyle\frac{E}{\sqrt{\xi(r)}}-m_{0}\,c^{2}\right)\;f(r)\\
\end{array}%
\right.
\end{eqnarray}
Pour un système physique donné, on remplace la fonction $\xi(r)$ relative au problème
en question. La résolution du système d'équations différentielles couplées va
permettre de déterminer la partie radiale de l'état d'une particule de spin $1/2$.

Attirons l'attention sur le fait que C.C.Barros, dans \cite{barros}, ne retrouve pas
exactement le système (\ref{sysdiff}). Il retrouve plutôt le système d'équations
suivant:
\begin{eqnarray}\label{sysdiffbarros}
\left\{%
\begin{array}{ll}
    \sqrt{\xi(r)}\;\,\displaystyle\frac{df}{dr}+\displaystyle\frac{(1+x)}{r}\;f=
    \displaystyle\frac{1}{\hbar\,c}
    \left(\displaystyle\frac{E}{\sqrt{\xi(r)}}+m_{0}\,c^{2}\right)\;g(r)
    \\
    \\
    \sqrt{\xi(r)}\;\,\displaystyle\frac{dg}{dr}+\displaystyle\frac{(1-x)}{r}\;g=
    -\displaystyle\frac{1}{\hbar\,c}
    \left(\displaystyle\frac{E}{\sqrt{\xi(r)}}-m_{0}\,c^{2}\right)\;f(r)\\
\end{array}%
\right.
\end{eqnarray}
On remarque qu'il y a un coefficient supplémentaire, par rapport à (\ref{sysdiff}),
dans chaque équation différentielle. C'est le facteur $\sqrt{\xi(r)}$.

\section{Conclusion}
Dans ce chapitre, deux équations d'ondes quantiques, pour une
particule sans spin et une particule de spin $1/2$, ont été
établies. Ces particules sont soumises à un potentiel non
gravitationnel affectant la métrique de l'espace-temps. Dans ce
travail, la métrique de Schwarzschild a été adoptée. Dans cette
nouvelle approche, au lieu d'introduire l'interaction de la
particule avec le champ par des considérations de symétrie
(couplage minimal), l'interaction est contenue dans la structure
de l'espace-temps. On montre, par la suite, qu'on retrouve des
équations identiques à celles obtenues par couplage minimal, après
avoir effectué certaines approximations. Dans le cas stationnaire,
une méthode de résolution, basée sur une séparation de variables,
a été proposée pour les deux équations d'ondes quantiques.

Le prochain chapitre sera consacré à l'application de la nouvelle théorie à deux
systèmes: l'atome d'hydrogène et l'oscillateur harmonique isotrope à 3 dimensions. Ces
deux systèmes offrent l'avantage d'être totalement décrits par la théorie de la
mécanique quantique, ce qui va permettre de vérifier la validité de la nouvelle
approche.

\newpage
\pagestyle{fancy} \lhead{chapitre\;5}\rhead{Application}
\chapter{Application}
Dans ce chapitre, une application à l'atome d'hydrogène est envisagée \cite{barros}.
Ce système physiques offre l'avantage d'être totalement décrit par la théorie de la
mécanique quantique habituelle, ce qui va permettre de tester la validité de la
nouvelle approche, proposée par Barros. Le test consiste à décrire une particule de
spin $1/2$, par le biais de l'équation d'onde quantique (\ref{eqdiracourbe}). Une
telle particule est plongée dans un potentiel non gravitationnel affectant la métrique
de l'espace-temps.

\section{ Application à l'atome d'hydrogène}
Dans cette section, nous étudions la validité de l'approche de Barros  à un système
pour lequel la solution est parfaitement établie, c'est l'atome d'hydrogène.
L'originalité de l'approche de Barros réside dans le fait de supposer que
l'interaction non gravitationnelle affecte la structure de l'espace temps. Ainsi, dans
cette application la courbure de l'espace-temps n'est pas due à l'interaction
gravitationnelle qui s'exerce entre l'électron et le proton, mais elle est due à
l'interaction électromagnétique. En fait l'interaction gravitationnelle est négligée
dans ce cas, car les masses de l'électron ($m_{e}$) et du proton ($m_{p}$) sont très
petites. L'électron est soumis à un potentiel coulombien à symétrie sphérique:
\begin{equation}\label{coulombpotent}
    V(r)=-\;\displaystyle\frac{\alpha Z}{r},
\end{equation}
où $Z$ est le numéro atomique (dans le cas de l'atome d'hydrogène $Z=1$) et la
constante $\alpha$ est le produit de la charge du proton par la charge de l'électron
par $(1/4\pi\varepsilon_{0})$, tel que $\alpha=e^{2}/4\pi\varepsilon_{0}$.

\A partir de l'expression de l'énergie potentielle précédente, on
détermine l'expression de la fonction $\xi(r)$, qui s'écrit
conformément à (\ref{xi}) dans le cas de l'atome d'hydrogène,
\begin{equation}\label{xiep}
    \xi_{ep}(r)=\bigg(1-\displaystyle\frac{\alpha Z}{m_{e}c^{2}r}\bigg)^{2}.
\end{equation}
C'est à travers cette fonction que l'interaction coulombienne électron-proton $V(r)$
affecte la structure de l'espace-temps.

On appelle rayon de Schwarzschild $r_{s}$, le rayon pour lequel la métrique de
Schwarzschild est "brisée" (ou n'est plus valable). Le rayon de Schwarzschild est
défini par la condition $\xi(r_{s})=0$, ce qui conduit à une infinité dans
l'expression de la métrique (\ref{metrique}). Alors que dans la théorie de la
relativité générale, le rayon de Schwarzschild est toujours négligeable par rapport
aux dimensions de la source du champ ( par exemple pour le soleil
$r_{s}=2.95\textrm{Km}\ll r_{\textrm{Soleil}}$), par contre pour l'atome d'hydrogène,
il ne peut plus être négligé. En effet \cite{barros},
\begin{equation}\label{l}
 \xi_{ep}(r)=0\;\Rightarrow \; r_{s}=\displaystyle\frac{\alpha}{m_{e}c^{2}}=2.818
 \;\textrm{fm},
 \nonumber
\end{equation}
le rayon de Schwarzschild de l'atome d'hydrogène  est comparable
au rayon classique de l'électron. Cette circonstance est exploitée
pour étudier le confinement des quarks \cite{barrosquark}. Pour
étudier le cas intérieur $r<r_{s}$, il faut tenir compte du
tenseur énergie-impulsion $T_{\mu\nu}$, déterminé par la charge et
la distribution de la matière à l'intérieur de la source du champ
\cite{barrosbis,barrosstability}.

Pour déterminer les niveaux d'énergie de l'atome d'hydrogène, on
s'intéresse dans ce qui suit au cas $r>r_{s}$. Rappelons qu'en
théorie quantique relativiste "habituelle", ces niveaux d'énergie
sont donnés par l'expression \cite{landauquantrelativ}:
\begin{equation}\label{niveauenrjlandau}
\displaystyle\frac{E_{n}}{m_{e}\,c^{2}}=
\left[1+\displaystyle\frac{\left(\frac{Z\,\alpha}{\hbar\,c}\right)^{2}}
{\left(\;\sqrt{x^{2}-\left(\frac{Z\,\alpha}{\hbar\,c}\right)^{2}}
+n\right)^{2}}\;\right]^{-1/2},
\end{equation}
telle que $n$ est un entier et $x$ est défini par
(\ref{xcondense}).

Le test consiste à décrire l'électron évoluant sous l'influence du
potentiel coulombien (\ref{coulombpotent}), crée par le proton du
noyau. Une telle description se fait par le biais de l'équation
d'onde quantique (\ref{eqdiracourbe}), et il faudrait rechercher
les niveaux d'énergie et les comparer à (\ref{niveauenrjlandau}).

L'électron est décrit par le spineur à 4 composantes
(\ref{spineurlandau}). L'injection d'une telle forme de solution
dans l'équation quantique (\ref{eqdiracourbe}), permet de
retrouver un système d'équation différentielles couplées
(\ref{sysdiff}). La résolution d'un tel système d'équations permet
de déterminer les fonctions $f(r)$ et $g(r)$, qui définissent
complètement la partie radiale de la fonction d'onde de
l'électron, et permet aussi de déterminer les énergies de l'atome
d'hydrogène. Dans le but de résoudre le système d'équation
différentielles (\ref{sysdiff}), écrit pour le cas de l'atome
d'hydrogène, on effectue un certain nombre d'approximations qui
nous permettent d'aboutir à des équations identiques à celles
obtenues pour l'étude de l'atome d'hydrogène par l'équation de
Dirac. \A partir de ce moment, on propose une méthode basée sur la
résolution par des séries pour retrouver les niveaux d'énergie
(\ref{niveauenrjlandau}).

\subsection{Approximations}
En tenant compte de l'expression de la fonction $\xi_{ep}(r)$
(\ref{xiep}), alors le système d'équations différentielle
(\ref{sysdiff}) s'écrit, dans le cas de l'atome d'hydrogène, sous
la forme suivante:
\begin{eqnarray}
\displaystyle\frac{df}{dr}+\displaystyle\frac{(1+x)}{r}\,f=\frac{1}{\hbar\,c}\;
\left[\displaystyle\frac{E}
{1+\frac{V(r)}{E_{0}}}+m_{e}\,c^{2}\right]g(r),&&\label{eqdif1}\\
\displaystyle\frac{dg}{dr}+\displaystyle\frac{(1-x)}{r}\,g=-\frac{1}{\hbar\,c}\;
\left[\displaystyle\frac{E}{1+\frac{V(r)}{E_{0}}}-m_{e}\,c^{2}\right]f(r),&&\label{eqdif2}
\end{eqnarray}
où $E_{0}=m_{e}c^{2}$ représente l'énergie au repos de l'électron.

L'approximation qu'on va effectuer est basée sur le fait que
l'énergie potentielle $V(r)$ est très petite par rapport à
l'énergie au repos $E_{0}$, de sorte que
$$\varepsilon=\frac{V(r)}{E_{0}}\ll1.$$
Il est donc possible de faire un développement limité en puissances de $\varepsilon$.
On se contente à un développement à l'ordre 1,
$$\frac{1}{1-\varepsilon}\simeq 1+\varepsilon+\mathcal{O}(\varepsilon^{2}).$$
En utilisant un tel développement, l'équation (\ref{eqdif1}) s'écrit comme suit:
\begin{eqnarray}
\displaystyle\frac{df}{dr}+\displaystyle\frac{(1+x)}{r}\,f&\simeq&\frac{1}{\hbar\,c}\;
\Bigg[E\bigg(1-\frac{V(r)}{E_{0}}\bigg)+m_{e}\,c^{2}\Bigg]g(r)\nonumber\\
&\simeq&\frac{1}{\hbar\,c}\;\bigg(E-\frac{E\,V(r)}{E_{0}}+m_{e}\,c^{2}\bigg)g(r).
\label{eqdif11}
\end{eqnarray}
De même, l'équation (\ref{eqdif2}) s'écrit aussi sous la forme:
\begin{eqnarray}
\displaystyle\frac{dg}{dr}+\displaystyle\frac{(1-x)}{r}\,g&\simeq&-\frac{1}{\hbar\,c}\;
\Bigg[E\bigg(1-\frac{V(r)}{E_{0}}\bigg)-m_{e}\,c^{2}\Bigg]f(r)\nonumber\\
&\simeq&-\frac{1}{\hbar\,c}\;\bigg(E-\frac{E\,V(r)}{E_{0}}-m_{e}\,c^{2}\bigg)f(r).
\label{eqdif22}
\end{eqnarray}
Dans la théorie de Dirac, on s'intéresse à des énergies $E$
d'ordre de grandeur comparable à l'énergie au repos $E_{0}$, de
sorte que $E/E_{0}\sim 1$ (lower moments). Les équations
(\ref{eqdif11}) et (\ref{eqdif22}) se réduisent alors pour former
le système d'équations différentielles suivant:
\begin{eqnarray}\label{sysdiffdirac}
\left\{%
\begin{array}{ll}
    \displaystyle\frac{df}{dr}+\displaystyle\frac{(1+x)}{r}\,f=\frac{1}{\hbar\,c}\;
    \bigg(E-V(r)+m_{e}\,c^{2}\bigg)g(r),\\
    \displaystyle\frac{dg}{dr}+\displaystyle\frac{(1-x)}{r}\,g=-\frac{1}{\hbar\,c}\;
    \bigg(E-V(r)-m_{e}\,c^{2}\bigg)f(r),\\
\end{array}%
\right.
\end{eqnarray}
Ce système d'équations est identique au système d'équations obtenu
pour décrire l'électron de l'atome d'hydrogène par le biais de la
théorie de Dirac. (Voir \cite{landauquantrelativ})

\A partir de ce moment, on peut affirmer que le test de la
nouvelle approche de Barros pour la description de l'atome
d'hydrogène est concluant. En effet, moyennant certaines
approximations, on a pu retrouver les mêmes équations que celles
retrouvées par le biais de la théorie de Dirac. On peut, ainsi,
conclure que la nouvelle approche de Barros permet de décrire
correctement l'atome d'hydrogène.

\subsection{Résolution du système d'équations différentielles couplées}
Dans le but de déterminer les niveaux d'énergie de l'atome
d'hydrogène, il faut résoudre le système d'équations
différentielles couplées (\ref{sysdiffdirac}). Pour ce faire, nous
optons pour une résolution par la méthode des séries. Avant de le
faire, effectuons le changement de variable suivant,
\begin{equation}
 \rho =\beta\;r \label{changvar},
\end{equation}
où $\beta$ est un paramètre ajustable qui sera déterminé
ultérieurement. Avec ce changement de variable, le système
d'équation (\ref{sysdiffdirac}) s'écrit sous forme:
\begin{eqnarray}
\displaystyle\frac{df}{d\rho}+\left(1+x\right)\frac{f}{\rho}=\left(\frac
{E+m_{e}\,c^{2}}{\hbar\,c\,\beta}\right)g-
\left(\frac{V(\rho)}{\hbar\,c\,\beta}\right)g,&&\label{eqdifro1}
\\
\displaystyle\frac{dg}{d\rho}+\left(1+x\right)\frac{g}{\rho}=-\left(\frac
{E-m_{e}\,c^{2}}{\hbar\,c\,\beta}\right )f-\left(\frac{
V(\rho)}{\hbar\,c\,\beta}\right)f.\label{eqdifro2}
\end{eqnarray}

Proposons des solutions sous forme de séries de type Frobinius \cite{barros}:
\begin{eqnarray}
f=\sum_{n=0}^{N}a_{n}\;\rho^{n+s}\;e^{-\rho},\label{frobinius1}\\
g=\sum_{n=0}^{N}b_{n}\;\rho^{n+s}\;e^{-\rho},\label{frobinius2}
\end{eqnarray}
où $s$ est un autre paramètre ajustable, à déterminer
ultérieurement. En fait, l'introduction des paramètres $\beta$ et
$s$ permet d'avoir un éventail de solutions très large. Pour
pouvoir exploiter les solutions précédentes dans les équations
(\ref{eqdifro1}) et (\ref{eqdifro2}), calculons d'abord les
dérivées premières. Elles sont données par les expressions
suivantes:
\begin{eqnarray}
\frac{df}{d\rho}&=&\displaystyle
\sum_{n=0}^{N}a_{n}\left[(n+s)\;\rho^{n+s-1}-\rho^{n+s}
\right]e^{-\rho},\label{fprim}\\
\frac{dg}{d\rho}&=&\displaystyle
\sum_{n=0}^{N}b_{n}\left[(n+s)\;\rho^{n+s-1}-\rho^{n+s} \right]e^{-\rho}.\label{gprim}
\end{eqnarray}
En tenant compte de (\ref{frobinius1}), (\ref{frobinius2}) et
(\ref{fprim}), alors l'équation différentielle (\ref{eqdifro1})
s'écrit sous la forme algébrique suivante,
\begin{eqnarray}
    &&\displaystyle\sum_{n=0}^{N}a_{n}\Bigg[(n+s)\;\rho^{n+s-1}-\rho^{n+s}\Bigg]
    e^{-\rho}+(1+x)\displaystyle\sum_{n=0}^{N}a_{n}\;\rho^{n+s-1}\;e^{-\rho}\nonumber\\
    &&\hspace{1.5cm}=\left(\displaystyle\frac{E+m_{e}\,c^{2}}{\hbar\,c\,\beta}\right)\displaystyle\sum_{n=0}^{N}b_{n}\;
    \rho^{n+s}\;e^{-\rho}+\left(\frac{\alpha\,Z}{\hbar\,c}\right)\;
    \displaystyle\sum_{n=0}^{N}b_{n}\;\rho^{n+s-1}\;e^{-\rho}.
\end{eqnarray}
En regroupant les termes de même puissance en exponentielle,
l'expression précédente se met sous la forme,
\begin{eqnarray}
&&\displaystyle\sum_{n=0}^{N}\rho^{n+s-1}\Bigg[a_{n}\left[(n+s)+(1+x)\right]-b_{n}
\left(\frac{\alpha\,Z}{\hbar\,c}\right)\Bigg]\nonumber\\
&&\hspace{5cm}-\displaystyle\sum_{n=0}^{N}\rho^{n+s}\Bigg[a_{n}+b_{n}
\left(\displaystyle\frac{E+m_{e}\,c^{2}}{\hbar\,c\,\beta}\right)\Bigg]=0\;.\label{algebrdif1}
\end{eqnarray}
De la même façon, en tenant compte de (\ref{frobinius1}),
(\ref{frobinius2}) et (\ref{gprim}), l'équation différentielle
(\ref{eqdifro2}) s'écrit sous la forme algébrique suivante,
\begin{eqnarray}
    &&\displaystyle\sum_{n=0}^{N}b_{n}\Bigg[(n+s)\;\rho^{n+s-1}-\rho^{n+s}\Bigg]
    e^{-\rho}+(1-x)\displaystyle\sum_{n=0}^{N}b_{n}\;\rho^{n+s-1}\;e^{-\rho}\nonumber\\
    &&\hspace{1.5cm}=-\left(\frac{E-m_{e}\,c^{2}}{\hbar\,c\,\beta}\right)
    \displaystyle\sum_{n=0}^{N}a_{n}\;
    \rho^{n+s}\;e^{-\rho}-\left(\frac{\alpha\,Z}{\hbar\,c}\right)
    \displaystyle\sum_{n=0}^{N}a_{n}\;\rho^{n+s-1}\;e^{-\rho},
\end{eqnarray}
ou encore,
\begin{eqnarray}
&&\displaystyle\sum_{n=0}^{N}\rho^{n+s-1}\Bigg[b_{n}\left[(n+s)+(1-x)\right]+a_{n}
\left(\frac{\alpha\,Z}{\hbar\,c}\right)\Bigg]\nonumber\\
&&\hspace{5cm}-\displaystyle\sum_{n=0}^{N}\rho^{n+s}\Bigg[b_{n}-a_{n}
\left(\frac{E-m_{e}\,c^{2}}{\hbar\,c\,\beta}\right)\Bigg]=0\;.\label{algebrdif2}
\end{eqnarray}
Donc le système d'équations différentielles formé par
(\ref{eqdifro1}) et (\ref{eqdifro2}) s'écrit sous la forme
algébrique suivante:
\begin{eqnarray}\label{sysalg1}
    \displaystyle\sum_{n=0}^{N}\rho^{n+s-1}\Bigg[a_{n}\left[(n+s)+(1+x)\right]-b_{n}
\left(\frac{\alpha\,Z}{\hbar\,c}\right)\Bigg]
-\displaystyle\sum_{n=0}^{N}\rho^{n+s}\Bigg[a_{n}+b_{n}
\left(\displaystyle\frac{E+m_{e}\,c^{2}}{\hbar\,c\,\beta}\right)\Bigg]=0\;,\hspace{0.4cm}
\label{eqalg1}
\\
\nonumber
\\
\nonumber
\\
\displaystyle\sum_{n=0}^{N}\rho^{n+s-1}\Bigg[b_{n}\left[(n+s)+(1-x)\right]+a_{n}
\left(\frac{\alpha\,Z}{\hbar\,c}\right)\Bigg]
-\displaystyle\sum_{n=0}^{N}\rho^{n+s}\Bigg[b_{n}-a_{n}
\left(\frac{E-m_{e}\,c^{2}}{\hbar\,c\,\beta}\right)\Bigg]=0\;.\hspace{0.4cm}\label{eqalg2}
\end{eqnarray}
Pour pouvoir "connecter" les deux séries
$\displaystyle\sum_{n=0}^{N}\rho^{n+s-1}(...)$ et
$\displaystyle\sum_{n=0}^{N}\rho^{n+s}(...)$, qui figurent dans les deux équations
(\ref{eqalg1}) et (\ref{eqalg2}), effectuons le changement d'indices qui suit:
\begin{eqnarray}
\left\{%
\begin{array}{ll}
    n=n^{'}-1 \Longrightarrow n^{'}=n+1, \nonumber\\
    \rho^{n+s}=\rho^{n^{'}+s-1}\nonumber, \\
    n:0\rightarrow N \Longrightarrow n^{'}:1\rightarrow N+1. \nonumber\\
\end{array}%
\right.
\end{eqnarray}
Avec ces transformations, les équations (\ref{eqalg1}) et
(\ref{eqalg2})  s'écrivent sous la forme,
\begin{eqnarray}
&&\displaystyle\sum_{n=0}^{N}\rho^{n+s-1}\Bigg[a_{n}\left[(n+s)+(1+x)\right]-b_{n}
\left(\frac{\alpha\,Z}{\hbar\,c}\right)\Bigg]\nonumber\\
&&\hspace{4cm}-\displaystyle\sum_{n^{'}=1}^{N+1}\rho^{n^{'}+s-1}\Bigg[a_{n^{'}-1}+b_{n^{'}-1}
\left(\displaystyle\frac{E+m_{e}\,c^{2}}{\hbar\,c\,\beta}\right)\Bigg]=0\;,\label{algebrprim1}
\\
\nonumber
\\
\nonumber
\\
&&\displaystyle\sum_{n=0}^{N}\rho^{n+s-1}\Bigg[b_{n}\left[(n+s)+(1-x)\right]+a_{n}
\left(\frac{\alpha\,Z}{\hbar\,c}\right)\Bigg]\nonumber\\
&&\hspace{4cm}-\displaystyle\sum_{n^{'}=1}^{N+1}\rho^{n^{'}+s-1}\Bigg[b_{n^{'}-1}-a_{n^{'}-1}
\left(\frac{E-m_{e}\,c^{2}}{\hbar\,c\,\beta}\right)\Bigg]=0\;.\label{algebrprim2}
\end{eqnarray}
Il est maintenant possible d'écrire chacune des deux équations
précédentes sous forme d'une seule série. Pour ce faire, il faut
écrire les termes supplémentaires de chaque série; le terme en
$n=0$ pour la première série et le terme en $n^{'}=N+1$ pour la
deuxième série. Les équations (\ref{algebrprim1}) et
(\ref{algebrprim2}) se mettent finalement sous la forme:
\begin{eqnarray}
&&\rho^{s-1}\Bigg[a_{0}\left[s+(1+x)\right]-b_{0}\left(\frac{\alpha\,Z}{\hbar\,c}\right)
 \Bigg]-\rho^{N+s}\Bigg[a_{N}+b_{N}\left(\frac{E+m_{e}\,c^{2}}{\hbar\,c\,\beta}\right)
 \Bigg]\nonumber\\
&&+\displaystyle\sum_{n=1}^{N}\rho^{n+s-1}\Bigg[a_{n}\left[(n+s)+(1+x)\right]-b_{n}
\left(\frac{\alpha\,Z}{\hbar\,c}\right)-a_{n-1}-b_{n-1}
\left(\frac{E+m_{e}\,c^{2}}{\hbar\,c\,\beta}\right)\Bigg]=0,\label{algebrfinal1}\hspace{1.2cm}
\\
\nonumber
\\
\nonumber
\\
&&\rho^{s-1}\Bigg[b_{0}\left[s+(1-x)\right]+a_{0}\left(\frac{\alpha\,Z}{\hbar\,c}\right)
\Bigg]-\rho^{N+s}\Bigg[b_{N}-a_{N}
\left(\frac{E-m_{e}\,c^{2}}{\hbar\,c\,\beta}\right)\Bigg]\nonumber\\
&&+\displaystyle\sum_{n=1}^{N}\rho^{n+s-1}\Bigg[b_{n}\left[(n+s)+(1-x)\right]+a_{n}
\left(\frac{\alpha\,Z}{\hbar\,c}\right)-b_{n-1}+a_{n-1}
\left(\frac{E-m_{e}\,c^{2}}{\hbar\,c\,\beta}\right)\Bigg]=0.\label{algebrfinal2}\hspace{1.2cm}
\end{eqnarray}

Les équations (\ref{algebrfinal1}) et (\ref{algebrfinal2}) sont
vraies si et seulement si tous les coéfficients qui accompagnent
les puissances de $\rho$ sont nuls. On déduit, ainsi d'une part,
les conditions initiales suivantes:
\begin{eqnarray}
a_{0}\bigg[s+(1+x)\bigg]-b_{0}\left(\frac{\alpha\,Z}{\hbar\,c}\right)&=&0,\label{aobo}\\
b_{0}\bigg[s+(1-x)\bigg]+a_{0}\left(\frac{\alpha\,Z}{\hbar\,c}\right)&=&0,\label{boao}
\end{eqnarray}
\begin{eqnarray}
a_{N}+b_{N}\left(\frac{E+m_{e}\,c^{2}}{\hbar\,c\,\beta}\right)&=&0,\label{anbn}\\
b_{N}-a_{N}\left(\frac{E-m_{e}\,c^{2}}{\hbar\,c\,\beta}\right)&=&0,\label{bnan}
\end{eqnarray}
et d'autre part, les relations de récurrences suivantes:
\begin{eqnarray}
a_{n}\bigg[(n+s)+(1+x)\bigg]-b_{n}
\left(\frac{\alpha\,Z}{\hbar\,c}\right)-a_{n-1}-b_{n-1}
\left(\frac{E+m_{e}\,c^{2}}{\hbar\,c\,\beta}\right)&=&0,\label{recurence1}
\\
b_{n}\bigg[(n+s)+(1-x)\bigg]+a_{n}
\left(\frac{\alpha\,Z}{\hbar\,c}\right)-b_{n-1}+a_{n-1}
\left(\frac{E-m_{e}\,c^{2}}{\hbar\,c\,\beta}\right)&=&0,\label{recurence2}
\end{eqnarray}
tel que $n=1,..,N$.

La prochaine étape consiste à déterminer les paramètres $s$ et
$\beta$, introduits dans le but d'avoir une plus grande
flexibilité dans le choix de la forme des solutions.

\subsubsection{a. Détermination du paramètre $s$}
La détermination du paramètre $s$ figurant dans les solutions
(\ref{frobinius1}) et (\ref{frobinius2}), se fait par le biais du
système d'équations linéaires (\ref{aobo}) et (\ref{boao}). Un tel
système n'admet de solution non triviale que si et seulement si le
déterminant du système est nul,
\begin{eqnarray}
\left|
\begin{array}{cc}
s+(1+x) & -\left(\frac{\alpha\,Z}{\hbar\,c}\right)\\
\nonumber\\
\left(\frac{\alpha\,Z}{\hbar\,c}\right) & s+(1-x)
\end{array}
\right|=0.
\end{eqnarray}
Donc, pour assurer que $a_{0}\neq0$ et $b_{0}\neq0$, le paramètre
$s$ doit vérifier l'équation du second degré en $s$ suivante,
\begin{equation}\label{zcarre}
(s+1)^{2}-x^{2}+\left(\frac{\alpha\,Z}{\hbar\,c}\right)^{2}=0.
\end{equation}
Pour avoir une fonction d'onde convergente au voisinage de 0,
$\psi(r\sim0)=0$, il faut que les fonctions d'ondes radiales
$f(\rho\sim0)$ et $g(\rho\sim0)$ soient également convergentes.
Pour ce faire, on ne retient que la valeur
\begin{equation}
s=-1+\sqrt{x^{2}-\left(\frac{\alpha\,Z}{\hbar\,c}\right)^{2}}.\label{sracine}
\end{equation}
En fait, la valeur négative
$s=-1-\sqrt{x^{2}-\left(\frac{\alpha\,Z}{\hbar\,c}\right)^{2}}$, conduit à une
infinité,
$f(\rho\sim0)=(a_{0}\;\rho^{s}+a_{1}\;\rho^{s+1}+...)\;e^{-\rho}\longrightarrow\infty.$

\subsubsection{b. Détermination du paramètre $\beta$}
Le paramètre $(\beta)$ a été introduit dans le changement de
variable (\ref{changvar}) et il sera choisi de telle sorte à avoir
$a_{N}\neq0$ et $b_{N}\neq0$. Pour ce faire, remarquons que le
système d'équations linéaires formé par (\ref{anbn}) et
(\ref{bnan}), n'admet de solution non triviale que si et seulement
si son déterminant est nul,
\begin{eqnarray}
\left|
\begin{array}{cc}
1 & \left(\frac{E+m_{e}\,c^{2}}{\hbar\,c\,\beta}\right)\\
-\left(\frac{E-m_{e}\,c^{2}}{\hbar\,c\,\beta}\right) & 1\\
\end{array}
\right|=0.
\end{eqnarray}
Cette condition conduit à la relation,
\begin{equation}\label{betacarre}
\beta^{2}=\frac{\left(\,m_{e}\,c^{\,2}\,\right)^{2}-E^{2}}{(\hbar\,c)^{2}}.
\end{equation}
Comme $r>0$, on voudrait bien que la nouvelle variable soit aussi positive $\rho>0$,
alors on opte pour la valeur positive,
\begin{equation}\label{betaracine}
\beta=\frac{1}{\hbar\,c}\;\sqrt{\left(\,m_{e}\,c^{\,2}\,\right)^{2}-E^{2}}.
\end{equation}

\subsubsection{c. Détermination de l'énergie $E_{N}$}
Considérons les relations de récurrence (\ref{recurence1}) et
(\ref{recurence2}). Connaissant les coefficients $a_{0}$ et
$b_{0}$ (donnés par un choix arbitraire), il est possible de
déterminer les coefficients $a_{1}$ et $b_{1}$. De manière plus
générale, la connaissance des coefficients $(a_{n-1},b_{n-1})$
permet de déterminer les coefficients $(a_{n},b_{n})$. Pour
déterminer $a_{n}$ et $b_{n}$ en fonction de $a_{n-1}$ et
$b_{n-1}$, réécrivons les relations de récurrences
(\ref{recurence1}) et (\ref{recurence2}) sous la forme d'un
système d'équations linéaires,
\begin{eqnarray}\label{recsyslineaire}
\left(
\begin{array}{cc}
\big[(n+s)+(1+x)\big] & -\left(\frac{\alpha\,Z}{\hbar\,c}\right)\\
\\
+\left(\frac{\alpha\,Z}{\hbar\,c}\right) & \big[(n+s)+(1-x)\big]\\
\end{array}
\right) \left(
\begin{array}{l}
a_{n} \\
\\
b_{n}\\
\end{array}
\right)=\left(
\begin{array}{l}
a_{n-1}+b_{n-1}\left(\frac{E+m_{e}\,c^{2}}{\hbar\,c\,\beta}\right)\\
\\
b_{n-1}-a_{n-1}\left(\frac{E-m_{e}\,c^{2}}{\hbar\,c\,\beta}\right)\\
\end{array}
\right),\nonumber\\
\end{eqnarray}
tel que $n=1,..,N$. La condition d'existence d'une solution unique
pour un tel système d'équations est que le déterminant soit non
nul (condition d'inversion d'une matrice):
\begin{eqnarray}
\Delta=\left|
\begin{array}{cc}
\big[(n+s)+(1+x)\big] & -\left(\frac{\alpha\,Z}{\hbar\,c}\right)\\
\\
+\left(\frac{\alpha\,Z}{\hbar\,c}\right) & \big[(n+s)+(1-x)\big]\\
\end{array}
\right|\neq 0.
\end{eqnarray}
Cette condition s'écrit de manière explicite comme suit:
\begin{eqnarray}
\Delta&=&\big[(n+s+1)+x\big]\;\big[(n+s+1)-x\big]+
\left(\frac{\alpha\,Z}{\hbar\,c}\right)^{2}\nonumber\\
&=&(n+s+1)^{2}-x^{2}+\left(\frac{\alpha\,Z}{\hbar\,c}\right)^{2}\nonumber \\
&=&n^{2}+2n(s+1)+\left[(s+1)^{2}-x^{2}+\left(\frac{\alpha\,Z}{\hbar\,c}\right)^{2}
\right].\nonumber
\end{eqnarray}
Conformément à (\ref{zcarre}), on aboutit aux valeurs interdites,
\begin{equation}\label{deltaneqzero}
    \Delta=n\left[n+2(s+1)\right]\neq 0.
\end{equation}
Dans le cas où (\ref{deltaneqzero}) est vérifiée, la solution du
système linéaire (\ref{recsyslineaire}) s'écrit comme suit:
\begin{eqnarray}\label{abig}
&&a_{n}=\left[
\begin{array}{l}
\displaystyle\frac{\big[(n+s)+(1-x)\big]}{n\big[n+2(s+1)\big]}-
\displaystyle\frac{\left(\frac{\alpha\,Z}{\hbar\,c}\right)
\left(\frac{E-m_{e}\,c^{2}}{\hbar\,c\,\beta}\right)}{n\big[n+2(s+1)\big]}
\end{array}
\right]a_{n-1}\nonumber\\
\nonumber\\
&&\hspace{2cm}+\left[
\begin{array}{l}
\displaystyle\frac{\left(\frac{E+m_{e}\,c^{2}}{\hbar\,c\,\beta}\right)\big[(n+s)+(1-x)\big]}
{n\big[n+2(s+1)\big]}+\displaystyle\frac{\left(\frac{\alpha\,Z}{\hbar\,c}\right)}
{n\big[n+2(s+1)\big]}
\end{array}
\right]b_{n-1}
\end{eqnarray}

\begin{eqnarray}\label{bbig}
&&b_{n}=-\left[
\begin{array}{l}
\displaystyle\frac{\left(\frac{E-m_{e}\,c^{2}}{\hbar\,c\,\beta}\right)
\big[(n+s)+(1+x)\big]}
{n\big[n+2(s+1)\big]}+\displaystyle\frac{\left(\frac{\alpha\,Z}{\hbar\,c}\right)}
{n\big[n+2(s+1)\big]}
\end{array}
\right]a_{n-1}\nonumber\\
\nonumber\\
&&\hspace{4cm}+\left[
\begin{array}{l}
\displaystyle\frac{\big[(n+s)+(1+x)\big]}{n\big[n+2(s+1)\big]}
-\displaystyle\frac{\left(\frac{E+m_{e}\,c^{2}}{\hbar\,c\,\beta}\right)
\left(\frac{\alpha\,Z}{\hbar\,c}\right)}{n\big[n+2(s+1)\big]}
\end{array}
\right]b_{n-1}
\end{eqnarray}

La prochaine étape consiste à vérifier la condition
(\ref{bnan})(ou bien (\ref{anbn})), par le biais des relations de
récurrence (\ref{abig}) et (\ref{bbig}), pour la valeur $n=N$.
Pour ce faire, réécrivons la condition (\ref{bnan}) sous la forme,
\begin{equation}\label{zourbous}
b_{N}=a_{N}\left(\frac{E-m_{e}\,c^{2}}{\hbar\,c\,\beta}\right),
\end{equation}
ensuite exprimons chacun des deux termes de l'équation
(\ref{zourbous}), par le biais des relations de récurrence
(\ref{abig}) et (\ref{bbig}), pour $n=N$. D'une part, le premier
terme s'exprime comme suit:
\begin{eqnarray}\label{nabig}
&&b_{N}=-\left[
\begin{array}{l}
\displaystyle\frac{\left(\frac{E-m_{e}\,c^{2}}{\hbar\,c\,\beta}\right)
\big[(N+s)+(1+x)\big]}
{N\big[N+2(s+1)\big]}+\displaystyle\frac{\left(\frac{\alpha\,Z}{\hbar\,c}\right)}
{N\big[N+2(s+1)\big]}
\end{array}
\right]a_{N-1}\nonumber\\
\nonumber\\
&&\hspace{3cm}+\left[
\begin{array}{l}
\displaystyle\frac{\big[(N+s)+(1+x)\big]}{N\big[N+2(s+1)\big]}
-\displaystyle\frac{\left(\frac{E+m_{e}\,c^{2}}{\hbar\,c\,\beta}\right)
\left(\frac{\alpha\,Z}{\hbar\,c}\right)}{N\big[N+2(s+1)\big]}
\end{array}
\right]b_{N-1},
\end{eqnarray}
et d'autre part, le second membre s'exprime aussi sous la forme:
\begin{eqnarray}\label{nbbig}
&&a_{N}\left(\frac{E-m_{e}\,c^{2}}{\hbar\,c\,\beta}\right)=\left[
\begin{array}{l}
\displaystyle\frac{\left(\frac{E-m_{e}\,c^{2}}{\hbar\,c\,\beta}\right)
\big[(N+s)+(1-x)\big]}{N\big[N+2(s+1)\big]}-
\displaystyle\frac{\left(\frac{\alpha\,Z}{\hbar\,c}\right)
\left(\frac{E-m_{e}\,c^{2}}{\hbar\,c\,\beta}\right)^{2}}{N\big[N+2(s+1)\big]}
\end{array}
\right]a_{N-1}\nonumber\\
\nonumber\\
&&\hspace{1cm}+\left[
\begin{array}{l}
\displaystyle\frac{\overbrace{\left(\frac{E^{2}-(m_{e}\,c^{2})^{2}}{(\hbar\,c\,\beta)^{2}}
\right)}^{-1}\big[(N+s)+(1-x)\big]}
{N\big[N+2(s+1)\big]}+\displaystyle\frac{\left(\frac{\alpha\,Z}{\hbar\,c}\right)
\left(\frac{E-m_{e}\,c^{2}}{\hbar\,c\,\beta}\right)}
{N\big[N+2(s+1)\big]}
\end{array}
\right]b_{N-1}.
\end{eqnarray}

En tenant compte de (\ref{betacarre}), l'identification membre à
membre de (\ref{nabig}) et (\ref{nbbig}) nous conduit aux deux
équations suivantes:
\begin{eqnarray}
 &&-\left(\frac{E-m_{e}\,c^{2}}{\hbar\,c\,\beta}\right)\big[(N+s)+(1-x)\big]
  -\left(\frac{\alpha\,Z}{\hbar\,c}\right)=\nonumber\\
  &&\hspace{2cm}\left(\frac{E-m_{e}\,c^{2}}{\hbar\,c\,\beta}\right)\big[(N+s)+(1-x)\big]
  -\left(\frac{\alpha\,Z}{\hbar\,c}\right)
  \left(\frac{E-m_{e}\,c^{2}}{\hbar\,c\,\beta}\right)^{2},\label{idnetif1}
  \\
  \nonumber
  \\
  \nonumber
  \\
  &&\big[(N+s)+(1+x)\big]
  -\left(\frac{E+m_{e}\,c^{2}}{\hbar\,c\,\beta}\right)
\left(\frac{\alpha\,Z}{\hbar\,c}\right)=\nonumber\\
  &&\hspace{4.5cm}\;-\big[(N+s)+(1-x)\big]
  +\left(\frac{\alpha\,Z}{\hbar\,c}\right)
\left(\frac{E-m_{e}\,c^{2}}{\hbar\,c\,\beta}\right).\label{idnetif2}
\end{eqnarray}
Ces deux équations sont identiques. En effet, en multipliant les
deux membres de (\ref{idnetif1}) par
$(E+m_{e}\,c^{2})/(\hbar\,c\,\beta)$, tout en tenant compte de
(\ref{betacarre}), on retrouve l'équation (\ref{idnetif2}). Donc,
pour déterminer les énergies $E_{N}$, on utilise l'une des
équations précédentes. L'équation (\ref{idnetif2}) nous permet
d'écrire:
\begin{eqnarray}
[(N+s)+(1+x)]+[(N+s)+(1-x)]&=&\left(\frac{E+m_{e}\,c^{2}}{\hbar\,c\,\beta}\right)
\left(\frac{\alpha\,Z}{\hbar\,c}\right)+\left(\frac{\alpha\,Z}{\hbar\,c}\right)
\left(\frac{E-m_{e}\,c^{2}}{\hbar\,c\,\beta}\right)\nonumber
\\
(N+s+1)&=&\left(\frac{\alpha\,Z}{\hbar\,c}\right)\times
\left(\frac{E}{\hbar\,c\,\beta}\right).\label{implicitexplicit}
\end{eqnarray}
L'énergie apparaît de manière explicite dans le second membre de
(\ref{implicitexplicit}), de plus elle est aussi contenue dans le
paramètre $\beta$. En élevant au carré les deux membres de cette
équation et en tenant compte de (\ref{betacarre}),
\begin{equation}
(N+s+1)^{2}=\left(\frac{\alpha\,Z}{\hbar\,c}\right)^{2}\times
\left(\frac{E^{2}}{(m_{e}\,c^{2})^{2}-E^{2}}\right),\nonumber
\end{equation}
il est possible de d'isoler le terme d'énergie. En effet,
\begin{eqnarray}
\left[\left(\frac{\alpha\,Z}{\hbar\,c}\right)^{2}
+(N+s+1)^{2}\right]\;E^{2}&=&(m_{e}\,c^{2})^{2}\;(N+s+1)^{2}\nonumber\\
E^{2}&=& (m_{e}\,c^{2})^{2}\;\left[\displaystyle\frac{(N+s+1)^{2}}
{\left(\frac{\alpha\,Z}{\hbar\,c}\right)^{2}+(N+s+1)^{2}}\right]\nonumber\\
E^{2}&=&(m_{e}\,c^{2})^{2}\;\left[
1+\displaystyle\frac{\left(\frac{\alpha\,Z}{\hbar\,c}\right)^{2}}{(N+s+1)^{2}}\right]^{-1}.
\end{eqnarray}
Les niveaux d'énergie $E=E_{N}$ sont donnés par l'expression,
\begin{eqnarray}
E_{N}&=&(m_{e}\,c^{2})\;\left[
1+\displaystyle\frac{\left(\frac{\alpha\,Z}{\hbar\,c}\right)^{2}}
{(N+s+1)^{2}}\right]^{-1/2}.\nonumber
\end{eqnarray}
En remplaçant la valeur de $s$, conformément à (\ref{sracine}),
les niveaux d'énergie de l'atome d'hydrogène s'écrivent
finalement:
\begin{eqnarray}
    E_{N}=(m_{e}\,c^{2})\;\left[1+
\displaystyle\frac{\left(\displaystyle\frac{\alpha\,Z}{\hbar\,c}\right)^{2}}
{\left(\;N+\displaystyle\sqrt{x^{2}-\left(\frac{\alpha\,Z}{\hbar\,c}\right)^{2}}\;
\right)^{2}}\right]^{-1/2}.\label{nivnrjatomhydrog}
\end{eqnarray}

Ce résultat est identique à (\ref{niveauenrjlandau}). C'est une preuve que l'électron
d'un tel atome est correctement décrit par l'équation d'onde quantique
(\ref{eqdiracourbe}). Au terme de cette application, on peut affirmer donc que la
nouvelle approche a permis de reproduire le spectre discret de l'atome d'hydrogène
\cite{barros}.

Dans son article, C.C.Barros prétend apporter une correction aux
niveaux d'énergie de l'atome d'hydrogène, déjà établis par le
biais de la théorie de Dirac. En résolvant le système d'équations
différentielles (\ref{sysdiffbarros}), écrit pour l'électron de
l'atome d'Hydrogène et sans faire d'approximations, il retrouve le
spectre d'énergie suivant \cite{barros}:
\begin{equation}\label{spectrebarros}
    E_{N}=m_{0}\,c^{2}\;\displaystyle\sqrt{\frac{1}{2}-\frac{N^{2}}{8\alpha^{2}}
    +\frac{N}{4\alpha}\;\displaystyle\sqrt{\frac{N^{2}}{4\alpha^{2}}+2}}.
\end{equation}
Bien qu'il affirme que ses résultats sont plus réalistes (ils
s'accordent mieux avec les résultats expérimentaux). L'écart des
niveaux d'énergie (\ref{spectrebarros}) avec l'expérience est de
l'ordre de $0,005\%$ \cite{barros}, alors que l'écart des niveaux
d'énergie de Dirac (\ref{nivnrjatomhydrog}) est de l'ordre de
$0.027\%$ \cite{barros}.
\section{Conclusion}
Dans ce chapitre, une application d'une nouvelle approche,
proposée par C.C.Barros, à l'atome d'hydrogène a été envisagée.
L'électron d'un tel atome évolue sous l'action d'un potentiel
coulombien. Pour déterminer les niveaux d'énergie il fallait
écrire l'équation d'onde quantique stationnaire pour l'électron de
spin $1/2$ qui évolue sous l'action du potentiel coulombien. Une
solution à variables séparées a été introduite, de sorte que la
partie angulaire de la fonction d'onde est déterminée par des
considérations physiques (conservation du moment cinétique), alors
que la partie radiale est définie par deux fonctions qui vérifient
un système d'équation d'équations différentielles couplées. Une
méthode de résolution par des séries a été proposée pour
déterminer les énergies propres. On a pu reproduire le spectre
d'énergie relativiste de l'atome d'hydrogène. La différence qui
existe entre la théorie de Dirac et la nouvelle approche est que
dans la première, l'interaction électromagnétique est introduite
par des considérations de symétrie (couplage minimal), alors que
dans la seconde, l'interaction est contenue dans la structure même
de l'espace-temps. Avec un simple développement limité, on a
retrouvé exactement les mêmes équations de la théorie de Dirac.
Bien que nous ne retrouvions pas les mêmes résultats que
C.C.Barros, l'approche proposée par ce dernier a le mérite de
reproduire les résultats de Dirac.

\newpage
\pagestyle{fancy} \lhead{chapitre\;6}\rhead{Conclusion générale}
\chapter{Conclusion générale}
Dans ce travail, une proposition récente, avancée par C.C.Barros, a été analysée.
Cette proposition stipule que d'une façon analogue à l'interaction gravitationnelle,
les trois autres interactions peuvent aussi se manifester à travers la structure de
l'espace-temps. L'idée fondamentale consiste à décrire une particule évoluant dans une
région plongée dans un potentiel non gravitationnel qui affecterait la métrique de
l'espace-temps. En raison de la symétrie sphérique du champ central considéré, une
métrique similaire à celle de Schwarzschild est adoptée.

Dans le cadre de ce mémoire, on s'est fixé l'objectif d'analyser,
d'une part, les fondements et les hypothèses sur lesquelles
s'appuie la nouvelle proposition, et d'autre part, de vérifier et
d'expliciter tous les calculs nécessaires, dans le but de
s'assurer de la validité de la nouvelle proposition de C.C.Barros.
En effet, avant de pouvoir adopter ou même explorer toutes les
implications et perspectives qu'offre l'application d'une telle
proposition originale, il fallait dans un premier temps
entreprendre une critique aussi bien des idées que du formalisme.
Précisons que dans son travail original, l'auteur n'a pas détaillé
ses calculs; il ne donne que les grandes lignes et les résultats
de son travail. Dans ce mémoire, tous les calculs sont explicités.
Nous pensons que ce travail s'inscrit dans l'effort de la
compréhension des origines de la relativité générale et de la
mécanique quantique afin de réussir un rapprochement de ces deux
piliers de la Physique Moderne.

Avant de pouvoir discuter les fondements physiques de la nouvelle proposition, citée
ci-dessus, il fallait rechercher les arguments clés qui ont incité Einstein à décrire
l'interaction gravitationnelle en terme de structure d'espace-temps. Ainsi, le second
chapitre a été consacré à quelques rappels sur la théorie de la relativité générale.
L'accent a été mis sur le contenu physique du postulat d'équivalence.

\A première vue, il n'y a apparemment aucun argument physique en
faveur de cette nouvelle proposition. En effet, la description du
champ gravitationnel en terme de structure d'espace-temps découle
de l'application du postulat d'équivalence d'Einstein. L'argument
crucial sur lequel s'appuie ce postulat est l'égalité des masses
inertielle et gravitationnelle. Egalité qui se traduit
concrètement par le fait que tous les corps se déplaçant
uniquement sous l'influence d'un champ gravitationnel seront
soumis à la même accélération, indépendamment de leurs masses, de
leurs substances et de leurs états physiques. Cette propriété
fondamentale caractérise exclusivement le champ gravitationnel,
c'est-à-dire que des corps différents se mouvant sous l'action
d'un champ non gravitationnel ne seront pas soumis à la même
accélération. C'est pour cette raison qu'en théorie de la
relativité générale, on n'envisage pas de décrire les
interactions: électromagnétique, faible et forte, en terme de
structure d'espace-temps.

L'auteur fait abstraction de toutes les considérations physiques
qui ont motivé le choix d'Einstein pour interpréter
géométriquement l'interaction gravitationnelle. Il adopte comme
point de départ une métrique similaire à celle de Schwarzschild,
et de façon analogue à ce qui se fait en théorie de la relativité
générale, il va introduire l'interaction non gravitationnelle dans
l'expression de la métrique. C'est à travers la fonction $\xi(r)$
que le champ non gravitationnel affecte la métrique.


Dans le troisième chapitre, une particule évoluant dans un espace courbe, doté d'une
métrique de Schwarzschild, est décrite. Dans un premier temps, les principes de
correspondance de l'énergie et de l'impulsion, initialement établis dans le cas d'un
espace plat de Minkowski, ont été généralisés pour le cas d'un espace courbe. En
second lieu, les principales grandeurs dynamiques, relatives à une particule évoluant
dans un espace doté d'une métrique de Schwarzschild, ont été déterminées, à savoir:
l'énergie, les impulsions généralisées, l'expression de l'invariant relativiste. La
combinaison des résultats obtenus dans les deux étapes précédentes a permis de
déterminer les opérateurs $(E,\overrightarrow{p},\overrightarrow{p}^{2})$,
indispensables à l'écriture d'équations d'ondes quantiques. Finalement, une connexion
entre le potentiel non gravitationnel, et la structure de l'espace-temps est rendue
possible grâce à l'établissement d'une formule reliant une fonction $\xi(r)$, figurant
dans l'expression de la métrique, et le potentiel $V(r)$.

Dans le quatrième chapitre, deux équations d'ondes quantiques ont
été établies dans le cas d'une particule évoluant dans un espace
courbe. La première est l'équation d'onde pour une particule de
spin $0$, alors que la seconde est une équation pour une particule
de spin $1/2$. La procédure utilisée pour l'écriture de ces
équations d'ondes quantiques est inspirée, d'une part de la
procédure de Klein-Gordon et d'autre part de la procédure de
Dirac, initialement définies pour écrire les équations d'ondes
pour une particule de spin $0$ et une particule de spin $1/2$,
dans l'espace-temps plat de Minkowski. Pour garantir que le nombre
de particule ne change pas (une seule particule), on suppose que
la particule est incapable d'interagir avec le champ non
gravitationnel avec une énergie plus importante que l'énergie de
seuil d'une quelconque autre particule. Vue qu'on s'intéresse à
des particules de masses très petites, l'interaction
gravitationnelle est alors négligée (à petite échelle
l'interaction électromagnétique est plus importante que
l'interaction gravitationnelle), ainsi la courbure de
l'espace-temps dont il est question ici, n'est pas due au champ de
gravitation. Elle est principalement causée par l'interaction non
gravitationnelle. Dans cette nouvelle approche, au lieu
d'introduire l'interaction entre la particule et le champ par des
considérations de symétrie (couplage minimal), l'interaction est
contenue dans la structure de l'espace-temps. Dans le cas
stationnaire, une méthode de résolution, basée sur une séparation
de variables, a été proposée pour les deux équations d'ondes
quantiques.

Le cinquième chapitre est une application de la nouvelle approche
à l'atome d'hydrogène. Ce système physique offre l'avantage d'être
totalement décrit par la théorie quantique habituelle, ce qui
permet de tester la validité de la proposition de Barros pour le
potentiel coulombien (interaction électromagnétique) qui s'exerce
entre l'électron et le proton du noyau. La symétrie sphérique du
potentiel coulombien motive le choix d'une métrique similaire à
celle de Schwarzschild. On suppose que ce potentiel affecte la
structure de l'espace-temps. La description de l'électron se fait
par le biais de l'équation d'onde quantique et stationnaire pour
une particule de spin $1/2$, déterminée au chapitre quatre. La
résolution de l'équation d'onde précédente a permis de reproduire
le spectre d'énergie, déjà connu de l'atome d'hydrogène, ce qui
prouve que l'électron de l'atome d'hydrogène semble être
correctement décrit par la nouvelle approche. En fait, la
différence entre l'approche de Dirac et cette nouvelle approche
est que dans la première, l'interaction électromagnétique est
introduite grâce à une transformation de symétrie, dite couplage
minimal, alors que dans la seconde l'interaction non
gravitationnelle (électromagnétique) est contenue dans
l'expression de la métrique de l'espace-temps. En exploitant le
fait que l'énergie potentielle coulombienne est négligeable devant
l'énergie au repos de l'électron, il a été possible d'effectuer un
développement limité qui a permis de retrouver des équations
identiques aux équations de Dirac. Cette approche est donc une
autre façon de voir les choses, mais qui offre l'avantage
d'expliquer de manière plus intuitive l'origine de l'interaction.
Celle-ci semble être contenue dans la structure même de
l'espace-temps. Précisons que C.C.Barros \cite{barros} a retrouvé
des résultats légèrement différents de ceux de Dirac. Il prétend
apporter une correction à cette théorie, mais qui est
insignifiante. Bien que nous ne retrouvions pas les mêmes
résultats que ceux de Barros, car on résout un système d'équations
différentielles différent, on a montré qu'avec les hypothèses de
départ, il était possible de reproduire les résultats de Dirac.

Au terme de ce travail, on peut affirmer que la nouvelle proposition de Barros semble
être remarquablement vérifiée pour l'interaction coulombienne s'exerçant entre
l'électron et le proton de l'atome d'hydrogène. Il est clair que pour concrétiser son
idée, C.C.Barros s'est appuyé à la fois sur le formalisme de deux théorie très
puissantes, qui sont la mécanique quantique et de la relativité générale. La question
qui se pose, à présent, est la suivante: \textbf{Est ce que le fait que l'atome
d'hydrogène soit correctement décrit par cette nouvelle approche n'est pas dû au fait
de la grande similitude qui existe entre le potentiel coulombien et le potentiel
gravitationnel, $V(r)\sim r^{-1}$, parfaitement décrit par la théorie de la relativité
générale?}

Pour explorer cette hypothèse, il faut proposer de nouveaux tests de cette approche,
avec des potentiels non proportionnels à $r^{-1}$. Il faut dire qu'on a d'ores et déjà
essayé de proposer une nouvelle application. Le système physique considéré est
l'oscillateur harmonique isotrope à 3 dimensions. Celui-ci offre le double avantage de
permettre la vérification de l'hypothèse précédente, et de permettre de tester
l'équation quantique (\ref{kgordon}) pour une particule de spin 0.

Des indices nous amènent à penser qu'un tel système n'est pas correctement décrit par
cette nouvelle approche. En fait, on s'est rendu compte que le potentiel quadratique
de l'oscillateur harmonique isotrope à trois dimensions ne répond pas à l'exigence
d'être convergent à l'infini. Il n'est donc pas possible d'espérer, quand
$r\rightarrow+\infty$, que la métrique de Schwarzschild (\ref{metrique}) se réduise à
la métrique plate de Minkowski. En fait, un système physique ne peut être modélisé par
un oscillateur harmonique qu'au voisinage de sa position d'équilibre stable.

La nouvelle proposition de C.C.Barros est vraiment très
prometteuse, elle nous amène à nous poser des questions sur le
rôle du principe d'Equivalence dans les fondements de la théorie
de la relativité générale, alors qu'Einstein le présente comme un
puissant argument en faveur du postulat de la relativité générale.
Autrement dit, bien que ses succès soient nombreux et
remarquables, avons-nous cerné les concepts de base sur lesquels
est bâtie la théorie de la relativité générale ? Il est vrai que
le fait qu'elle soit vérifiée pour le potentiel coulombien de
l'atome d'hydrogène n'est sûrement pas une preuve de sa validité
pour d'autres potentiels. C'est pourquoi, il faut dans un premier
temps multiplier les tests pour les trois interactions non
gravitationnelles (électromagnétique, faible et forte). \A long
terme, disons que si la proposition s'avérerait être vérifiée,
alors il va falloir penser à "reformuler" le postulat
d'équivalence d'Einstein, en l'élargissant aux interactions:
électromagnétique, faible et forte. Faut-il rappeler que dans la
version déterministe actuelle de la mécanique quantique, Faraggi
et Matone ont récemment proposé un nouveau postulat, qui ressemble
étrangement au postulat d'équivalence d'Einstein, dans lequel ils
affirment qu'il est toujours possible de connecter deux états
quantiques différents par des transformations de coordonnées. La
proposition de C.C.Barros pourrait conduire à une meilleure
compréhension du lien qui puisse exister entre ces deux postulat,
ce qui permettrait, sans nul doute, de faire une avancée
considérable dans l'effort de formulation d'une version
déterministe de la mécanique quantique, piste très prometteuse
dans la course à l'unification des quatres interactions
fondamentales.

\chapter*{Appendices}
\appendix
\addcontentsline{toc}{chapter}{\hspace{5cm}Appendices}
\newpage
\pagestyle{fancy} \lhead{Appendices}\rhead{Rappels sur l'analyse tensorielle}
\chapter{Rappels sur quelques notions de l'analyse tensorielle}

Les lois de la nature doivent répondre à deux exigences \cite{weinbergs}:
\begin{enumerate}
    \item Une fois formulées dans un système de
coordonnées particulier, elles doivent avoir la même forme quand on passe à un système
de coordonnée quelconque. Autrement dit, les lois de la nature doivent être
covariantes sous n'importe quelle changement de coordonnées.
    \item En l'absence de gravitation, le tenseur métrique $g_{\mu\nu}$ tend à être égal
    au tenseur de Minkowski $\eta_{\mu\nu}$ et on retombe sur les
    lois déjà établies dans le cadre de la théorie de la
    relativité restreinte.
\end{enumerate}

Le formalisme tensoriel est un outil indispensable dans la théorie de la relativité
générale, il permet, non seulement, de formuler les lois d'une manière compacte et
élégante, mais aussi il offre l'avantage qu'une lois écrite sous forme tensorielle
possède nécessairement une forme indépendante du système de coordonnées.

\section{Variance}
Soit une base quelconque $\{\overrightarrow{e_{i}}\}$ d'un espace vectoriel euclidien
$E$, de dimension $n$, $i\in[1,n]$. On appelle composantes contravariantes d'un
vecteur $\overrightarrow{A}\in E$, les quantités: $\{A^{i}\}_{i=1,n}$, tel que:
\begin{equation}\label{composcontavdef}
    \overrightarrow{A}=\displaystyle\sum_{i=1}^{n}A^{i}\overrightarrow{e_{i}}.
\end{equation}

On appelle composantes covariantes d'un vecteur $\overrightarrow{A}\in E$, les
quantités: $\{A_{i}\}_{i=1,n}$, tel que:
\begin{eqnarray}\label{composcovdef}
    \left\{%
\begin{array}{ll}
    A_{i}=\overrightarrow{A}.\overrightarrow{e_{i}},\\
    i=1\rightarrow n.\\
\end{array}%
\right.
\end{eqnarray}

Pour déterminer, comment ces grandeurs se transforment sous l'action d'une
transformation de coordonnées quelconque: $\{x^{\mu}\}\longrightarrow\{x^{'\mu}\}$, il
faut d'abord définir les coordonnées curvilignes.

\section{Coordonnées curvilignes} Considérons un point $M$ d'un espace vectoriel
euclidien $E$ et un système de coordonnées $\{x_{i}\},i\in[1,n]$. On associe au point
$M$ un repère naturel, admettant $M$ pour origine et
$\{\overrightarrow{e_{i}}\},i\in[1,n]$ pour base, tel que:
\begin{eqnarray}\label{deei}
\left\{%
\begin{array}{ll}
    \overrightarrow{e_{i}}=\displaystyle\frac{\partial \,\overrightarrow{OM}}
    {\partial x^{i}},\\
     i=1\rightarrow n,\\
\end{array}%
\right.
\end{eqnarray}
d'une manière générale:
\begin{eqnarray}
    d\,\overrightarrow{OM}&=&\displaystyle\sum_{i=1}^{n}\displaystyle\frac{\partial \,\overrightarrow{OM}}
    {\partial x^{i}}\;dx^{i}\nonumber\\
    &=&\displaystyle\sum_{i=1}^{n}\overrightarrow{e_{i}}\;dx^{i}.\nonumber
\end{eqnarray}

Pour pouvoir faire le passage d'un système de coordonnées $\{x^{i}\}$ à un autre
système de coordonnées $\{x^{'i}\}$, déterminons comment se transforment les vecteurs
de la base $\{\overrightarrow{e_{i}}\}\rightarrow\{\overrightarrow{e_{i}}^{'}\}$. On
suppose que les deux systèmes de coordonnées sont reliés par une transformation
inversible, de la forme:
\begin{eqnarray}
\left\{%
\begin{array}{ll}
    x^{'1}=x^{'1}(x^{1},x^{2},...,x^{n}),\\
    x^{'2}=x^{'}(x^{1},x^{2},...,x^{n}), \\
    \vdots\\
    x^{'n}=x^{'n}(x^{1},x^{2},...,x^{n}), \\
\end{array}%
\right.\hspace{0.5cm}\Leftrightarrow\hspace{0.5cm}\left\{%
\begin{array}{ll}
    x^{'i}=x^{'i}(x^{j}),  \\
    1\leq i\leq n\;\;,\;\;1\leq j\leq n \\
\end{array}%
\right.
\end{eqnarray}
En utilisant la définition (\ref{deei}), nous avons:
\begin{eqnarray}\label{ejprimei}
    \overrightarrow{e_{j}}^{'}&=&\displaystyle\frac{\partial \,\overrightarrow{OM}}
    {\partial x^{'j}}\nonumber\\
    &=&\displaystyle\sum_{i=1}^{n}\displaystyle\frac{\partial \,\overrightarrow{OM}}{\partial x^{i}}
    \;\displaystyle\frac{\partial x^{i}}{\partial
    x^{'j}}\nonumber\\
    &=&\displaystyle\sum_{i=1}^{n}\displaystyle\frac{\partial x^{i}}{\partial
    x^{'j}}\;\overrightarrow{e_{i}},
\end{eqnarray}
et de manière analogue, il est possible de montrer que:
\begin{equation}\label{eiejprim}
     \overrightarrow{e_{i}}=\displaystyle\sum_{j=1}^{n}\displaystyle\frac{\partial x^{'j}}{\partial
    x^{i}}\;\overrightarrow{e_{j}}^{'}.
\end{equation}
Les équations (\ref{ejprimei}) et (\ref{eiejprim}) permettent de faire le passage
entre les deux bases $\{\overrightarrow{e_{i}}\}$ et $\{\overrightarrow{e_{i}}^{'}\}$

\section{Transformation des composantes d'un vecteur}
\subsection{Composantes contravariantes} Déterminons comment se transforment les
composantes contravariantes d'un vecteur $\overrightarrow{A}$, lors d'une
transformation de coordonnées $\{x^{i}\}\rightarrow\{x^{'i}\}$ et vis versa:
\begin{eqnarray}
    \overrightarrow{A}&=&\displaystyle\sum_{i=1}^{n}A^{i}\;\overrightarrow{e_{i}},\hspace{1cm}
    \textrm{dans la base} \{\overrightarrow{e_{i}}\},\nonumber\\
    &=&\displaystyle\sum_{j=1}^{n}A^{'j}\;\overrightarrow{e_{j}}^{'},\hspace{1cm}
    \textrm{dans la base} \{\overrightarrow{e_{j}}^{'}\}.\nonumber
\end{eqnarray}
En utilisant (\ref{ejprimei}), nous avons:
\begin{eqnarray}
    \overrightarrow{A}&=&\displaystyle\sum_{j=1}^{n}A^{'j}\;\Bigg(\displaystyle\sum_{i=1}^{n}
    \displaystyle\frac{\partial x^{i}}{\partial
    x^{'j}}\;\overrightarrow{e_{i}}\Bigg)\nonumber\\
    &=&\displaystyle\sum_{i=1}^{n}\Bigg(\displaystyle\sum_{j=1}^{n}A^{'j}\;
    \displaystyle\frac{\partial x^{i}}{\partial
    x^{'j}}\Bigg)\;\overrightarrow{e_{i}}\nonumber\\
    &=&\displaystyle\sum_{i=1}^{n}A^{i}\;\overrightarrow{e_{i}},\nonumber
\end{eqnarray}
une identification membre à membre nous permet d'écrire:
\begin{equation}\label{aiaprimj}
       A^{i}=\displaystyle\sum_{j=1}^{n}A^{'j}\;
    \displaystyle\frac{\partial x^{i}}{\partial
    x^{'j}}.
\end{equation}
En utilisant (\ref{eiejprim}), il est possible de montrer de manière analogue que:
\begin{equation}\label{aprimjai}
       A^{'j}=\displaystyle\sum_{i=1}^{n}A^{i}\;
    \displaystyle\frac{\partial x^{'j}}{\partial
    x^{i}}.
\end{equation}
Les formules (\ref{aiaprimj}) et (\ref{aprimjai}) montrent comment se transforment les
composantes contravariantes d'un vecteur $\overrightarrow{A}$.

\subsection{Composantes covariantes} Pour déterminer comment se transforment les
composantes covariantes, il faut écrire la définition (\ref{composcovdef}) dans les
deux bases et utiliser (\ref{eiejprim}) et (\ref{ejprimei}):
\begin{eqnarray}
    \left\{%
\begin{array}{ll}
    A_{i}=\overrightarrow{A}.\overrightarrow{e_{i}}=\overrightarrow{A}.\displaystyle\sum_{j=1}^{n}\displaystyle\frac{\partial x^{'j}}{\partial
    x^{i}}\;\overrightarrow{e_{j}}^{'}=\displaystyle\sum_{j=1}^{n}\displaystyle\frac{\partial x^{'j}}{\partial
    x^{i}}\;A_{j}^{'} \\
    A_{j}^{'}=\overrightarrow{A}.\overrightarrow{e_{j}}^{'}=\overrightarrow{A}.\displaystyle\sum_{i=1}^{n}\displaystyle\frac{\partial x^{i}}{\partial
    x^{'j}}\;\overrightarrow{e_{i}}=\displaystyle\sum_{i=1}^{n}\displaystyle\frac{\partial x^{i}}{\partial
    x^{'j}}\;A_{i}\\
\end{array}%
\right.
\end{eqnarray}

\section{Définition d'un tenseur}
On appelle composante «$p$ fois contravariante» et «$q$ fois covariante» d'un tenseur
mixte d'ordre $p+q$, toute quantité: $A_{\ell_{1}...\ell_{q}}^{k_{1}...k_{p}}$ se
transformant comme le produit de $p$ composantes contravariantes et $q$ composantes
covariantes d'un vecteur, lors d'un changement de coordonnée:
$\{x^{i}\}\rightarrow\{x^{'i}=x^{'i}(x^{j})\}$. La transformation se fait comme:
\begin{equation}\label{tranformtanseur}
    A_{j_{1}...j_{q}}^{'i_{1}...i_{p}}=\displaystyle\sum_{k_{1}=1}^{n}...
    \displaystyle\sum_{k_{p}=1}^{n}\displaystyle\sum_{\ell_{1}=1}^{n}
    ...\displaystyle\sum_{\ell_{p}=1}^{n}\;\displaystyle\frac{\partial x^{'i_{1}}}
    {\partial x^{k_{1}}}...\displaystyle\frac{\partial x^{'i_{p}}}{\partial
    x^{k_{p}}}\;
    \displaystyle\frac{\partial x^{\ell_{1}}}{\partial x^{'j_{1}}}
    ...\displaystyle\frac{\partial x^{\ell_{q}}}{\partial
    x^{'j_{q}}}\;A_{\ell_{1}...\ell_{q}}^{k_{1}...k_{p}},
\end{equation}
et la transformation inverse:
\begin{equation}\label{invtranformtanseur}
    A_{j_{1}...j_{q}}^{i_{1}...i_{p}}=\displaystyle\sum_{k_{1}=1}^{n}...
    \displaystyle\sum_{k_{p}=1}^{n}\displaystyle\sum_{\ell_{1}=1}^{n}
    ...\displaystyle\sum_{\ell_{p}=1}^{n}\;\displaystyle\frac{\partial x^{i_{1}}}
    {\partial x^{'k_{1}}}...\displaystyle\frac{\partial x^{i_{p}}}{\partial
    x^{'k_{p}}}\;
    \displaystyle\frac{\partial x^{'\ell_{1}}}{\partial x^{j_{1}}}
    ...\displaystyle\frac{\partial x^{'\ell_{q}}}{\partial
    x^{j_{q}}}\;A_{\ell_{1}...\ell_{q}}^{'k_{1}...k_{p}}.
\end{equation}

Exemples:
\begin{enumerate}
    \item Soit un tenseur $T_{\;\;\nu}^{\mu\;\lambda}$ à deux indices
contravariants (2 indices en haut) et un indice covariant (1 indice en bas).
conformément à (\ref{tranformtanseur}), une telle se transforme par:
\begin{equation}\label{expppun}
         T_{\;\;\nu}^{'\mu\;\lambda}=\displaystyle\sum_{k=1}^{n}
         \displaystyle\sum_{\rho=1}^{n}\displaystyle\sum_{\sigma=1}^{n}
         \displaystyle\frac{\partial x^{'\mu}}{\partial x^{k}}\;
         \displaystyle\frac{\partial x^{\rho}}{\partial x^{'\nu}}
   \displaystyle\frac{\partial x^{'\lambda}}
    {\partial x^{\sigma}}T_{\;\;\rho}^{k\;\sigma},
\end{equation}
    \item Dans l'espace à 4 dimensions, soit un tenseur $R_{i\;lj}^{\;s}$ (tenseur de
    courbure).
    Conformément à (\ref{invtranformtanseur}), sa transformation
    inverse s'écrit:
    \begin{equation}\label{expppdeux}
            R_{i\;\;lj}^{\;s}=\displaystyle\sum_{q=1}^{4}
            \displaystyle\sum_{p=1}^{4}\displaystyle\sum_{m=1}^{4}
            \displaystyle\sum_{n=1}^{4}
            \displaystyle\frac{\partial x^{s}}{\partial x^{'q}}
            \;\displaystyle\frac{\partial x^{'p}}{\partial x^{i}}\;
            \displaystyle\frac{\partial x^{'m}}{\partial x^{l}}
            \;\displaystyle\frac{\partial x^{'n}}{\partial
            x^{j}}\;R_{p\;\;mn}^{'\;q}
    \end{equation}
\end{enumerate}

\section{Tenseur métrique} Considérons un système orthonormé $\{x^{(0)i}\}$ et un
système curviligne $\{x^{i}\}$. Soit $A^{(0)i}$ et $B^{(0)i}$ les composantes
contravariantes respectives dans $\{x^{(0)i}\}$ des deux vecteurs $\overrightarrow{A}$
et $\overrightarrow{B}$, et $A^{i}$ et $B^{i}$ leurs composantes contravariantes dans
$\{x^{i}\}$. Conformément à (\ref{aprimjai}), nous avons:
\begin{eqnarray}\label{aokai}
    \left\{%
\begin{array}{ll}
    A^{(0)k}=\displaystyle\sum_{i=1}^{n}\displaystyle\frac{\partial x^{(0)k}}{\partial x^{i}}
    \;A^{i},\\
    B^{(0)k}=\displaystyle\sum_{i=1}^{n}\displaystyle\frac{\partial x^{(0)k}}{\partial x^{i}}
    \;B^{i}.\\
\end{array}%
\right.
\end{eqnarray}
Calculons le produit scalaire:
\begin{eqnarray}
    \overrightarrow{A}.\overrightarrow{B}&=&\Bigg(\displaystyle\sum_{k=1}^{n}
    A^{(0)k}\;\overrightarrow{e_{k}}^{0}\Bigg).\Bigg(\displaystyle\sum_{\ell=1}^{n}
    A^{(0)\ell}\;\overrightarrow{e_{\ell}}^{0}\Bigg)\nonumber\\
    &=&\displaystyle\sum_{k=1}^{n}\displaystyle\sum_{\ell=1}^{n}A^{(0)k}\;A^{(0)\ell}
    \underbrace{(\overrightarrow{e_{k}}^{0}.\;\overrightarrow{e_{\ell}}^{0})}_{\delta_{k\ell}}\nonumber\\
    &=&\displaystyle\sum_{k=1}^{n}\displaystyle\sum_{\ell=1}^{n}\delta_{k\ell}\;
    \Bigg(\displaystyle\sum_{i=1}^{n}\displaystyle\frac{\partial x^{(0)k}}{\partial
    x^{i}}\;A^{i}\Bigg)\;\Bigg(\displaystyle\sum_{j=1}^{n}\displaystyle
    \;\displaystyle\frac{\partial x^{(0)k}}{\partial x^{j}}\;B^{j}\Bigg)\nonumber\\
    &=&\displaystyle\sum_{i=1}^{n}\displaystyle\sum_{j=1}^{n}\Bigg(\displaystyle\sum_{k=1}^{n}
    \displaystyle\sum_{\ell=1}^{n}\delta_{k\ell}\;\displaystyle\frac{\partial x^{(0)k}}{\partial
    x^{i}}\;\displaystyle\frac{\partial x^{(0)k}}{\partial x^{j}}\Bigg)\;A^{i}\,B^{j},
\end{eqnarray}
et posons:
\begin{equation}\label{gijdelta}
    g_{ij}=\displaystyle\sum_{k=1}^{n}
    \displaystyle\sum_{\ell=1}^{n}\delta_{k\ell}\;\displaystyle\frac{\partial x^{(0)k}}{\partial
    x^{i}}\;\displaystyle\frac{\partial x^{(0)k}}{\partial x^{j}}.
\end{equation}
Le produit scalaire s'écrit, d'une part:
\begin{equation}\label{produiscalair1}
   \overrightarrow{A}.\overrightarrow{B}=\displaystyle\sum_{i=1}^{n}
   \displaystyle\sum_{j=1}^{n}
   g_{ij}\;A^{i}\,B^{j},
\end{equation}
et d'autre part, il s'exprime dans le système $\{x^{i}\}$ par:
\begin{eqnarray}\label{produiscalair2}
   \overrightarrow{A}.\overrightarrow{B}&=&\Bigg(\displaystyle\sum_{i=1}^{n}A^{i}\;
   \overrightarrow{e_{i}}\Bigg).\Bigg(\displaystyle\sum_{j=1}^{n}B^{j}\;
   \overrightarrow{e_{j}}\Bigg)\nonumber\\
   \overrightarrow{A}.\overrightarrow{B}&=&\displaystyle\sum_{i=1}^{n}\displaystyle\sum_{j=1}^{n}
   (\overrightarrow{e_{i}}\,.\overrightarrow{e_{j}})A^{i}\;B^{j}.
\end{eqnarray}
La comparaison entre (\ref{produiscalair1}) et (\ref{produiscalair2}) conduit à poser:
\begin{equation}\label{tensmetriqueiej}
    g_{ij}=\overrightarrow{e_{i}}\,.\overrightarrow{e_{j}}.
\end{equation}

D'après (\ref{gijdelta}) et (\ref{tensmetriqueiej}), il est clair que les $g_{ij}$
forment les composantes d'un tenseur symétrique d'ordre deux, dit: tenseur métrique.
Einstein a eu l'intuition de décrire le champ de gravitation par le tenseur métrique.

\section{Passage entre composantes covariantes et contravariantes} Soit un vecteur
$\overrightarrow{A}\in E$, alors une composante covariante de celui-ci est:
\begin{eqnarray}
    A_{i}&=&\overrightarrow{A}.\overrightarrow{e_{i}}
         =\Bigg(\displaystyle\sum_{j=1}^{n}A^{j}\;\overrightarrow{e_{j}}\Bigg).\;
         \overrightarrow{e_{i}}\nonumber\\
         &=&\displaystyle\sum_{j=1}^{n}(\underbrace{\overrightarrow{e_{j}}.
         \overrightarrow{e_{i}}}_{g_{ji}})
         \;A^{j}
         =\displaystyle\sum_{j=1}^{n}(\underbrace{\overrightarrow{e_{i}}.
         \overrightarrow{e_{j}}}_{g_{ii}})
         \;A^{j}\nonumber\\
    A_{i}&=&\displaystyle\sum_{j=1}^{n}g_{ij}\;A^{j},
\end{eqnarray}
le tenseur métrique permet d'élever l'indice.\\

Soit $G$ la matrice dont les éléments sont les $g_{ij}$:
$$G=(g_{ij}),$$
et notons $g^{ij}$ les éléments de la matrice inverse:
$$G^{-1}=(g^{ij}),$$
alors, par définition: $$G\;G^{-1}=1,$$ ce qui conduit à:
\begin{equation}
    \displaystyle\sum_{k=1}^{n}g_{ik}\;g^{kj}=\delta_{i}^{j}.
\end{equation}
Calculons l'expression:
\begin{eqnarray}
    \displaystyle\sum_{i=1}^{n}g^{\ell\,i}\;A_{i}&=&\displaystyle\sum_{i=1}^{n}g^{\ell\,i}\;
    \Bigg(\displaystyle\sum_{j=1}^{n}g_{ij}\;A^{j}\Bigg)
    =\displaystyle\sum_{j=1}^{n}\Bigg(\displaystyle\sum_{i=1}^{n}g^{\ell\,i}\;
    g_{ij}\Bigg)\;A^{j}=\displaystyle\sum_{j=1}^{n}\delta_{j}^{\ell}\;A^{j}
    =A^{\ell}.\nonumber
\end{eqnarray}
Donc:
\begin{equation}
    A^{i}=\displaystyle\sum_{j=1}^{n}g^{ij}\;A_{j},
\end{equation}
les éléments $g^{ij}$ du tenseur inverse permettent d'abaisser l'indice.

D'une manière générale, pour élever ou abaisser les indices des composantes d'un
tenseur, on utilise autant de fois, respectivement, les $g_{ij}$ ou $g^{ij}$:
\begin{eqnarray}
    A_{i_{1}...i_{p}}&=&\displaystyle\sum_{j_{1}=1}^{n}g_{i_{1}j_{1}}\;
    A_{i_{2}...i_{p}}^{j_{1}}\nonumber\\
    &=&\displaystyle\sum_{j_{1}=1}^{n}\displaystyle\sum_{j_{2}=1}^{n}
    g_{i_{1}j_{1}}\;g_{i_{2}j_{2}}\;A_{i_{3}...i_{p}}^{j_{1}j_{2}}\nonumber\\
    A_{i_{1}...i_{p}}&=&\displaystyle\sum_{j_{1}=1}^{n}...\displaystyle\sum_{j_{p}=1}^{n}
    g_{i_{1}j_{1}}...g_{i_{p}j_{p}}\;A^{j_{1}...j_{p}},
\end{eqnarray}
de même:
\begin{eqnarray}
    A^{i_{1}...i_{p}}&=&\displaystyle\sum_{j_{1}=1}^{n}g^{i_{1}j_{1}}\;
    A^{i_{2}...i_{p}}_{j_{1}}\nonumber\\
    &=&\displaystyle\sum_{j_{1}=1}^{n}\displaystyle\sum_{j_{2}=1}^{n}
    g^{i_{1}j_{1}}\;g^{i_{2}j_{2}}\;A^{i_{3}...i_{p}}_{j_{1}j_{2}}\nonumber\\
    A^{i_{1}...i_{p}}&=&\displaystyle\sum_{j_{1}=1}^{n}...\displaystyle\sum_{j_{p}=1}^{n}
    g^{i_{1}j_{1}}...g^{i_{p}j_{p}}\;A_{j_{1}...j_{p}}.
\end{eqnarray}
C'est aussi valable même si les indices sont mélangés. Par exemple, pour un tenseur mixte:
\begin{eqnarray}
    A^{i_{1}i_{2}...i_{p}}_{j_{1}j_{2}...j_{q}}&=&\displaystyle\sum_{\ell_{1}=1}^{n}
    g^{i_{1}\ell_{1}}\; A^{i_{2}...i_{p}}_{\ell_{1}j_{1}j_{2}...j_{q}}
\end{eqnarray}

\section{Les symboles de Christoffel} Considérons un point $M$ de l'espace et soit
$(M,\{\overrightarrow{e_{i}}\}_{i=1,n})$ le repère naturel attaché à un tel point. En
un point infiniment voisin $M^{'}$, repéré par:
$\overrightarrow{OM}^{'}=\overrightarrow{OM}+d\,\overrightarrow{OM}$, le repère
naturel associé est défini:
$$(M^{'},\{\overrightarrow{e_{i}}^{'}\}_{i=1,n})=(M^{'},\{\overrightarrow{e_{i}}+
d\overrightarrow{e_{i}}\}_{i=1,n}),$$ tel que:
\begin{equation}
     d\,\overrightarrow{e_{i}}=\displaystyle\sum_{k=1}^{n}\displaystyle
     \frac{\partial \overrightarrow{e_{i}}}{\partial x^{k}}\;dx^{k}.
\end{equation}
Le développement de $\partial \overrightarrow{e_{i}}/\partial x^{k}$ dans la
base $\{\overrightarrow{e_{i}}\}_{i=1,n}$ s'écrit:
\begin{equation}
    \displaystyle\frac{\partial \overrightarrow{e_{i}}}{\partial x^{k}}\equiv
    \partial_{k}\overrightarrow{e_{i}}=
     \displaystyle\sum_{j=1}^{n}\Gamma_{i\,k}^{j}\;\overrightarrow{e_{j}},
\end{equation}
les coefficients du développement $\Gamma_{i\,k}^{j}$ sont les symboles de Christoffel
de seconde espèce. Ainsi:
\begin{equation}
     d\overrightarrow{e_{i}}=\displaystyle\sum_{k=1}^{n}
     \displaystyle\sum_{j=1}^{n}\Gamma_{i\,k}^{j}\;dx^{k}\;\overrightarrow{e_{j}}.
\end{equation}

Pour exprimer les symboles $\Gamma_{i\,k}^{j}$ en fonction de la métrique, calculons
d'une part:
\begin{eqnarray}\label{dgij1}
    dg_{ij}&=&d(\overrightarrow{e_{i}}.\overrightarrow{e_{j}})\nonumber\\
    &=&d\overrightarrow{e_{i}}.\overrightarrow{e_{j}}+\overrightarrow{e_{i}}.
    d\overrightarrow{e_{j}}\nonumber\\
    &=&\displaystyle\sum_{k=1}^{n}
     \displaystyle\sum_{\ell=1}^{n}\Gamma_{i\,k}^{\ell}\;dx^{k}\;\overrightarrow{e_{\ell}}
     .\overrightarrow{e_{j}}+\overrightarrow{e_{i}}.\displaystyle\sum_{k=1}^{n}
     \displaystyle\sum_{\ell=1}^{n}\Gamma_{j\,k}^{\ell}\;dx^{k}\;\overrightarrow{e_{\ell}}
     \nonumber\\
     &=&\displaystyle\sum_{k=1}^{n}
     \displaystyle\sum_{\ell=1}^{n}\Big(\Gamma_{i\,k}^{\ell}\;g_{\ell\,j}+
     \Gamma_{j\,k}^{\ell}\;g_{i\,\ell}\Big)\;dx^{k}.
\end{eqnarray}
D'autre part, nous avons:
\begin{equation}\label{dgij2}
    dg_{ij}=\displaystyle\sum_{k=1}^{n}\displaystyle\frac{\partial g_{ij}}{\partial
    x^{k}}\;dx^{k}=\displaystyle\sum_{k=1}^{n}\partial_{k}g_{ij}\;dx^{k}.
\end{equation}
D'après (\ref{dgij1}) et (\ref{dgij2}), nous déduirons:
\begin{equation}\label{dkgij}
    \partial_{k}g_{ij}=\displaystyle\sum_{\ell=1}^{n}\Big(\Gamma_{i\,k}^{\ell}\;g_{\ell\,j}
    +\Gamma_{j\,k}^{\ell}\;g_{i\,\ell}\Big).
\end{equation}
Posons
\begin{equation}\label{christoffel1}
    \Gamma_{i\,k\,,j}\equiv \displaystyle\sum_{m=1}^{n}g_{mj}\;\Gamma_{i\,k}^{m},
\end{equation}
en adoptant cette nouvelle définition,(\ref{dkgij}) s'écrit:
\begin{equation}
    \partial_{k}g_{ij}=\Gamma_{i\,k\,,j}+
     \Gamma_{j\,k\,,i}.
\end{equation}
Les $\Gamma_{i\,k\,,j}$ sont les symboles de Christoffel de première espèce.

En utilisant la propriété $\Gamma_{i\,k\,,j}=\Gamma_{k\,i\,,j}$,
il est possible de montrer que:
\begin{equation}\label{importantegij}
   \Gamma_{i\,j}^{k}=\frac{1}{2}\;\displaystyle\sum_{m=1}^{n}g^{km}
   \Big(\partial_{i}g_{mj}+\partial_{j}g_{im}-
   \partial_{m}g_{ij}\Big)
\end{equation}

\A présent, déterminons une expression importante de la somme
$\displaystyle\sum_{i=1}^{n}\Gamma_{i\,j}^{i}$ en fonction de
$det(g_{ij})$. En premier lieu, d'après (\ref{importantegij}):
\begin{equation}\label{compar1}
    \displaystyle\sum_{i=1}^{n}\Gamma_{i\,j}^{i}=\frac{1}{2}\displaystyle\sum_{i=1}^{n}
    \displaystyle\sum_{m=1}^{n}g^{im}\partial_{j}g_{im}
\end{equation}
Ensuite, considérons une matrice arbitraire $M(x)$ et intéressons-nous à la variation
de $\ln\big(Det(M)\big)$, due à la variation $\delta x^{\ell}$ \cite{weinbergs}:
\begin{eqnarray}
    \delta \ln[Det(M)]&\equiv& \ln\bigg[Det(M+\delta M)\bigg]-\ln\bigg[Det(M)\bigg]\nonumber\\
    &=& \ln \Bigg[\displaystyle\frac{Det(M+\delta
    M)}{Det(M)}\Bigg]\nonumber\\
    &=& \ln\bigg[[Det(M)]^{-1}Det(M+\delta M)\bigg]\nonumber\\
    &=&\ln\bigg[Det(M^{-1})Det(M+\delta M)\bigg]\nonumber\\
    &=&\ln\Bigg[Det \Big(\displaystyle\frac{M+\delta
    M}{M}\Big)\Bigg]\nonumber\\
    &=&\ln\bigg[Det(1+M^{-1}\delta
    M)\bigg],\nonumber
\end{eqnarray}
Si $\delta M$ est assez faible, on peut écrire à l'ordre 1:
\begin{equation}
   \ln\bigg[Det(1+M^{-1}\delta
    M)\bigg]\simeq \ln\bigg[1+Tr(M^{-1}\delta
    M)\bigg]\simeq Tr(M^{-1}\delta
    M),
\end{equation}
finalement:
\begin{equation}\label{deltalog}
    \delta \ln[Det(M)]\simeq Tr(M^{-1}\delta
    M).
\end{equation}
Ce résultat nous permet de calculer:
\begin{equation}
    \displaystyle\frac{\delta \ln[Det(M)]}{\delta
    x^{\ell}}=\displaystyle\frac{Tr(M^{-1}\delta
    M)}{\delta
    x^{\ell}},
\end{equation}
et comme $(M^{-1}\delta M)$ est linéaire en $\delta M_{ij}$, il est possible d'écrire:
\begin{equation}\label{derivtrace}
    \displaystyle\frac{Tr(M^{-1}\delta M)}{\delta
    x^{\ell}}\simeq Tr\Bigg(M^{-1}\displaystyle\frac{\delta M}{\delta
    x^{\ell}}\Bigg).
\end{equation}
Quand $\delta x^{\ell}\rightarrow 0$, alors:
\begin{equation}\label{derivrondtrace}
    \displaystyle\frac{\partial \ln[Det(M)]}{\partial
    x^{\ell}}\simeq Tr\Bigg(M^{-1}\displaystyle\frac{\partial M}{\partial
    x^{\ell}}\Bigg).
\end{equation}
Dans le cas $M=(g_{ij})$, en considérant l'élément de matrice:
$$\Bigg(M^{-1}\displaystyle\frac{\partial M}{\partial x^{\ell}}\Bigg)^{i}_{j}=
\displaystyle\sum_{k=1}^{n}g^{ik}\;\displaystyle\frac{\partial\, g_{kj}}{\partial
x^{\ell}},$$ nous déduirons:
\begin{eqnarray}
    Tr\Bigg(M^{-1}\displaystyle\frac{\partial M}{\partial
    x^{\ell}}\Bigg)&\equiv& \displaystyle\sum_{i=1}^{n}\Bigg(M^{-1}
    \displaystyle\frac{\partial M}{\partial
    x^{\ell}}\Bigg)^{i}_{i}\nonumber\\
    &=&\displaystyle\sum_{i=1}^{n}\displaystyle\sum_{k=1}^{n}g^{ik}\;
    \displaystyle\frac{\partial\,
g_{ki}}{\partial x^{\ell}}.
\end{eqnarray}
En posant $Det(g_{ij})=g$, on peut écrire, conformément à (\ref{derivrondtrace})
\begin{eqnarray}\label{compar}
    \displaystyle\frac{\partial \ln g}{\partial
    x^{\ell}}&=& \displaystyle\frac{1}{g}\;\displaystyle\frac{\partial g}{\partial
    x^{\ell}}\nonumber\\
    &=& \displaystyle\frac{2}{\sqrt{|g|}}\;\displaystyle\frac{\partial (\sqrt{|g|})}{\partial
    x^{\ell}}\nonumber\\
    &=&\displaystyle\sum_{i=1}^{n}\displaystyle\sum_{k=1}^{n}g^{ik}\;
    \displaystyle\frac{\partial\,
g_{ki}}{\partial x^{\ell}}.
\end{eqnarray}
Finalement, en comparant (\ref{compar}) avec (\ref{compar1}):
\begin{equation}\label{gammaiji}
    \displaystyle\sum_{i=1}^{n}\Gamma_{ij}^{i}=\displaystyle\frac{1}{\sqrt{|g|}}\;
    \displaystyle\frac{\partial(\sqrt{|g|})}{\partial
    x^{\ell}}
\end{equation}

\section{Dérivation covariante}
\subsection{Tenseur dérivée covariante d'un vecteur $A^{i}$} Soit un vecteur
$\overrightarrow{A}\in E$. Dans la base $\{\overrightarrow{e_{i}}\}_{i=1,n}$,
celui-ci se développe
$$\overrightarrow{A}=\displaystyle\sum_{i=1}^{n}A^{i}\;\overrightarrow{e_{i}}.$$
Calculons
\begin{eqnarray}
    d\overrightarrow{A}&=&\displaystyle\sum_{i=1}^{n}(dA^{i}\;\overrightarrow{e_{i}}+
       A^{i}\;d\overrightarrow{e_{i}})\nonumber\\
       &=&\displaystyle\sum_{i=1}^{n}dA^{i}\;\overrightarrow{e_{i}}+
       \displaystyle\sum_{j=1}^{n}
       A^{j}\;d\overrightarrow{e_{j}}\nonumber\\
       &=&\displaystyle\sum_{i=1}^{n}dA^{i}\;\overrightarrow{e_{i}}+
       \displaystyle\sum_{j=1}^{n}A^{j}\Bigg(\displaystyle\sum_{i=1}^{n}
       \displaystyle\sum_{k=1}^{n}\Gamma_{j\,k}^{i}\;dx^{k}
       \;\overrightarrow{e_{i}}\Bigg)\nonumber\\
       &=&\displaystyle\sum_{i=1}^{n}\Bigg(dA^{i}+\displaystyle\sum_{j=1}^{n}
       \displaystyle\sum_{k=1}^{n}A^{j}\;\Gamma_{j\,k}^{i}\;dx^{k}\Bigg)
       \overrightarrow{e_{i}}
\end{eqnarray}
Définissons les composantes contravariantes de $d\overrightarrow{A}$, par: $$DA^{i}= dA^{i}+\displaystyle\sum_{j=1}^{n}
       \displaystyle\sum_{k=1}^{n}A^{j}\;\Gamma_{j\,k}^{i}\;dx^{k},$$
de telle sorte que:
\begin{equation}\label{da}
    d\overrightarrow{A}=\displaystyle\sum_{i=1}^{n}DA^{i}.\;\overrightarrow{e_{i}}.
\end{equation}
Il est possible de réécrire $DA^{i}$ par:
\begin{equation}\label{daicontrav}
   DA^{i}=\displaystyle\sum_{k=1}^{n}\partial_{k}A^{i}\;dx^{k}+
   \displaystyle\sum_{j=1}^{n}
       \displaystyle\sum_{k=1}^{n}A^{j}\;\Gamma_{j\,k}^{i}\;dx^{k},
\end{equation}
En définissant
\begin{equation}\label{dkaicov}
    D_{k}A^{i}=\partial_{k}A^{i}+A^{j}\;\Gamma_{j\,k}^{i},
\end{equation}
Nous pouvons écrire:
$$DA^{i}=\displaystyle\sum_{k=1}^{n} D_{k}A^{i}\;dx^{k}.$$
Il est possible de montrer que les $D_{k}A^{i}$ sont les composantes d'un tenseur
d'ordre 2, appelées: Dérivée covariante de $\overrightarrow{A}$.

\subsection{Tenseur dérivée covariante d'un vecteur $A_{i}$} On va utiliser un
artifice de calcul qui consiste à considérer un champ $\overrightarrow{B}$ uniforme,
tel que $d\overrightarrow{B}=0$. D'après (\ref{da}), on a:
\begin{eqnarray}
     d\overrightarrow{B}&=&\displaystyle\sum_{i=1}^{n}DB^{i}\;\overrightarrow{e_{i}}=
     \overrightarrow{0}\nonumber\\
     &\Rightarrow&\hspace{1cm}DB^{i}=0\hspace{0.5cm}\textrm{pour:}\;\;i=1\rightarrow
     n\nonumber\\
     &\Rightarrow&\hspace{1cm}dB^{i}+\displaystyle\sum_{j=1}^{n}
       \displaystyle\sum_{k=1}^{n}B^{j}\;\Gamma_{j\,k}^{i}\;dx^{k}=0\hspace{0.5cm}
       \textrm{pour:}\;\;i=1
       \rightarrow n\nonumber\\
       &\Rightarrow&\hspace{1cm}dB^{i}=-\displaystyle\sum_{j=1}^{n}
       \displaystyle\sum_{k=1}^{n}B^{j}\;\Gamma_{j\,k}^{i}\;dx^{k}\hspace{0.5cm}
       \textrm{pour:}\;\;i=1
       \rightarrow n
\end{eqnarray}

Soit un vecteur quelconque $\overrightarrow{A}$, alors d'une part
\begin{eqnarray}\label{daicov1}
    \overrightarrow{B}\,.\,d\overrightarrow{A}&=&d(\overrightarrow{B}.\overrightarrow{A})
    -\underbrace{d\overrightarrow{B}}_{=0}.\overrightarrow{A}\nonumber\\
    &=&d\Bigg[\Big(\displaystyle\sum_{i=1}^{n}B^{i}\;\overrightarrow{e_{i}}\Big).
       \Big(\displaystyle\sum_{j=1}^{n}A^{j}\;\overrightarrow{e_{j}}\Big)\Bigg]\nonumber\\
       &=&d\Bigg[\displaystyle\sum_{i=1}^{n}\displaystyle\sum_{j=1}^{n}g_{ij}B^{i}\;A^{j}\Bigg]
       =d\Bigg[\displaystyle\sum_{i=1}^{n}B^{i}\;A_{i}\Bigg]
       \nonumber\\
       &=&\displaystyle\sum_{i=1}^{n}\Big(dA_{i}\;B^{i}+
       A_{i}\;dB^{i}\big)
       =\displaystyle\sum_{i=1}^{n}dA_{i}\;B^{i}+\displaystyle\sum_{j=1}^{n}
       A_{j}\;dB^{j}\nonumber\\
       &=&\displaystyle\sum_{i=1}^{n}dA_{i}\;B^{i}+\displaystyle\sum_{j=1}^{n}
       A_{j}\;\Big(-\displaystyle\sum_{i=1}^{n}
       \displaystyle\sum_{k=1}^{n}B^{i}\;\Gamma_{i\,k}^{j}\;dx^{k}\Big)\nonumber\\
       &=&\displaystyle\sum_{i=1}^{n}B^{i}\big(dA_{i}-\displaystyle\sum_{j=1}^{n}
       \displaystyle\sum_{k=1}^{n}\Gamma_{i\,k}^{j}\;dx^{k}\;
       A_{j}\big),
\end{eqnarray}
et d'autre part:
\begin{eqnarray}\label{daicov2}
   \overrightarrow{B}\,.\,d\overrightarrow{A}&=&\Bigg(\displaystyle\sum_{i=1}^{n}B^{i}\;
   \overrightarrow{e_{i}}\Bigg).\Bigg(\displaystyle\sum_{j=1}^{n}DA^{j}\;\overrightarrow{e_{j}}
   \Bigg)\nonumber\\
   &=&\displaystyle\sum_{i=1}^{n}\displaystyle\sum_{j=1}^{n}B^{i}\;DA^{j}\;g_{ij}
   =\displaystyle\sum_{i=1}^{n}\displaystyle\sum_{j=1}^{n}B^{i}\Bigg(
   \displaystyle\sum_{k=1}^{n}D_{k}A^{j}\;dx^{k}\Bigg)g_{ij}\nonumber\\
   &=&\displaystyle\sum_{i=1}^{n}\displaystyle\sum_{k=1}^{n}B^{i}\Bigg(
   \displaystyle\sum_{j=1}^{n}D_{k}A^{j}\;g_{ij}\Bigg)dx^{k}
   =\displaystyle\sum_{i=1}^{n}\displaystyle\sum_{k=1}^{n}B^{i}\Big(
   D_{k}A_{i}\Big)dx^{k}\nonumber\\
   &=&\displaystyle\sum_{i=1}^{n}B^{i}\Bigg(
   \displaystyle\sum_{k=1}^{n}D_{k}A_{i}\;dx^{k}\Bigg)\nonumber\\
   &=&\displaystyle\sum_{i=1}^{n}B^{i}\;DA_{i},
\end{eqnarray}
où $DA_{i}=\displaystyle\sum_{k=1}^{n}D_{k}A_{i}\;dx^{k}.$
En comparant les équations (\ref{daicov1}) et (\ref{daicov2}), on déduit:
\begin{equation}\label{davor}
    DA_{i}=dA_{i}-\displaystyle\sum_{j=1}^{n}
       \displaystyle\sum_{k=1}^{n}\Gamma_{i\,k}^{j}\;dx^{k}\;
       A_{j},
\end{equation}
ou encore:
\begin{equation}
    DA_{i}=\displaystyle\sum_{k=1}^{n}\Bigg(\partial_{k}A_{i}-\displaystyle\sum_{j=1}^{n}
       \Gamma_{i\,k}^{j}\;
       A_{j}\Bigg)dx^{k},
\end{equation}
ce qui permet de retrouver la relation recherchée:
\begin{equation}\label{daicovx}
  D_{k}A_{i}=\partial_{k}A_{i}-\displaystyle\sum_{j=1}^{n}\Gamma_{i\,k}^{j}\;A_{j}.
\end{equation}
Il est possible de montrer que les $D_{k}A_{i}$ sont les composantes d'un tenseur.

\section{Analyse vectorielle en coordonnées orthogonales}\label{a9}

\subsection{Coordonnées ordinaires}

Soit  un  système  de coordonnées spatiales à 3 dimensions $(x^{1},x^{2},x^{3})$,
caractérisé  par  un tenseur métrique $(g_{ij})$ diagonal, c'est à dire que ses
éléments sont sous la forme:
\begin{eqnarray}\label{gij}
\left\{%
\begin{array}{ll}
    g_{ij}=h_{i}^{2}\,\delta_{ij}\\
    i,j=1,2,3, \\
\end{array}%
\right.
\end{eqnarray}
où $h_{i}$ sont fonction des coordonnées $(x^{1},x^{2},x^{3})$. Les éléments du
tenseur métrique inverse sont alors:
\begin{eqnarray}\label{gijinverse}
\left\{%
\begin{array}{ll}
    g^{ij}=h_{i}^{-2}\,\delta_{ij} \\
    i,j=1,2,3, \\
\end{array}%
\right.
\end{eqnarray}
et la longueur propre invariante s'écrit:
\begin{eqnarray}\label{dl2}
    ds^{2}&=&\displaystyle\sum_{i=1}^{3}\displaystyle\sum_{j=1}^{3}
    g_{ij}\,dx^{i}dx^{j},\nonumber\\
          &=&h_{1}^{2}(dx^{1})^{2}+h_{2}^{2}(dx^{2})^{2}+h_{3}^{2}(dx^{3})^{2}.
\end{eqnarray}
L'élément de volume invariant est
\begin{eqnarray}\label{dv}
    dV&=&\sqrt{Det (g_{ij})}\,dx^{1}dx^{2}dx^{3}\nonumber\\
      &=&(h_{1}h_{2}h_{3})\,dx^{1}dx^{2}dx^{3}
\end{eqnarray}

On définit “les composantes ordinaires” d'un vecteur
$\overrightarrow{V}$ par les grandeurs \cite{weinbergs}
\begin{eqnarray}\label{coordonneesordinaires}
\left\{%
\begin{array}{ll}
    \overline{V_{i}}\equiv h_{i} V^{i}=h_{i}^{-1} V_{i},\\
    i=1,2,3,\\
\end{array}%
\right.
\end{eqnarray}
de telle sorte que le produit scalaire de deux vecteur donne:
\begin{equation}\label{produitscalaire}
    \overrightarrow{V}.\overrightarrow{U}=\displaystyle\sum_{i=1}^{3}
    \overline{V_{i}}.\overline{U_{i}}.
\end{equation}
En effet,
\begin{eqnarray}
    \overrightarrow{V}.\overrightarrow{U}&=&\displaystyle\sum_{i=1}^{3}\displaystyle\sum_{j=1}^{3}
    (V^{i} \overrightarrow{e_{i}}).(U^{j}
    \overrightarrow{e_{j}})=\displaystyle\sum_{i=1}^{3}\displaystyle\sum_{j=1}^{3}
    V^{i}U^{j}(\overrightarrow{e_{i}}.\overrightarrow{e_{j}})\nonumber\\
    &=&\displaystyle\sum_{i=1}^{3}\displaystyle\sum_{j=1}^{3}
    V^{i}U^{j}g_{ij}=\displaystyle\sum_{i=1}^{3}\displaystyle\sum_{j=1}^{3}
    V^{i}U^{j}(h_{i})^{2}\delta_{ij}\nonumber\\
    &=&\displaystyle\sum_{i=1}^{3}V^{i}U^{i}(h_{i})^{2}=\displaystyle\sum_{i=1}^{3}(h_{i}V^{i})(h_{i}U^{i})\nonumber\\
    &=&\displaystyle\sum_{i=1}^{3}\overline{V_{i}}\, \overline{U_{i}}.\nonumber\\
\end{eqnarray}

\subsection{Gradient d'un scalaire}

Les composantes du gradient d'un scalaire $S$ sont définies par \cite{weinbergs}:
\begin{eqnarray}\label{grdientscalcov}
    (\overrightarrow{\nabla}S)_{i}&=&\overline{D_{i}S}\nonumber\\
       &=& h_{i}^{-1}\,D_{i}S,
\end{eqnarray}
$D_{i}$ étant la dérivée covariante. Comme la dérivée covariante d'un scalaire
n'est autre que la dérivée ordinaire, les composantes du gradient du scalaire $S$
s'écrivent finalement:
\begin{eqnarray}\label{gradientscalaire}
\left\{%
\begin{array}{ll}
    (\overrightarrow{\nabla}S)_{i}=h_{i}^{-1}\displaystyle\frac{\partial S}{\partial x^{i}}\\
    i=1,2,3, \\
\end{array}%
\right.
\end{eqnarray}

\subsection{Rotationnel d'un vecteur}
Soit $\overrightarrow{V}$ un vecteur à 3 composantes, alors les composantes du
rotationnel d'un tel vecteur sont \cite{weinbergs}:
\begin{eqnarray}\label{compostrotationnel}
    \left\{%
\begin{array}{ll}
    (\overrightarrow{\nabla}\times \overrightarrow{V})_{1}=\displaystyle
    \frac{h_{1}}{\sqrt{|g|}}\;(D_{2}V_{3}-D_{3}V_{2}),\\
    \\
    (\overrightarrow{\nabla}\times \overrightarrow{V})_{2}=\displaystyle
    \frac{h_{2}}{\sqrt{|g|}}\;(D_{3}V_{1}-D_{1}V_{3}),\\
    \\
    (\overrightarrow{\nabla}\times \overrightarrow{V})_{3}=\displaystyle
    \frac{h_{3}}{\sqrt{|g|}}\;(D_{1}V_{2}-D_{2}V_{1}),\\
\end{array}%
\right.
\end{eqnarray}
où $$g=Det(g_{ij}).$$ D'après (\ref{daicovx}), on peut écrire:
\begin{equation}\label{djvk}
    D_{j}V_{k}=\partial_{j}V_{k}-\displaystyle\sum_{m=1}^{3}\Gamma_{kj}^{m}\,V_{m},
\end{equation}
\begin{equation}\label{dkvj}
    D_{k}V_{j}=\partial_{k}V_{j}-\displaystyle\sum_{m=1}^{3}\Gamma_{j\,k}^{m}\,V_{m}.
\end{equation}
comme $\Gamma_{kj}^{m}=\Gamma_{j\,k}^{m}$, alors une soustraction membre à membre de
(\ref{djvk}) et (\ref{dkvj}) donne
\begin{equation}\label{difderivcov}
     D_{j}V_{k}-D_{k}V_{j}=\partial_{j}V_{k}-\partial_{k}V_{j},
\end{equation}
et la $i^{\textrm{ème}}$ composante du rotationnel de ce vecteur s'écrit
\cite{weinbergs}:
\begin{equation}\label{rotationnel}
    (\overrightarrow{\nabla} \times \overrightarrow{V})_{i}\equiv h_{i}
    \sum_{j=1}^{3}\sum_{k=1}^{3}\,\displaystyle\frac{1}{\sqrt{|g|}}\;\epsilon^{ij\,k}\;\frac{\partial V_{k}}{\partial
    x^{j}},
\end{equation}
où $\epsilon^{ij\,k}$ est le tenseur complètement antisymétrique de Levi-Cevita, avec
la convention $\epsilon^{123}=1$. En utilisant (\ref{coordonneesordinaires}), alors la
définition (\ref{rotationnel}) s'écrit finalement:
\begin{equation}\label{rotationnelbis}
   (\overrightarrow{\nabla} \times \overrightarrow{V})_{i}= h_{i} \sum_{j=1}^{3}\sum_{k=1}^{3}\,
   \displaystyle\frac{1}{\sqrt{|g|}}\;\epsilon^{ij\,k}\;\frac{\partial}{\partial
   x^{j}}(h_{k}\,\overline{V_{k}}\,).
\end{equation}
Les trois composantes s'expriment de façon explicite :
\begin{eqnarray}
    (\overrightarrow{\nabla} \times \overrightarrow{V})_{1}&=& h_{1} \sum_{j=1}^{3}\sum_{k=1}^{3}\,\displaystyle\frac{1}{\sqrt{|g|}}\;\epsilon^{1j\,k}\;\frac{\partial}{\partial
   x^{j}}(h_{k}\,\overline{V_{k}}\,)\nonumber\\
   &=&h_{1}\Bigg[(h_{1}h_{2}h_{3})^{-1}\,\underbrace{\epsilon^{123}}_{=1}\frac{\partial}{\partial
   x^{2}}(h_{3}\overline{V_{3}}\,)+(h_{1}h_{2}h_{3})^{-1}\,\underbrace{\epsilon^{132}}_{=-1}\frac{\partial}{\partial
   x^{3}}(h_{2}\overline{V_{2}}\,)\Bigg]\nonumber\\
   &=&\frac{1}{h_{2}h_{3}}\Bigg[\frac{\partial}{\partial
   x^{2}}(h_{3}\overline{V_{3}})-\frac{\partial}{\partial
   x^{3}}(h_{2}\overline{V_{2}})\Bigg]\nonumber\\
   (\overrightarrow{\nabla} \times \overrightarrow{V})_{2}&=&\frac{1}{h_{3}h_{1}}\Bigg[\frac{\partial}{\partial
   x^{3}}(h_{1}\overline{V_{1}})-\frac{\partial}{\partial
   x^{1}}(h_{3}\overline{V_{3}})\Bigg]\nonumber\\
   (\overrightarrow{\nabla} \times \overrightarrow{V})_{3}&=&\frac{1}{h_{1}h_{2}}\Bigg[\frac{\partial}{\partial
   x^{1}}(h_{2}\overline{V_{2}})-\frac{\partial}{\partial
   x^{2}}(h_{1}\overline{V_{1}})\Bigg]\nonumber
\end{eqnarray}

\subsection{Divergence covariante d'un vecteur}

L'expression de la la divergence covariante d'un vecteur $\overrightarrow{V}$ à trois
composantes est déduite à partir de l'expression de l'opérateur divergence en
coordonnées cartésiennes, en remplaçant les dérivées ordinaires $\partial_{i}$ par des
dérivées covariantes, de sorte que
\begin{equation}\label{divcovremplac}
    \overrightarrow{\nabla}.\overrightarrow{V}=\displaystyle\sum_{i=1}^{3}D_{i}V^{i}.
\end{equation}
En utilisant (\ref{dkaicov}) et (\ref{gammaiji}), on a
\begin{eqnarray}
         \overrightarrow{\nabla}.\overrightarrow{V}&=&\displaystyle\sum_{i=1}^{3}\Bigg[\partial_{i}V^{i}
         +\displaystyle\sum_{j=1}^{3}\Gamma_{ij}^{i}\,V^{j}\Bigg]\nonumber\\
         &=&\displaystyle\sum_{i=1}^{3}\partial_{i}V^{i}+\displaystyle\sum_{j=1}^{3}
         \Bigg[\displaystyle\sum_{i=1}^{3}\Gamma_{ij}^{i}\Bigg]V^{j}\nonumber\\
         &=&\displaystyle\sum_{i=1}^{3}\partial_{i}V^{i}+\displaystyle\sum_{j=1}^{3}
         \Bigg[\displaystyle\frac{1}{\sqrt{|g|}}\;
         \displaystyle\frac{\partial\,\big(\sqrt{|g|}\,\big)}{\partial
         x^{j}}\Bigg]V^{j}\nonumber\\
         &=&\displaystyle\sum_{i=1}^{3}\Bigg[\partial_{i}V^{i}+
         V^{i}\;\displaystyle\frac{1}{\sqrt{|g|}}\;
         \displaystyle\frac{\partial\,\big(\sqrt{|g|}\,\big)}{\partial
         x^{i}}\Bigg]\nonumber\\
         &=&\displaystyle\frac{1}{\sqrt{|g|}}\;\displaystyle\sum_{i=1}^{3}
          \Bigg[\sqrt{|g|}\;(\partial_{i}V^{i})+V^{i}\;\displaystyle\frac{\partial\,
          \big(\sqrt{|g|}\,\big)}{\partial x^{i}}\Bigg]\nonumber\\
          &=&\displaystyle\frac{1}{\sqrt{|g|}}\;\displaystyle\sum_{i=1}^{3}
             \displaystyle\frac{\partial\,
          \big(\sqrt{|g|}\;V^{i}\big)}{\partial x^{i}}.\nonumber
\end{eqnarray}
Finalement, en tenant compte de (\ref{coordonneesordinaires}), on retrouve la
définition \cite{weinbergs}
\begin{equation}
     \overrightarrow{\nabla}.\overrightarrow{V}=\displaystyle\frac{1}{\sqrt{|g|}}\;\displaystyle\sum_{i=1}^{3}
             \displaystyle\frac{\partial}{\partial
             x^{i}}\;\big(\sqrt{|g|}\;h_{i}^{-1}\,\overline{V_{i}}\big)?
\end{equation}
qui s'écrit de manière explicite
\begin{equation}\label{divcovariante}
  \overrightarrow{\nabla}.\overrightarrow{V}=(h_{1}h_{2}h_{3})^{-1}\Bigg[\frac{\partial}{\partial
  x^{1}}(h_{2}h_{3}\,\overline{V_{1}}\,)+\frac{\partial}{\partial
  x^{2}}(h_{1}h_{3}\,\overline{V_{2}}\,)+\frac{\partial}{\partial
  x^{3}}(h_{1}h_{2}\,\overline{V_{3}}\,)\Bigg].
\end{equation}

\subsection{Le laplacien d'un scalaire}

Soit $S$ un scalaire. Alors le laplacien d'un tel scalaire est définit comme étant la
divergence de son gradient. Soit donc:
\begin{eqnarray}
    \overrightarrow{\nabla}^{2}S&=&\overrightarrow{\nabla}.(\overrightarrow{\nabla}
    S)\nonumber
\end{eqnarray}

La première étape consiste à appliquer (\ref{gradientscalaire}):
\begin{eqnarray}
    \overrightarrow{\nabla}^{2}S&=&\overrightarrow{\nabla}.\Bigg[\displaystyle\sum_{i=1}^{3}\overrightarrow{e_{i}}\,(\overrightarrow{\nabla} S)_{i}\Bigg]\nonumber\\
    &=&\overrightarrow{\nabla}.\Bigg[\displaystyle\sum_{i=1}^{3}\overrightarrow{e_{i}}\,h_{i}^{-1}\,\frac{\partial S}{\partial
                      x^{i}}\Bigg]\nonumber\\
    &=&\overrightarrow{\nabla}.\Bigg[\underbrace{\frac{1}{h_{1}}\frac{\partial S}{\partial
    x^{1}}}_{\overline{W_{1}}}\overrightarrow{e_{1}}+\underbrace{\frac{1}{h_{2}}\frac{\partial S}{\partial
    x^{2}}}_{\overline{W_{2}}}\overrightarrow{e_{2}}+\underbrace{\frac{1}{h_{3}}\frac{\partial S}{\partial
    x^{3}}}_{\overline{W_{3}}}\overrightarrow{e_{3}}\Bigg].\nonumber
\end{eqnarray}
La seconde étape consiste à appliquer (\ref{divcovariante}):
\begin{equation}
  \overrightarrow{\nabla}^{2}S=(h_{1}h_{2}h_{3})^{-1}\Bigg[\frac{\partial}{\partial
  x^{1}}(h_{2}h_{3}\,\overline{W_{1}}\,)+\frac{\partial}{\partial
  x^{2}}(h_{1}h_{3}\,\overline{W_{2}}\,)+\frac{\partial}{\partial
  x^{3}}(h_{1}h_{2}\,\overline{W_{3}}\,)\Bigg].\nonumber
\end{equation}
Finalement \cite{weinbergs}:
\begin{equation}\label{laplaciencourbe}
  \overrightarrow{\nabla}^{2}S=(h_{1}h_{2}h_{3})^{-1}\Bigg[\frac{\partial}{\partial
  x^{1}}\Bigg(\frac{h_{2}h_{3}}{h_{1}}\,\frac{\partial S}{\partial
  x^{1}}\Bigg)+\frac{\partial}{\partial
  x^{2}}\Bigg(\frac{h_{1}h_{3}}{h_{2}}\,\frac{\partial S}{\partial
  x^{2}}\Bigg)+\frac{\partial}{\partial
  x^{3}}\Bigg(\frac{h_{1}h_{2}}{h_{3}}\,\frac{\partial S}{\partial
  x^{3}}\Bigg)\Bigg]
\end{equation}

\newpage
\pagestyle{fancy} \lhead{Appendices}\rhead{Fonction d'onde des états stationnaires
d'une particule de spin: 1/2 et...}
\chapter{Fonction d'onde relativiste et stationnaire,
pour une particule
 de spin $1/2$ et de moment cinétique
orbital $\ell$}

Dans le cas non relativiste, une particule de spin $(s)$ peut se trouver dans $(2s+1)$ états
différents, caractérisés par les projections de spin suivantes $s_{z}=-s,-s+1,...,s-1,+s$.
Pour pouvoir rendre compte de tous ces états, la fonction d'onde d'une telle particule
s'écrit sous forme d'un spineur à ($2s+1$) composantes,
\begin{equation}
    \psi^{nr}=\left(%
\begin{array}{c}
  \psi^{nr}_{1} \\
  \vdots \\
  \psi^{nr}_{2s+1}\\
\end{array}%
\right).
 \end{equation}
Cette fonction d'onde peut être regardée, à un instant donné,
comme une fonction de la variable spatiale $\overrightarrow{r}$ et
de la variable de spin $s$, de sorte qu'elle représente un certain
vecteur d'état $|\psi^{nr}(t)\;\rangle$ appartenant à l'espace des
états $\varepsilon$. Un tel espace est formé par le produit
tensoriel suivant:
$$\varepsilon=\varepsilon^{(r)}\otimes\varepsilon^{(s)},$$
où $\varepsilon^{(r)}$ est l'espace des variables spatiales et
$\varepsilon^{(s)}$ est l'espace des variables de spin. La
fonction d'onde s'écrit comme la projection de l'état
$|\psi^{nr}(t)\rangle$ dans une base appropriée,
$$\psi^{nr}(\overrightarrow{r},s,t)\equiv\psi^{nr}_{s}(\overrightarrow{r},t)=\langle \;r\;s|
\psi^{nr}(t)\;\rangle.$$ La densité de probabilité de présence en
un point est définie comme la somme des modules au carrée de
toutes les composantes $\psi^{nr}_{i}$. Elle s'exprime par la
formule:
$$P(\overrightarrow{r},t)=\displaystyle\sum_{i=1}^{2s+1}|\psi^{nr}_{i}|^{2}.$$

Il est clair, ainsi, que la fonction d'onde non relativiste d'une
particule de spin $1/2$ est un spineur à 2 composantes, par contre
dans le cas relativiste, il en est tout autrement. En effet, en
théorie  quantique  relativiste,  la fonction d'onde d'une telle
particule vérifie l'équation (\ref{eqdiracourbe}), dans laquelle
figure des matrices d'ordre $4\times4$. Ceci est une preuve que la
fonction d'onde décrivant la  particule  de  spin  $1/2$,  dans le
cas relativiste, est un spineur à 4 composantes. Dans ce cas, la
fonction d'onde s'écrit sous la forme condensée suivante,
\begin{eqnarray}\label{spineurquatre}
    \psi=\left(%
\begin{array}{c}
  \psi^{1} \\
  \psi^{2} \\
  \psi^{3} \\
  \psi^{4} \\
\end{array}%
\right)=\left(%
\begin{array}{c}
  \varphi \\
  \chi \\
\end{array}%
\right),
\end{eqnarray}
où $\varphi$ et $\chi$ sont deux spineurs à deux composantes, qui
décrivent aussi bien  la particule que l'antiparticule associée.

La dépendance angulaire de l'hamiltonien (\ref{ss}), d'une particule sans spin, est
exactement identique à l'expression du moment cinétique orbital. Il est donc possible
de montrer la commutation des deux opérateurs suivants,
$[H^{2},\overrightarrow{L}^{2}]=0$. Cette circonstance est exploitée dans la
détermination de la partie angulaire de la fonction d'onde d'une particule sans spin
(\ref{sss}). Une telle dépendance est contenue dans des harmoniques sphériques d'ordre
$\ell$, (voir (\ref{sepdevar})). Comme chacune des quatre composantes du spineur
(\ref{spineurquatre}), décrivant la particule de spin $1/2$, vérifie l'équation d'onde
correspondante à une particule de même masse et de spin $0$ (\ref{ss}), alors il est
naturel de conclure que la dépendance angulaire du spineur à quatre composantes est
une certaine combinaison linéaire des harmoniques sphériques. En fait, pour déterminer
la dépendance angulaire du spineur (\ref{spineurquatre}), on aura recours à deux
arguments:
\begin{enumerate}
    \item La conservation du moment orbital permet de montrer que les deux spineurs
    $\varphi$ et $\chi$ sont proportionnels à des spineurs à deux composantes, formés
    à partir d'une certaine combinaison linéaire d'harmoniques sphériques et des vecteurs
    propres des matrices de Pauli. Ces
    spineurs sont dits "spineurs sphériques", et sont notés $\Omega_{j,\ell,m}$.
    La fonction d'onde $\psi$ s'écrit, en fonction de
    ces spineurs sphériques, sous la forme
    \begin{equation}\label{spineurquatrespherique}
          \psi=\left(%
\begin{array}{c}
  \varphi \sim \Omega_{j,\ell,m} \\
  \chi \sim \Omega_{j,\ell^{\;'},m} \\
\end{array}%
\right),
    \end{equation}
    \item La conservation de la parité d'état va permettre de déterminer le lien entre
    les deux spineurs sphériques $\Omega_{j,\ell,m}$ et $\Omega_{j,\ell^{\;'},m}$.
\end{enumerate}

Après avoir complètement déterminé la dépendance angulaire de la solution, par des
considérations de symétrie, il ne reste qu'à déterminer la partie radiale de la
solution.

\section{Spineurs sphériques}\label{zeta}
 En supposant un système physique constitué de deux parties, en faible interaction et
 ayant respectivement des moments cinétiques
 $\overrightarrow{j_{1}}$ et
 $\overrightarrow{j_{2}}$, alors conformément au modèle vectoriel d'addition de deux
 moments cinétiques\footnote{Rigoureusement, en négligeant totalement l'interaction,
 l'énergie totale du système
 est la somme des énergies respectives des deux parties indépendantes. De même, le moment
 cinétique total est
 $\overrightarrow{J}=\overrightarrow{j_{1}}+\overrightarrow{j_{2}}$ est aussi une
 constante de mouvement. Le modèle vectoriel permet de
 coupler de moments cinétiques, en supposant que les carrés des moments cinétiques sont
 invariants.
 Classiquement, les deux moments $\overrightarrow{j_{1}}$
 et $\overrightarrow{j_{2}}$ tournent autour de la direction de $\overrightarrow{J}$ \cite{landauquantique}.}, il est
  possible de déterminer le moment cinétique total, ainsi
  que la fonction d'onde du système.

  En supposant que les deux parties sont décrites, respectivement, par les fonctions
  d'ondes $\psi_{j_{1}\,m_{1}}^{(1)}$, $\psi_{j_{2}\,m_{2}}^{(2)}$, alors la fonction
  d'onde du système total est donnée par l'expression suivante \cite{landauquantique}
\begin{equation}\label{fonctiononde3j}
    \psi_{J\,M}=(-1)^{j_{1}-j_{2}-M}\sqrt{2J+1}\displaystyle\sum_{m_{1}=-j_{1}}^{+j_{1}}
    \sum_{m_{2}=-j_{2}}^{+j_{2}}\left(%
\begin{array}{ccc}
  j_{1} & j_{2} & J \\
  m_{1} & m_{2} & -M \\
\end{array}%
\right)\,\psi_{j_{1}\,m_{1}}^{(1)}\,\psi_{j_{2}\,m_{2}}^{(2)},
\end{equation}
tel que, $$M=m_{1}+m_{2}\hspace{0.5cm} \textrm{et}\hspace{0.5cm} |j_{1}-j_{2}|\leq
J\leq j_{1}+j_{2}.$$

Les coefficients du développement sont les symboles «$3j$» de Wigner; ils sont reliés
aux coefficients de Clebsh-Gordon par la formule suivante \cite{landauquantique}
\begin{equation}\label{clebsh3j}
    < j_{1}\,j_{2}\,J\,M\mid  j_{1}\,m_{1}\,j_{2}\,m_{2}>=(-1)^{j_{1}-j_{2}+M}\sqrt{2J+1}
    \left(%
\begin{array}{ccc}
  j_{1} & j_{2} & J \\
  m_{1} & m_{2} & -M \\
\end{array}%
\right).
\end{equation}
Rappelons que les coefficients de Clebsh-Gordon sont les éléments
de matrice d'une transformation unitaire\footnote{Les
transformations unitaires conservent les normes, donc les
probabilités.}, permettant, lors de l'addition de deux moments
cinétiques $\overrightarrow{j_{1}}$ et $\overrightarrow{j_{2}}$,
de passer de la base couplée $\{\,\mid j_{1}\,j_{2}\,J\,M>\,\}$ à
la base découplée $\{\,\mid
j_{1}\,m_{1}>\otimes\mid j_{2}\,m_{2}>\,\}$ (ou vis versa).\\

Pour ce qui nous concerne, on veut déterminer la fonction d'onde
d'une particule de spin $1/2$ et de moment orbital $\ell$. En
considérant que le couplage spin-orbite (interaction entre le spin
de la particule et son moment orbital) est négligeable, alors
conformément au modèle vectoriel d'addition des moments
cinétiques, le moment cinétique total est $J=\ell\pm 1/2$. Dans ce
cas, en remplaçant dans (\ref{fonctiononde3j}) les symboles «$3j$»
correspondants (ou les coefficients de Clebsh-Gordon), les
fonctions d'ondes (non relativistes) d'un tel système s'expriment
comme suit \cite{landauquantrelativ}:
\begin{eqnarray}\label{spineurspherique1}
    \Omega_{\ell+1/2,\;\ell,\;M}(\theta,\phi)=\sqrt{\displaystyle\frac{J+M}{2J}}\;\;\chi(+1/2)\;\;
    Y_{\ell}^{M-1/2}(\theta,\phi)\nonumber\\
    +\;\displaystyle\sqrt{\displaystyle\frac{J-M}{2J}}\;\;\chi(-1/2)\;\;
    Y_{\ell}^{M+1/2}(\theta,\phi),
\end{eqnarray}

\begin{eqnarray}\label{spineurspherique2}
    \Omega_{\ell-1/2,\;\ell,\;M}(\theta,\phi)=-\;\sqrt
    {\displaystyle\frac{J-M+1}{2J+2}}\;\;\chi(+1/2)\;\;Y_{\ell}^{M-1/2}(\theta,\phi)\nonumber\\
    +\displaystyle\sqrt{\displaystyle\frac{J+M+1}{2J+2}}\;\;\chi(-1/2)\;\;
    Y_{\ell}^{M+1/2}(\theta,\phi),
\end{eqnarray}
où les spineurs
\begin{eqnarray}
    \chi(+1/2)=\left(%
\begin{array}{c}
  1 \\
  0 \\
\end{array}%
\right),\label{xhplus}\\
\chi(-1/2)=\left(%
\begin{array}{c}
  0 \\
  1 \\
\end{array}%
\right),\label{xhmoins}
\end{eqnarray}
décrivent la particule dans les états, respectifs, de spin $+1/2$
et $-1/2$. Ce sont des vecteurs propres des matrices de Pauli,
reliés au spin de la particule par la relation,
\begin{equation}\label{best}
    \overrightarrow{S}=\hbar\;\frac{\overrightarrow{\sigma}}{2}.
\end{equation}
De plus, les fonctions décrivant une particule de moment orbital
$\ell$ sont les harmoniques sphériques. Ils sont définis, dans
notre cas, par la formule \cite{landauquantrelativ}:
\begin{equation}\label{harmoniquespherique}
    Y_{\ell}^{M}(\theta,\phi)=(-1)^{\frac{M+|
    M|}{2}}\;i^{\ell}\;\sqrt{\displaystyle\frac{(2\ell+1)(\ell-|M|)!}{4\pi(\ell+|
    M|)!}}\;\;P_{\ell}^{|M|}(\cos \theta)\;e^{iM\phi}.
\end{equation}
En tenant compte de (\ref{xhplus}), (\ref{xhmoins}) et (\ref{harmoniquespherique}), alors les
 spineurs sphériques pour les deux valeurs de moment cinétique total $J=\ell\pm1/2$
 s'expriment, finalement sous la forme:
 \begin{eqnarray}
     \Omega_{\ell+1/2,\;\ell,\;M}(\theta,\phi)&=&\left(%
\begin{array}{c}
  \sqrt{\displaystyle\frac{J+M}{2J}}\;\;Y_{\ell}^{M-1/2}(\theta,\phi)\\
  \sqrt{\displaystyle\frac{J-M}{2J}}\;\;Y_{\ell}^{M+1/2}(\theta,\phi)\\
\end{array}%
\right),\label{montre1}\\
\Omega_{\ell-1/2,\;\ell,\;M}(\theta,\phi)&=&\left(%
\begin{array}{c}
  -\sqrt
    {\displaystyle\frac{J-M+1}{2J+2}}\;\;Y_{\ell}^{M-1/2}(\theta,\phi)\\
  \sqrt{\displaystyle\frac{J+M+1}{2J+2}}\;\;Y_{\ell}^{M+1/2}(\theta,\phi)\\
\end{array}%
\right).\label{montre2}
\end{eqnarray}

Ces spineurs sphériques sont normalisés par la condition
\cite{landauquantique},
\begin{equation}\label{normalspherique}
    \int_{0}^{\pi}\sin\theta\; d\theta \int_{0}^{2\pi}
    d\phi\; \Omega_{J\,\ell\,M}^{*}\;\Omega_{J^{'}\,\ell^{'}\,M^{'}}=\delta_{J\,J^{'}}\;\delta_{\ell\,\ell^{'}}\;\delta_{M\,M^{'}}
\end{equation}

\section{Parité d'état}
Les lois de conservation permettent de simplifier considérablement
les problèmes physiques. Classiquement, l'homogénéité et
l'isotropie et de l'espace, impliquent respectivement l'invariance
de l'hamiltonien sous une translation et une rotation quelconque
du système. Ceci conduit à la conservation de l'impulsion et du
moment cinétique. De façon plus générale, selon le théorème de
Noether, chaque fois que le lagrangien du système physique est
invariant sous une transformation d'un paramètre continu, alors il
est possible de retrouver une grandeur conservée, associée à cette
transformation continue de symétrie.

En mécanique quantique, il est possible de généraliser le théorème
précédent, pour les transformations continues de symétrie. Il
existe, en plus, des transformations discrètes, sous lesquelles
l'hamiltonien du système reste invariant.

L'inversion de l'espace est une transformation qui se traduit par
le changement simultané des signes de toutes les coordonnées,
$$\overrightarrow{r}\longrightarrow-\overrightarrow{r}.$$
Il est à noter que cette transformation discrète laisse invariant
l'hamiltonien d'un système fermé ou soumis à un champ central
symétrique \cite{landauquantique}, et que classiquement, elle
n'aboutit à aucune lois de conservation. Par contre en mécanique
quantique, la grandeur qui se conserve dans ce cas est la parité
d'état.

Soit $\Pi$ l'opérateur d'inversion spatiale, ou simplement
opérateur parité. Son action sur une fonction quelconque des
coordonnées, particulièrement sur la fonction d'onde, est donnée
par la définition suivante,
\begin{equation}\label{inversionfontonde}
    \Pi\;\psi(\overrightarrow{r_{1}},\overrightarrow{r_{2}},...,\overrightarrow{r_{n}})=
    \psi(\overrightarrow{-r_{1}},\overrightarrow{-r_{2}},...,\overrightarrow{-r_{n}}).
\end{equation}
L'invariance de l'hamiltonien sous l'action de l'opérateur parité $\Pi$ se traduit
par la relation de commutation,
$$[H,\Pi]=0.$$

Les valeurs propres de l'opérateur parité sont déterminées
par le biais de l'équation aux valeurs propres suivante,
$$\Pi\;\psi=\lambda\;\psi.$$
L'application deux fois de l'opérateur $\Pi$ sur une fonction quelconque
équivaut à l'application de la transformation d'identité, ce qui
conduit à deux valeurs propres possibles,
$$\lambda=\pm 1.$$
Donc, les fonctions propres de l'opérateur parité sont, ou bien
inchangées, «état pair», ou bien elles changent de signe, «état
impair». Un système physique, fermé ou soumis à un champ central
symétrique, conserve sa parité (valeur propre) au cours du temps.

Déterminons la parité d'état d'une particule de moment cinétique $\ell$. L'inversion d'espace
se traduit, en coordonnées cartésiennes, par la transformation:
\begin{eqnarray}
\left\{%
\begin{array}{ll}
    x\;\longrightarrow\;-x,\\
    y\;\longrightarrow\;-y,\\
    z\;\longrightarrow\;-z.\\
\end{array}%
\right.
\end{eqnarray}
En coordonnées sphériques, cette transformation s'exprime comme suit:
\begin{eqnarray}\label{paritspherique}
\left\{%
\begin{array}{ll}
    r\;\longrightarrow\;r,\\
    \theta\;\longrightarrow\;\pi-\theta,\\
    \phi\;\longrightarrow\;\pi+\phi.\\
\end{array}%
\right.
\end{eqnarray}
Dans le but de déterminer la parité d'une particule de moment cinétique $\ell$, appliquons
la transformation précédente (\ref{paritspherique}) sur la fonction d'onde décrivant une
telle particule. Celle-ci peut se mettre sous la forme à variables séparées suivante
\begin{equation}\label{zuker}
    \psi(r,\theta,\phi)=G(r)\;Y_{\ell}^{M}(\theta,\phi).
\end{equation}
Comme la partie radiale $G(r)$ est invariante sous parité, alors l'action de
l'opérateur $\Pi$ sur la fonction d'onde $\psi$ se réduit à son action sur
l'harmonique sphérique d'ordre $\ell$. En effet,
$$\Pi\;\Big(G(r)\;Y_{\ell}^{M}(\theta,\phi)\Big)=G(r)\;\Pi\;Y_{\ell}^{M}(\theta,\phi).$$
En tenant compte de (\ref{paritspherique}), alors on peut écrire,
$$\Pi\;Y_{\ell}^{M}(\theta,\phi)=Y_{\ell}^{M}(\pi-\theta,\pi+\phi).$$
Conformément à (\ref{harmoniquespherique}), il est clair
l'harmonique sphérique d'ordre $\ell$ est formée à partir d'un
polynôme de Legendre de 2eme espèce $P_{\ell}^{\mid M\mid}(\cos
\theta)$  et d'une exponentielle $e^{iM\phi}$,
   \begin{equation}\label{harmspher}
      Y_{\ell}^{M}(\theta,\phi)\sim P_{\ell}^{\mid M\mid}(\cos \theta)\;e^{iM\phi}.
  \end{equation}
Chacune des deux parties de l'harmonique sphérique se transforme, d'après
(\ref{paritspherique}), comme suit:
\begin{eqnarray}
    \left\{%
\begin{array}{ll}
  \phi\;\longrightarrow\;\pi+\phi\hspace{0.5cm}\Rightarrow\hspace{0.5cm}
e^{iM\phi}\;\longrightarrow\;(-1)^{M}e^{iM\phi}  ,  \\
  \theta\;\longrightarrow\;\pi-\theta\hspace{0.5cm}\Rightarrow\hspace{0.5cm}
 P_{\ell}^{M}(\cos \theta)\;\longrightarrow\; P_{\ell}^{M}(-\cos \theta)=(-1)^{\ell-M}P_{\ell}^{M}(\cos \theta), \\
\end{array}%
\right.
\end{eqnarray}
de sorte que l'harmonique sphérique se transforme finalement sous l'action de
l'opérateur parité conformément à l'expression suivante,
\begin{equation}\label{zukero}
    \Pi\;Y_{\ell}^{M}(\theta,\phi)=(-1)^{\ell}\;Y_{\ell}^{M}(\theta,\phi).
\end{equation}
En remplaçant (\ref{zukero}) dans (\ref{zuker}), on aboutit à une équation aux valeurs
propres:
$$\Pi\;\psi(r,\theta,\phi)=(-1)^{\ell}\;\psi(r,\theta,\phi).$$
La valeur propre associée à la fonction propre $\psi(r,\theta,\phi)$ est
$(-1)^{\ell}$. C'est la parité de l'état d'une particule de moment cinétique orbital
$\ell$.

\section{Fonction d’onde des états stationnaires d’une particule
de spin $1/2$ et de moment cinétique orbital $\ell$, dans le cas
relativiste}

Rappelons qu’on a établit que la fonction d’onde relativiste $\psi$, d’une particule
de moment cinétique orbital $\ell$ et de spin $1/2$, est un spineur à 4 composantes.
La conservation du moment cinétique nous a permis de conclure que la partie angulaire
d'un tel spineur est contenue dans des spineurs sphériques. Rappelons que $\psi$ est
de la forme,
\begin{equation}\label{deuxspinspher}
          \psi=\left(%
\begin{array}{c}
  \varphi \sim \Omega_{j,\,\ell,\,m} \\
  \chi \sim \Omega_{j,\,\ell^{\;'},m} \\
\end{array}%
\right).
    \end{equation}
Pour définir complètement la dépendance angulaire de ce spineur, il reste à déterminer
le lien qui existe entre les deux spineurs sphériques $\Omega_{j,\,\ell,\,m}$ et
$\Omega_{j,\,\ell^{\;'},m}$. Pour se faire, il faut, dans un premier temps établir la
relation qui existe entre $\ell$ et $\ell^{\;'}$, ensuite montrer que les spineurs
sphériques figurant dans (\ref{deuxspinspher}) sont proportionnels (reliés par
(\ref{yahya})).

\subsection{Relation entre $\ell$ et $\ell^{\;'}$}
Pour déterminer la relation entre les moments cinétiques $\ell$ et
$\ell^{\;'}$, la conservation de la parité de l’état est
exploitée. En effet, en appliquant sur le spineur $\psi$ défini
par (\ref{deuxspinspher}) l’opérateur parité $\Pi$, celui-ci va
agir sur les deux spineurs sphériques. Pour que l’état ait une
parité bien déterminée, il faut que sous l’action de l'opérateur
parité $\Pi$, toutes les composantes soient multipliées par un
même facteur \cite{landauquantrelativ}.

Dans la représentation standard, l’inversion  spatiale  agit sur
les deux composantes $\varphi$ et $\chi$ d'un spineur à 4
composantes, conformément aux relations \cite{landauquantrelativ}
\begin{equation}\label{paritespineurs}
    \Pi:\hspace{1cm}\varphi(\overrightarrow{r})\longrightarrow
    i\varphi(-\overrightarrow{r}),\hspace{1cm}\chi(\overrightarrow{r})\longrightarrow
    -i\chi(-\overrightarrow{r}).\nonumber
\end{equation}
D'autre part, pour pouvoir écrire les équations précédentes sous forme d'équations aux
valeurs propres, on exploite l'action de l'opérateur parité sur la partie angulaire
des spineurs $\varphi$ et $\chi$. La partie angulaire de $\varphi$ est un spineur
sphérique,
\begin{eqnarray}
    \left\{%
\begin{array}{ll}
    \varphi \sim \Omega_{j,\,\ell,\,m},\\
    \ell=j\pm1/2, \\
\end{array}%
\right.
\end{eqnarray}
L'action de l'opérateur parité sur un spineur sphérique se déduit de son action sur
les harmoniques sphériques \cite{landauquantrelativ}. En effet,
\begin{eqnarray}
    \Pi:\hspace{1cm}\Omega_{j,\,\ell,\,m}(\theta,\phi)=\Omega_{j,\,\ell,\,m}
    (\overrightarrow{n})\longrightarrow \Omega_{j,\,\ell,\,m}(\pi-\theta,\pi+\phi)&=&
    \Omega_{j,\,\ell,\,m}(-\overrightarrow{n})\nonumber\\
    &=&(-1)^{\ell}\;\Omega_{j,\,\ell,\,m}
    (\overrightarrow{n}).\nonumber
\end{eqnarray}
Finalement, on peut résumer les étapes précédentes, pour la composante $\varphi$,
comme suit:
\begin{eqnarray}
    \Pi:\hspace{1cm}\varphi(\overrightarrow{r})\longrightarrow
    i\;\varphi(\overrightarrow{-r})&\sim& i\;\Omega_{j,\,\ell,\,m}(-\overrightarrow{n})
    \nonumber\\
    &\sim& i(-1)^{\ell}\;\Omega_{j,\,\ell,\,m}(\overrightarrow{n})\nonumber\\
    &=&i(-1)^{\ell}\;\varphi(\overrightarrow{r}),
\end{eqnarray}
autrement dit,
$$\Pi\;\varphi(\overrightarrow{r})=i(-1)^{\ell}\;\varphi(\overrightarrow{r}).$$

On refait la même procédure pour la composante $\chi$. En effet,
\begin{eqnarray}
    \Pi:\hspace{1cm}\chi(\overrightarrow{r})\longrightarrow
    -i\;\chi(\overrightarrow{-r})&\sim& -i\;\Omega_{j,\,\ell^{\;'},m}(-\overrightarrow{n})
    \nonumber\\
    &\sim& -i(-1)^{\ell^{\;'}}\;\Omega_{j,\,\ell^{\;'},m}(\overrightarrow{n})\nonumber\\
    &=&-i(-1)^{\ell^{\;'}}\chi(\overrightarrow{r}),
\end{eqnarray}
autrement dit,
$$\Pi\;\chi(\overrightarrow{r})=-i(-1)^{\ell^{\;'}}\chi(\overrightarrow{r}).$$
La conservation de la parité d'état de $\psi$ impose que sous l'action de l'opérateur
$\Pi$, les composantes $\varphi$ et $\chi$ soient multipliées par un même facteur,
soit:
 \begin{eqnarray}
   i(-1)^{\ell}=-i(-1)^{\ell^{\;'}}\hspace{1cm}&\Rightarrow&
   \hspace{1cm}e^{i\pi(\ell-\ell^{\;'})}=e^{i\pi}\nonumber\\
   &\Rightarrow&\hspace{1cm}\pi(\ell-\ell^{\;'})=\pi+2k\pi,\;\;\textrm{tq:}\;k=0,\pm1,\pm2,...
\end{eqnarray}
pour $k=0\hspace{1cm}\;\;\;\Rightarrow\hspace{1cm}
\ell-\ell^{\;'}=1$, $\;\;\,$on retrouve le cas: $\ell=j+1/2$ et
    $\ell^{\;'}=j-1/2$.\\
pour $k=-1\hspace{1cm}\Rightarrow\hspace{1cm} \ell-\ell^{\;'}=-1$,
on retrouve le cas: $\ell=j-1/2$ et
    $\ell^{\;'}=j+1/2$.\\

Il est possible de réunir les deux écritures précédentes par la relation condensée,
$$\ell^{\;'}=2j-\ell.$$

\subsection{Relation entre $\Omega_{j,\,\ell,\,m}$ et $\Omega_{j,\,\ell^{\;'},m}$}
\A présent, exprimons le spineur sphérique
$\Omega_{j,\,\ell^{\;'},\,m}$ en fonction de
$\Omega_{j,\,\ell,\,m}$. Montrons qu'ils sont reliés par la
relation,

    \begin{equation}\label{pech}
        \Omega_{j,\,\ell^{\;'},\,m}=i^{\ell-\ell\,^{\;'}}\bigg(\overrightarrow{\sigma}.
        \frac{\overrightarrow{r}}{r}\bigg)\;\Omega_{j,\,\ell,\,m}.
    \end{equation}
Mais avant de le faire, attirons l'attention sur le fait que le
coefficient $i^{\ell-\ell^{\;'}}$ apparaît dans l'équation
précédente du fait qu'on a utilisé une base standard «réelle» du
moment cinétique orbital \cite{messiah},

\begin{eqnarray}
    Y_{\ell}^{m}(\theta,\phi)&=&i^{\ell}\;y_{\ell}^{m}(\theta,\phi), \\
    Y_{\ell^{\;'}}^{m}(\theta,\phi)&=&i^{\ell^{\;'}}\;y_{\ell^{\;'}}^{m}(\theta,\phi),
\end{eqnarray}
tel que \cite{zeep},
 \begin{eqnarray}
   &&y_{\ell}^{m}(\theta,\phi)=(-1)^{m}
   \displaystyle\sqrt{\displaystyle\frac{(2\ell+1)(\ell-m)!}{4\pi\;(\ell+m)!}}
   \;\;P_{\ell}^{m}(\cos \theta)\;e^{i\,m\,\phi},\hspace{1cm}m\geq0,\nonumber\\
   &&y_{\ell}^{m}(\theta,\phi)=
   \displaystyle\sqrt{\displaystyle\frac{(2\ell+1)(\ell-m)!}{4\pi\;(\ell+m)!}}
   \;\;P_{\ell}^{|m|}(\cos \theta)\;e^{i\,m\,\phi},\hspace{2cm}m<0.\nonumber
\end{eqnarray}

\subsubsection{a. Premier cas $\ell=j-1/2,\ell^{\,'}=j+1/2\;\;(\ell^{\,'}=\ell+1)$}

La preuve est basée sur un calcul explicite des deux membres de
l'équation (\ref{pech}) et une utilisation des propriétés des
harmoniques sphériques. En effet, le premier membre s'écrit,
$$\Omega_{j,\,\ell^{\;'},\,m}=\Omega_{\ell^{\;'}-1/2,\,\ell^{\;'},\,m}.\nonumber\\ $$
En utilisant la définition (\ref{montre2}), alors
\begin{eqnarray}
    \Omega_{j,\,\ell^{\;'},\,m}&=&-\;\sqrt{\displaystyle\frac{(\ell^{\;'}-1/2)-m+1}
    {2(\ell^{\;'}-1/2)+2}}\;\;
    Y_{\ell^{\;'}}^{m-1/2}(\theta,\phi)\;\;\chi(+1/2)\nonumber\\
    &&+\;\sqrt{\displaystyle\frac{(\ell^{\;'}-1/2)+m+1}{2(\ell^{\;'}-1/2)+2}}\;\;
    Y_{\ell^{\;'}}^{m+1/2}(\theta,\phi)\;\;\chi(-1/2)\nonumber\\
    &=&-\sqrt{\displaystyle\frac{\ell^{\;'}-m+1/2}{2\ell^{\;'}+1}}\;
    Y_{\ell^{\;'}}^{m-1/2}(\theta,\phi)\;\chi(+1/2)+
    \sqrt{\displaystyle\frac{\ell^{\;'}+m+1/2}{2\ell^{\;'}+1}}\;\;
    Y_{\ell^{\;'}}^{m+1/2}(\theta,\phi)\;\;\chi(-1/2)\nonumber\\
    &=&-\sqrt{\displaystyle\frac{\ell-m+3/2}{2\ell+3}}\;
    Y_{\ell+1}^{m-1/2}(\theta,\phi)\;\;\chi(+1/2)+
    \sqrt{\displaystyle\frac{\ell+m+3/2}{2\ell+3}}\;
    Y_{\ell+1}^{m+1/2}(\theta,\phi)\;\chi(-1/2).\nonumber
\end{eqnarray}
Donc, finalement, le premier membre de l'équation (\ref{pech}) s'écrit sous la forme,
\begin{equation}\label{membre1}
\Omega_{j,\,\ell^{\;'},\,m}=\left(%
\begin{array}{c}
 -\sqrt{\displaystyle\frac{\ell-m+3/2}{2\ell+3}}\;\;Y_{\ell+1}^{m-1/2}(\theta,\phi)\\
 \sqrt{\displaystyle\frac{\ell+m+3/2}{2\ell+3}}\;\;Y_{\ell+1}^{m+1/2}(\theta,\phi)\\
\end{array}%
\right).
\end{equation}
Exprimons, ensuite, le second membre de l'équation (\ref{pech}).
Pour se faire, rappelons que le vecteur unitaire, repéré par les
deux angles $\theta$ et $\phi$, s'exprime sous forme
\begin{equation}\label{vectunitaire}
    \overrightarrow{r}/r=\sin\theta\cos\phi\;\overrightarrow{e_{1}}+
    \sin\theta\sin\phi\;\overrightarrow{e_{2}}+
    \cos\theta\;\overrightarrow{e_{3}}=\displaystyle\sum_{i=1}^{3}n_{i}\;
    \overrightarrow{e_{i}}.
\end{equation}
En utilisant la relation précédente et les définitions des matrices de Pauli
(\ref{matpauli}), il est possible d'exprimer la projection de l'opérateur
$\overrightarrow{\sigma}$ sur la direction repérée par $(\theta,\phi)$. En effet,
\begin{equation}\label{paulispherique}
\bigg(\overrightarrow{\sigma}.\frac{\overrightarrow{r}}{r}\bigg)=
\left(%
\begin{array}{cc}
  n_{3} & n_{1}-in_{2} \\
  n_{1}+in_{2} & -n_{3} \\
\end{array}%
\right)=
\left(%
\begin{array}{cc}
  \cos\theta & \sin\theta\;e^{-i\phi} \\
  \sin\theta\;e^{i\phi} & -\cos\theta \\
\end{array}%
\right).
\end{equation}
pour déterminer l'action de l'opérateur précédent sur un spineur sphérique, il faut
calculer son action sur les spineurs $\chi(\pm1/2)$. En effet,
\begin{eqnarray}
   \bigg(\overrightarrow{\sigma}.\frac{\overrightarrow{r}}{r}\bigg)\;\chi(+1/2)&=&
   \left(%
   \begin{array}{cc}
     \cos\theta & \sin\theta\;e^{-i\phi} \\
     \sin\theta\;e^{i\phi} & -\cos\theta \\
   \end{array}%
   \right)\left(%
          \begin{array}{c}
            1 \\
            0 \\
          \end{array}%
          \right)=
       \left(%
       \begin{array}{c}
         \cos\theta \\
         \sin\theta\;e^{i\phi} \\
       \end{array}%
       \right), \label{projex1} \\
   \bigg(\overrightarrow{\sigma}.\frac{\overrightarrow{r}}{r}\bigg)\;\chi(-1/2)&=&
   \left(%
   \begin{array}{cc}
     \cos\theta & \sin\theta\;e^{-i\phi} \\
     \sin\theta\;e^{i\phi} & -\cos\theta \\
   \end{array}%
   \right)\left(%
          \begin{array}{c}
            0 \\
            1 \\
          \end{array}%
          \right)=
       \left(%
       \begin{array}{c}
         \sin\theta\;e^{-i\phi} \\
         \cos\theta \\
       \end{array}%
       \right). \label{projex2}
\end{eqnarray}
Ainsi, d'après (\ref{montre1}), le second membre de l'équation (\ref{pech}) s'écrit
come suit:

\begin{eqnarray}
&&i^{\ell-\ell^{\;'}}\bigg(\overrightarrow{\sigma}.\frac{\overrightarrow{r}}{r}\bigg)\;
\Omega_{j,\,\ell,\,m}=i^{\ell-\ell^{\;'}}
\bigg(\overrightarrow{\sigma}.\frac{\overrightarrow{r}}{r}\bigg)\;
\Omega_{\ell+1/2,\,\ell,\,m}\nonumber\\
&&\hspace{4cm}=\underbrace{i^{\ell-\ell^{\;'}}}_{(-i)}\;\bigg(\overrightarrow{\sigma}.
\frac{\overrightarrow{r}}{r}\bigg)\;
\Bigg[\sqrt{\displaystyle\frac{\ell+m+1/2}{2\ell+1}}\;\;Y_{\ell}^{m-1/2}(\theta,\phi)\;\;
\chi(+1/2)\nonumber\\
&&\hspace{5cm}+\sqrt{\displaystyle\frac{\ell-m+1/2}{2\ell+1}}\;\;Y_{\ell}^{m+1/2}(\theta,\phi)
\;\;\chi(-1/2)\Bigg].\nonumber\\
\end{eqnarray}
En utilisant (\ref{projex1}) et  (\ref{projex2}), finalement
\begin{eqnarray}\label{membre2}
&&i^{\ell-\ell^{\;'}}\bigg(\overrightarrow{\sigma}.\frac{\overrightarrow{r}}{r}\bigg)\;
\Omega_{j,\,\ell,\,m}=\;\nonumber\\
&&\hspace{0.5cm}
(-i)\left(%
\begin{array}{cc}
  \sqrt{\displaystyle\frac{\ell+m+1/2}{2\ell+1}}\;\cos\theta\;Y_{\ell}^{m-1/2}(\theta,\phi)\;
 +\sqrt{\displaystyle\frac{\ell-m+1/2}{2\ell+1}}\;\sin\theta\;e^{-i\phi}Y_{\ell}^{m+1/2}(\theta,\phi)\\
  \sqrt{\displaystyle\frac{\ell+m+1/2}{2\ell+1}}\;\sin\theta\;e^{i\phi}\;Y_{\ell}^{m+1/2}(\theta,\phi)
 -\sqrt{\displaystyle\frac{\ell-m+1/2}{2\ell+1}}\;\cos\theta\;Y_{\ell}^{m+1/2}(\theta,\phi) \\
\end{array}%
\right).\nonumber\\
\end{eqnarray}
Pour montrer que les deux spineurs (\ref{membre1}) et (\ref{membre2}) sont identiques, on a
recours aux propriétés suivantes des harmoniques sphériques \cite{zeep}:
\begin{eqnarray}
  &&\cos\theta\;\;y_{\ell}^{m}(\theta,\phi)=\sqrt{\displaystyle\frac{(\ell+m+1)(\ell-m+1)}
  {(2\ell+1)(2\ell+3)}}\;\;y_{\ell+1}^{m}(\theta,\phi)\nonumber\\
  &&\hspace{3cm}+\sqrt{\displaystyle\frac{(\ell+m)(\ell-m)}{(2\ell+1)(2\ell-1)}}\;\;
  y_{\ell-1}^{m}(\theta,\phi),
  \label{cosylm}\\
  &&e^{i\phi}\;\;\sin\theta\;y_{\ell}^{m}(\theta,\phi)=
  -\sqrt{\displaystyle\frac{(\ell+m+1)(\ell+m+2)}{(2\ell+1)(2\ell+3)}}\;\;
  y_{\ell+1}^{m+1}(\theta,\phi)\nonumber\\
  &&\hspace{4cm}+\sqrt{\displaystyle\frac{(\ell-m)(\ell-m-2)}{(2\ell+1)(2\ell+3)}}\;\;
  y_{\ell-1}^{m+1}(\theta,\phi),\label{plusiphisinylm}\\
  &&e^{-i\phi}\;\;\sin\theta\;y_{\ell}^{m}(\theta,\phi)=
  \sqrt{\displaystyle\frac{(\ell-m+1)(\ell-m+2)}{(2\ell+1)(2\ell+3)}}\;\;
  y_{\ell+1}^{m-1}(\theta,\phi)\nonumber\\
  &&\hspace{4cm}-\sqrt{\displaystyle\frac{(\ell+m)(\ell+m-1)}{(2\ell+1)(2\ell-1)}}\;\;
  y_{\ell-1}^{m-1}(\theta,\phi).\label{moinsiphisinylm}
\end{eqnarray}
En utilisant la base standard «réelle» du moment cinétique orbital, les expressions précédentes
s'écrivent comme suit:
 \begin{eqnarray}
   &&\cos\theta\;\;Y_{\ell}^{m}(\theta,\phi)=-i\Bigg[\;\sqrt{\displaystyle\frac{(\ell+m+1)
   (\ell-m+1)}
   {(2\ell+1)(2\ell+3)}}\;\;Y_{\ell+1}^{m}(\theta,\phi)\nonumber\\
   &&\hspace{3cm}+\sqrt{\displaystyle\frac{(\ell+m)(\ell-m)}{(2\ell+1)(2\ell-1)}}\;\;
   Y_{\ell-1}^{m}(\theta,\phi)\;\Bigg],
   \label{cosylmr}\\
   &&e^{i\phi}\;\;\sin\theta\;\;Y_{\ell}^{m}(\theta,\phi)=-i\Bigg[\;
   -\sqrt{\displaystyle\frac{(\ell+m+1)(\ell+m+2)}{(2\ell+1)(2\ell+3)}}\;\;
   Y_{\ell+1}^{m+1}(\theta,\phi)\nonumber\\
   &&\hspace{4cm}+\sqrt{\displaystyle\frac{(\ell-m)(\ell-m-2)}{(2\ell+1)(2\ell+3)}}\;\;
   Y_{\ell-1}^{m+1}(\theta,\phi)\;\Bigg],\label{plusiphisinylmr}\\
   &&e^{-i\phi}\;\;\sin\theta\;\;Y_{\ell}^{m}(\theta,\phi)=-i\Bigg[\;
   \sqrt{\displaystyle\frac{(\ell-m+1)(\ell-m+2)}{(2\ell+1)(2\ell+3)}}\;\;
   Y_{\ell+1}^{m-1}(\theta,\phi)\nonumber\\
   &&\hspace{4cm}-\sqrt{\displaystyle\frac{(\ell+m)(\ell+m-1)}{(2\ell+1)(2\ell-1)}}\;\;
   Y_{\ell-1}^{m-1}(\theta,\phi)\;\Bigg],\label{moinsiphisinylmr}
 \end{eqnarray}
\newpage
Conformément à (\ref{cosylmr}) et (\ref{moinsiphisinylmr}), la première composantes
du spineur (\ref{membre2}) s'écrit comme suit:
\begin{eqnarray}
&&(-i)\;\Bigg[\;\sqrt{\displaystyle\frac{\ell+m+1/2}{2\ell+1}}\;\cos\theta\;Y_{\ell}^{m-1/2}(\theta,\phi)\;
+\sqrt{\displaystyle\frac{\ell-m+1/2}{2\ell+1}}\;\sin\theta\;e^{-i\phi}
Y_{\ell}^{m+1/2}(\theta,\phi)\;\Bigg]\nonumber\\
&&\hspace{1cm}=\underbrace{(-i)\times(-i)}_{-1}\;\sqrt{\displaystyle\frac{\ell+m+1/2}{2\ell+1}}\;\Bigg[\;
\sqrt{\displaystyle\frac{(\ell+m+1/2)(\ell-m+3/2)}{(2\ell+1)(2\ell+3)}}\;
Y_{\ell+1}^{m-1/2}(\theta,\phi)\nonumber\\
&&\hspace{3cm}+\sqrt{\displaystyle\frac{(\ell+m-1/2)(\ell-m+1/2)}
{(2\ell+1)(2\ell-1)}}\;Y_{\ell-1}^{m-1/2}(\theta,\phi)\;\Bigg]\nonumber\\
&&\hspace{3cm}+\sqrt{\displaystyle\frac{\ell-m+1/2}{2\ell+1}}\;\Bigg[\;
\sqrt{\displaystyle\frac{(\ell-m+1/2)(\ell-m+3/2)}{(2\ell+1)(2\ell+3)}}\;
Y_{\ell+1}^{m-1/2}(\theta,\phi)\nonumber\\
&&\hspace{3cm}-\sqrt{\displaystyle\frac{(\ell+m+1/2)(\ell+m-1/2)}{(2\ell+1)(2\ell-1)}}\;
Y_{\ell-1}^{m-1/2}(\theta,\phi)\;\Bigg]\nonumber\\
&&\hspace{1cm}=-\;\Bigg\{\;\sqrt{
\displaystyle\frac{\ell-m+3/2}{2\ell+3}}\;\bigg[\;\displaystyle\frac{(\ell+m+1/2)+(\ell-m+1/2)}
{2\ell+1}\;\bigg]\;\Bigg\}\;Y_{\ell+1}^{m-1/2}(\theta,\phi)\nonumber\\
&&\hspace{1cm}=-\;\sqrt{\displaystyle\frac{\ell-m+3/2}{2\ell+3}}\;Y_{\ell+1}^{m-1/2}(\theta,\phi),
\end{eqnarray}
il est clair qu'une telle composante est identique à la première composante du
spineur (\ref{membre1}).

La prochaine étape consiste à refaire la même procédure pour comparer les deuxièmes
composantes des spineurs (\ref{membre1}) et (\ref{membre2}). En effet, conformément à
(\ref{plusiphisinylmr}) et (\ref{cosylmr}), la deuxième composante du spineur
(\ref{membre2}) s'écrit comme suit:
\newpage
\begin{eqnarray}
&&(-i)\;\Bigg[\;\sqrt{\displaystyle\frac{\ell+m+1/2}{2\ell+1}}\;\sin\theta\;e^{i\phi}\;Y_{\ell}^{m-1/2}(\theta,\phi)\;
-\sqrt{\displaystyle\frac{\ell-m+1/2}{2\ell+1}}\;\cos\theta\;
Y_{\ell}^{m+1/2}(\theta,\phi)\;\Bigg]\nonumber\\
&&\hspace{1cm}=\underbrace{(-i)\times(-i)}_{-1}\;\sqrt{\displaystyle\frac{\ell+m+1/2}{2\ell+1}}\;\Bigg[\;
-\sqrt{\displaystyle\frac{(\ell+m+1/2)(\ell+m+3/2)}{(2\ell+1)(2\ell+3)}}\;
Y_{\ell+1}^{m+1/2}(\theta,\phi)\nonumber\\
&&\hspace{3cm}+\sqrt{\displaystyle\frac{(\ell-m+1/2)(\ell-m-1/2)}
{(2\ell+1)(2\ell+3)}}\;Y_{\ell-1}^{m+1/2}(\theta,\phi)\;\Bigg]\nonumber\\
&&\hspace{3cm}-\sqrt{\displaystyle\frac{\ell-m+1/2}{2\ell+1}}\;\Bigg[\;
\sqrt{\displaystyle\frac{(\ell+m+3/2)(\ell-m+1/2)}{(2\ell+1)(2\ell+3)}}\;
Y_{\ell+1}^{m+1/2}(\theta,\phi)\nonumber\\
&&\hspace{3cm}+\sqrt{\displaystyle\frac{(\ell+m+1/2)(\ell-m-1/2)}{(2\ell+1)(2\ell-1)}}\;
Y_{\ell-1}^{m+1/2}(\theta,\phi)\;\Bigg]\nonumber\\
&&\hspace{1cm}=-\;\Bigg\{\;-\sqrt{
\displaystyle\frac{\ell+m+3/2}{2\ell+3}}\;\bigg[\;\displaystyle\frac{(\ell+m+1/2)+(\ell-m+1/2)}
{2\ell+1}\;\bigg]\;\Bigg\}\;Y_{\ell+1}^{m+1/2}(\theta,\phi)\nonumber\\
&&\hspace{1cm}=\sqrt{\displaystyle\frac{\ell+m+3/2}{2\ell+3}}\;Y_{\ell+1}^{m+1/2}(\theta,\phi),
\end{eqnarray}
De même, il est clair qu'une telle composante est identique à la deuxième composante
du spineur (\ref{membre1}). On est en mesure de conclure que dans le cas où
$\ell=j-1/2,\ell^{\,'}=j+1/2$, les deux spineurs $\Omega_{j,\,\ell,\,m}$ et
$\Omega_{j,\,\ell^{\;'},\,m}$ sont reliés par la formule,
$$\Omega_{j,\,\ell^{\;'},\,m}=i^{\ell-\ell^{\;'}}\;
\bigg(\overrightarrow{\sigma}.\frac{\overrightarrow{r}}{r}\bigg)\;
\Big(\Omega_{j,\,\ell,\,m}\Big)$$

\subsubsection{b. Deuxième cas $\ell=j+1/2,\ell^{\,'}=j-1/2\;\;(\ell^{\,'}=\ell-1)$}

Des calculs similaires à ceux effectués dans le cas précédent
permettent de vérifier aussi la relation (\ref{pech}), ce qui
achève la démonstration.\\
\\

Après avoir complètement déterminé la partie angulaire de la
solution, il ne reste que l’évaluation de la partie radiale (Voir
(chapitre \ref{chap4})). La solution peut s’écrire sous la forme
suivante \cite{landauquantrelativ},
\begin{equation}
    \psi=\Bigg(
\begin{array}{c}
\varphi\\
\chi\\
\end{array}
\Bigg)=\Bigg(
\begin{array}{c}
f(r)\,\Omega_{j\,\ell\,m}\\
(-1)^{\frac{1+\ell-\ell\,^{'}}{2}}g(r)\,\Omega_{j\,\ell\,^{'}\,m}\\
\end{array}
\Bigg),
\end{equation}
tel que $f(r)$ et $g(r)$ sont deux fonctions radiales à
déterminer.

\newpage
\pagestyle{fancy}\lhead{Bibliographie}\rhead{\;}

\end{document}